\documentclass[12pt, journal=pb]{iopart}

\usepackage{graphicx}

\expandafter\let\csname equation*\endcsname\relax
\expandafter\let\csname endequation*\endcsname\relax 

\usepackage{color}

\usepackage{subfigure}
\usepackage{amsmath}
\usepackage{fullpage}

\begin{document}

\title[Motility of radial centrosomal microtubule arrays]{Collective behavior of minus-ended motors in mitotic microtubule asters gliding towards DNA}

\author{Chaitanya A. Athale$^{1,2}$,  Ana Dinarina$^2$\footnote{Present address: Thermo Fisher Scientific, Germany.}, Francois Nedelec$^2$ and Eric Karsenti$^2$}

\address{$^1$ Div. of Biology, IISER Pune, Sai Trinity, Sutarwadi Road, Pashan, Pune 411021, India.}
\address{$^2$ Cell Biology and Biophysics Div., EMBL, Meyerhofstrasse 1, D-69117, Heidelberg, Germany.}
\ead{cathale@iiserpune.ac.in}

\maketitle

\submitto{Physical Biology}

\begin{abstract}
Microtubules (MTs) nucleated by centrosomes form star-shaped structures referred to as asters. Aster motility and dynamics is vital for genome stability, cell division, polarization and differentiation. Asters move either towards the cell center or away from it. Here, we focus on the centering mechanism in a membrane independent system of {\it Xenopus} cytoplasmic egg extracts. Using live microscopy and single particle tracking, we find that asters move towards chromatinized DNA structures. The velocity and directionality profiles suggest a random walk with drift directed towards DNA. We have developed a theoretical model that can explain this movement as a result of a gradient of MT length dynamics and MT gliding on immobilized dynein motors. In simulations, the antagonistic action of the motor species on the radial array of MTs leads to a tug-of-war purely due to geometric considerations and aster motility resembles a directed random-walk. Additionally our model predicts that aster velocities do not change greatly with varying initial distance from DNA. The movement of asymmetric asters becomes increasingly super-diffusive with increasing motor density, but for symmetric asters it becomes less super-diffusive. The transition of symmetric asters from superdiffusive to diffusive mobility is the result of number fluctuations in bound motors in the tug-of-war. Overall, our model is in good agreement with experimental data in {\it Xenopus} cytoplasmic extracts and predicts novel features of the collective effects of motor-MT interactions.
\end{abstract}

\section{Introduction}
Centrosomes nucleating microtubules (MTs) form an aster, with MT minus-ends inside and plus-ends pointing outwards. Smaller asters are also formed in the absence of centrosomes by MT-organizing centers (MTOCs) in plant and animal cells undergoing meiosis. These structures are responsible for a wide range of functions, such as nuclear positioning, fertilization, cell-division and polarization. The positioning of centrosomal asters has been shown to result from pushing by MT polymerization \cite{dogterom2005}, pulling by depolymerization \cite{Grishchuk2005, Tanaka2007} and pulling by dynein motors in the cytoplasm \cite{Kimura2005} and on membranes \cite{Grill2005, Burakov2003}. In live cells, asters typically move either towards the center of the cell or away from it \cite{Mitchison2012}. The centering movement occurs during fertilization, when male pronuclei are moved by centrosomal asters towards the center of the egg by cytoplasmic pulling forces acting on the longest microtubules, as seen in {\it Clypeaster japonicus} \cite{Hamaguchi1986} and {\it Caenorhabditis elegans} \cite{Kimura2005}. Similar centering movement of centrosomes towards DNA was seen during spindle assembly in meiotic {\it Xenopus} egg-extracts \cite{Carazo-Salas2003}. Even in the absence of centrosomes, smaller self-organized MTOCs show directional movement towards DNA in mouse oocyte meiotic spindle assembly \cite{schuh2007}. Thus, investigating this centering movement is important for a complete understanding of fertilization and meiosis. 

The mechanical origin of the pulling forces has been identified in {\it C. elegans} as dynein localized on intracellular organelles \cite{Kimura2007, Kimura2010}. Early experiments with {\it Xenopus} egg extracts have also reported dynein dependent movement of microtubules on coverslip surfaces \cite{Allan1991, Verde1991}. Most recently {\it in vitro} work showed that a combination of cortical end-on and surface immobilized dynein motors can center centrosomal asters \cite{Laan2012}. However, the inability to distinguish between membrane and cytoplasmic effects and the lack of control over DNA amount and geometry make it difficult to understanding the physical nature of the movement.

Theoretical work based on {\it C. elegans} centrosomal movement has validated the hypothesis that pulling forces were necessary for the centering movement of asters \cite{Kimura2005, Kimura2007}. However, the study ignored  collective motor effects. A subsequent study using anti-paralllel microtubule doublets gliding on multiple kinesin molecules, demonstrated that geometrical considerations result in a tug-of-war caused by multiple motors of the same kind (i.e. single-motor species) \cite{Leduc2010}. A motor model with force-dependent detachment and MT length asymmetries in the doublet can resolve this tug-of-war. However, for a complete picture MT length fluctuations characteristic of mitotic spindle assembly \textit{i.e.} the dynamic instability of MTs \cite{Mitchison1984}, also needs to be included. Additionally, MT dynamic instability has been shown to be biochemically regulated \cite{Verde1991} and show spatial inhomogeneity in mitotic \cite{Athale2008, Kalab2006, Wollman2005} and interphasic cells \cite{Niethammer2004, Wuhr2009}. Thus a theoretical model that includes the physical and biochemical aspects of centrosome centering motility will provide a more physically realistic understanding of the process.

Here, we quantify the centering movement of centrosomal asters towards micro-patterned DNA using {\it Xenopus} egg extracts in experiment. We further develop a theoretical model that combines centrosomal MT dynamics with asters gliding on multiple dynein motors. As a result of the anti-parallel radial orientation of MTs in asters, multiple dynein motors generate antagonistic forces resulting in a 2D single-motor species tug of war for a symmetric aster. A gradient aster asymmetry originating from the DNA resolves the tug-of-war and causes aster to move towards DNA. The movement statistics of the model qualitatively agree with experimental data and predict novel features of the system.

\section{Experimental and data analysis methods}

\subsection{Cytoplasmic Extract and DNA Micropattern Preparation} 
Meiosis II arrested {\it Xenopus} egg cytoplasmic extract was prepared as described before  \cite{Murray1989, Hannak2006} and flowed into a PDMS chamber with an inlet and outlet (Figure \ref{fig:pdms3D}). The chamber was sealed at the bottom with a cover glass with regularly spaced circular patterns of BSA-biotin $30 \mu m$ in diameter (Figure \ref{fig:pdmsxy}). This was used to immobilize bis-biotinylated DNA sandwiched between two layers of paramagnetic beads (DynaBeads, Invitrogen, USA) (Figure \ref{fig:pdmsyz}) as described previously \cite{Athale2008, Dinarina2009}. The presence of DNA on the patterns was confirmed by staining with Hoechst (Sigma, Germany) (Figure \ref{sfig:dnaPatt}). The bead attached chambers were pre-incubated with {\it Xenopus} egg extract to allow the loading of chromatin proteins. Anti-TPX2 antibodies were added to block chromatin-based nucleation of MTs \cite{Athale2008}. Fresh extract containing Cy3-labelled tubulin, anti-Tpx2 antibodies and centrosomes were flowed into the chamber and observed under the microscope.

\subsection{Microscopy and Image Analysis}
Samples were imaged using a Zeiss LSM 5 Live (Carl Zeiss, Jena, Germany) confocal fluorescence microscope. At the start of the experiment still images of the DNA pattern were acquired in the Hoechst channel (Figure \ref{sfig:dnaPatt}). This was followed by time-lapse imaging of Cy3-tubulin every 117 s over the same tiled area (Video \ref{sup:asterdna}). Image acquisition was stopped when most of the asters had coalesced into indistinguishable masses on the chromatin (Video \ref{sup:asterdna}) after $\sim 1$h of the start of the reaction. The acquired time series were divided into regions of interest (ROI) containing asters in the Cy3-tubulin channel with a nearby DNA pattern. These time-series of MT asters were tracked using a  home built single-particle tracking (SPT) program that filters the raw data and detects bright points based on an optimized intensity threshold. The bright points correspond to the centers of mass of the asters (Figure \ref{fig:asterSnap}). Aster centers were connected in time (tracking) based on threshold distances between detected asters in subsequent time frames. The coordinates of circular DNA patterns were automatically extracted by image thresholding and object detection. The image analysis routines were developed in MATLAB (Mathworks Inc., USA), and are available upon request from the authors.

\section{Theory}

\subsection{Model geometry}
The 2D model developed in this study treats the cytoplasmic extract as an aqueous medium with an effective viscosity similar to that measured in {\it Xenopus} egg extract cytoplasm  ($\eta = 0.05  pN\cdot s/\mu m^{2}=0.05 Pa\cdot s$) \cite{valentine2005, Nedelec2007}. The DNA pattern is modeled as a static circle of radius 15 $\mu m$. This DNA circle is the source of a radially symmetric spatial gradient of dynamic instability. 

\subsection{Centrosomal Microtubule Dynamics Model} 
In our model, microtubule dynamics is treated as a spatial variant of the four-parameter and two-state model of dynamic instability \cite{Mitchison1984, Verde1992}. Rapid transitions between growth and shrinkage are described by the frequencies of catastrophe ($f_{c}$: growth to shrinkage) and rescue ($f_{r}$: shrinkage to growth), and velocities of growth ($v_g$) and shrinkage ($v_s$) \cite{Verde1992}. These parameters affect only the plus-tips of the MT filaments. Initially,  all MTs are in a growing state, at a rate defined by the velocity of growth ($v_g$). When an MT shrinks, it shrinks at the rate given by the velocity of shrinkage ($v_s$). Those microtubules that reach length of zero are instantaneously re-nucleated at the next iteration step. The number of centrosomal nucleation sites is finite and constant (100) based on typical values from vertebrate centrosomes \cite{evans1985}. The frequency of catastrophe ($f_c$) and rescue ($f_r$) were modeled to be spatially inhomogeneous based on previous work that modeled the RanGTP dependent zone of `stabilization' around chromosomes \cite{Carazo-Salas2003, Wollman2005}. Within the radius of the `stabilization' zone ($r_c$), and around the chromatin of radius ($r_n$), the values of dynamic instability enhance MT growth ($f_c^{min}$, $f_r^{max}$), while outside it they attain cytoplasmic values  ($f_c^{max}$, $f_r^{min}$) \cite{Athale2008, channels2008spatregul, Wollman2005}. The simulation box is made large enough that no MTs leave it (Table \ref{tab:simpar}), and at 1/3 box-distance a  boundary is set such that microtubule catastrophes become instantaneous resulting in MTs being inside the box even in the absence of a confining boundary. Thus, the spatial frequencies of dynamic instabilitiy as a function of the 2D radial distance from chromatin surface ($r$) are described by:\\
\begin{eqnarray}
    f_r(r) = f_r^{min} +  \Delta f_r  \cdot \frac{e^{(d-r)/s}}{1+e^{(d-r)/s}}      \label{eq:gradres}\\
 f_c(r) = f_c^{max} +  \Delta f_c  \cdot \Big(1-\frac{e^{(d-r)/s}}{1+e^{(d-r)/s}} \Big) \label{eq:gradcat}
\end{eqnarray}
where $d= r_c + r_n$, the value $s$ is a steepness factor, $\Delta f_c = f_c^{max} - f_c^{min}$ and $\Delta f_r = f_r^{max} - f_r^{min}$. 
Experimental values were used to fit $r_n=15 \mu m$, $r_c=20 \mu m$, $s=10 \mu m$, $f_c$ and $f_r$ (Table \ref{tab:mtdynpar}, \cite{Wilde2001, Caudron2005}) producing an effective gradient as shown in Figure \ref{fig:simgrad}. 

\subsection{Molecular Motor Model:}  A cytoskeleton simulation tool that has been previously described \cite{Nedelec2007} was used to simulate the movement of microtubules. The simulation uses overdamped Langevin equations to represent the motion of elastic fibres in a viscous medium. Forces are generated due to Brownian motion and motor mechanics. Motor-MT interactions are modeled as Hookean springs, where the force exerted by a motor on a point of the fiber is given by $\vec{f_{ex}} = -k*\vec{\delta r}$, where $k$ is the stiffness and $\vec{\delta r}$ is the motor stretch, projected on the fibre. The vector $\vec{\delta r}$ is directed between the position of attachment of the motor base and the end of the projected stretch. Motors were characterized by their positions, attachment ($r_{attach}$) and detachment rates ($r_{detach}$), MT-tip detachment rates ($k_{detach}^{tip}$), speed of movement ($v_{mot}$), probability of stepping, direction of movement, magnitude of the stall force ($f_{0}$) and stiffness ($k$) of the induced links. Single-molecule experiments with vertebrate dynein have shown that opposing forces induce increasing rates of detachment \cite{Kunwar2011, Driver2011}, reduction in step-sizes (gear-like stepping) \cite{Mallik2004} and can even induce backward stepping behavior \cite{Gennerich2007}. We have modeled detachment rates to be dependent on the magnitude of the parallel force generated ($\vec{f_{ex}}$) based on Kramers theory \cite{Kramers1940}  and used previously in motor-microtubule models \cite{Leduc2010, channels2008spatregul, Foethke2009, mahajan2012}:

\begin{equation}
r_{detach}=r^{'}_{detach} \cdot \e^{(|f_{ex}|/f_{0})}
\label{eq:rdetach}
\end{equation}

Using the basal detachment rate $r^{'}_{detach}=1.5 s^{-1}$  and $f_0 = 1.7$ pN, when $f_{ex}$ increases to three-fold of $f_0$, the effective detachment rate increases by an order of magnitude. These parameters as well as motor speed ($v_{mot}$), stiffness (k), the maximal distance for motor-MT attachment ($d_{attach}$) and motor-MT attachment rate ($r_{attach}$) are chosen from experimentally measured values where possible (Table \ref{tab:motpar}). If MTs and motors are closer than $d_{attach}$, a random choice to attach is made by comparing to the probability of binding ($r_{attach}\cdot \delta t$, where $\delta t$ is the simulation time step). Hence, motor movement was modeled as a probabilistic process where the motor may either stay immobile, detach or take a step forwards or backwards \cite{Foethke2009}. When the motor is load free, the probability of forwards movement per unit time is $p_{forw} = 0.9$. In order to model the observed gear-like behaviour of dynein \cite{Mallik2004gear} we use a piece-wise linear approximation for the probability of forward stepping, while maintaining a fixed step size. The first ($f_1=5 pN$) and second ($f_2 = 8 pN$) force thresholds are compared to the parallel force $f_{ex}$ to calculate the probability of forward movement as follows:
\begin{equation}
    p_{forw} =
    \begin{cases}
	0.9 & \text{if } f_{ex} < f_1\\
	\e^{(-a\cdot (f_{ex} - f_{1}) )} - p_{back}  & \text{if } f_1\leq f_{ex} < f_2\\
	p_{back} & \text{if } f_{ex} \geq f_2.
      \end{cases}
	\label{eq:motorgear}
\end{equation}
where the exponent $a=\big[ \ln(100\cdot p_{forw}/p_{back} +1)\big]/(f_2-f_1)$ and $p_{back}=0.1$. The model of motor behavior when $f_{ex} \geq f_2$ is based on observation of backward stepping behavior of dynein in response to super-stall forces  \cite{Gennerich2007}.

\subsection{Simulation setup}
Simulations were performed with the parameters and box sizes described in Table \ref{tab:simpar}. Centrosomes with 100 nucleation sites (typical for vertebrate centrosomes \cite{evans1985}) were initialized at a specific distance away from the chromatin pattern. Microtubules were initially of uniform length (5 $\mu m$) and in a growing state. The microtubules fluctuate in length according to the dynamic instability model and motor binding occurs if an MT filament is within the search radius of a single motor and satisfies the rate of attachment. The simulation was performed for 60 min for independent non-interacting asters. The numbers reported are averages over all asters under the same initial conditions and parameters.

\section{Data Analysis}\label{sec:analysis}
The aster trajectories were analyzed using the following measures:\\

{\it Tortuosity:} The ratio of the net displacement ($d_{net}$) to path-length ($L$) for a given trajectory is the tortuosity $\chi=d_{net} / L$. The value of tortuosity $\chi$ is nearly 1 for a straight (directed) movement and it approaches 0 for a random walk.\\

{\it Directional motility coefficient (dmc):} The tendency of the aster to move towards the ÔtargetÕ DNA was quantified by $dmc=\Delta d_c / d_{net} $ where $\Delta d_c$ is the change in distance from chromatin from the start of the motion to its end. The value of dmc can vary between -1 to +1. Negative values indicate movement away from the chromatin, 0 can arise from rotation around the chromosome, a lack of movement or completely random movement, and 1 if it is moving directly towards the target.\\

{\it Trajectory cos($\theta$):} The angle made by the points target (DNA), origin of the trajectory ($t=0$) and its end ($t=N\cdot\delta t$) where t is the time, $\delta t$ is the time between each step and $N$ is the number of steps. If the angle is  large ($\theta > \pi/2$),  movement is random and ranges from $cos(\theta)$ from -1 to 0, but if the angle is small $cos(\theta)$ ranges from 0 to 1 and the movement is directed towards the target.\\

{\it Directed random-walk simulation:} In order to test the measures of directionality in a simple directed random-walk scenario, a 2D random-walk simulation was developed. The particles were initialized a certain distance away from the origin in the first-quadrant at an angle of $\pi/4$ and at each time step they undergo a random walk. The simulation uses radial coordinates where the new position of the particles is determined by $r$ and $\theta_{net}$. The mean radial step is a random number sampled from a normal distribution with mean $\langle r^2 \rangle = 4Dt$ where $D=1 \mu m^2/s$ is the diffusion coefficient and $t$ the time step. The angle of movement ($\theta_{net}$) is the circular mean\cite{fisher1995statistical} given by:
\begin{equation}
	\theta_{net} = \tan^{-1}\Big( \frac{W_d\cdot \sin(\theta_d) + (1-W_d)\cdot \sin(\theta_r) }{W_d\cdot \cos{\theta_d} + (1-W_d)\cdot \cos(\theta_r )} \Big)
\end{equation}
where $W_d$ is the weight of directionality and $\tan^{-1}$ is calculated from the numerical function {\it arctan2} (Mathworks Inc., USA) to return both the angle and the quadrant (so $\theta_{net}$ is between 0 and $2\pi$). The random angle $\theta_{r}$ is a uniform random number between 0 and $2\pi$ and the constant directed angle $\theta_{d}$ was fixed as $\pi/4$. The value of $W_d$ was varied between 0 (random) and 1 (directed) in steps of 0.1. The simulation was performed with 100 particles for a total of 100 s (time step 1 s). The resultant time-projected 2D tracks for increasing weight of directionality ($W_d$) show a transition from random to ballistic motion (Figure \ref{sup:randwk}\subref{sfig:rwplt}). The velocity of the simulated particles is variable at low $W_d$ and becomes increasingly uniform as $W_d$ increases, but the population mean remains constant (Figure \ref{sup:randwk}\subref{sfig:vel}). For such tracks moving at an average constant velocity, we compare $\chi$ (Figure \ref{sup:randwk}\subref{sfig:chi}), dmc (Figure \ref{sup:randwk}\subref{sfig:dmc}) and $\cos(\theta)$ (Figure \ref{sup:randwk}\subref{sfig:costheta}) as directionality measures.\\

{\it Mean square displacement to quantify effective diffusion:} The simulated aster movement is evaluated using the expression for the local mean square displacement ($\langle \Delta r^2 \rangle$, MSD) reported previously \cite{Arcizet2008, Otten2012}:

 \begin{equation}
	<\Delta r^2> = \langle \big[ r(t) - r(t  + \delta t)\big]^{2}  \rangle  = 4\cdot D_{eff} \cdot \delta t^{\alpha}
\label{eq:msd}
\end{equation}
where the displacement between position vectors $r$ at times $t$ and $(t+\delta t)$ are averaged over time intervals not exceeding 3/4th of the simulated time to avoid artifacts \cite{Qian1991, Michalet2010}. The MSD as a function of time is fit to the right hand side of Equation \ref{eq:msd}, where $D_{eff}$ is the effective diffusion coefficient and $\alpha$ is the coefficient of anomaly. The movement is considered to be effectively diffusive for $\alpha \sim 1$, super-diffusive for $\alpha > 1$ and sub-diffusive for $\alpha < 1$. The MSD profiles are fit using the Nelder-Mead simplex algorithm implemented in MATLAB (Mathworks Inc., USA).
\section{Results}

\subsection{Centrosomal microtubule asters in experiment move towards chromatin}
\begin{figure}[ht!]
\begin{center}
	\subfigure[]{		\label{fig:pdms3D}  \includegraphics[width=0.4\textwidth]{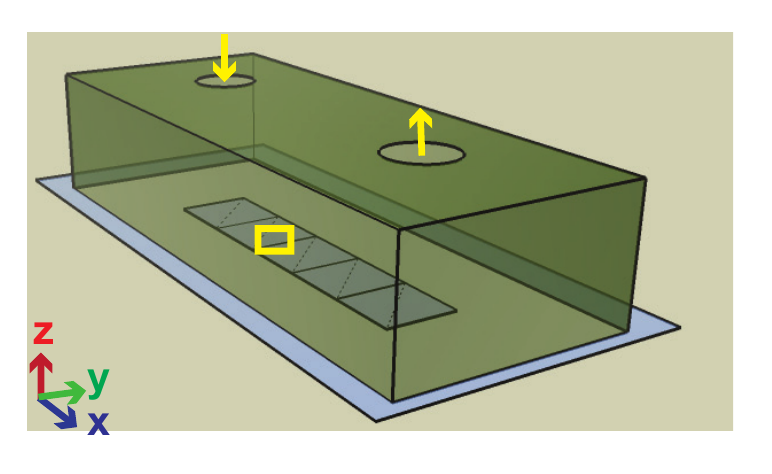} }
	\subfigure[]{		\label{fig:pdmsxy}  \includegraphics[width=0.2\textwidth]{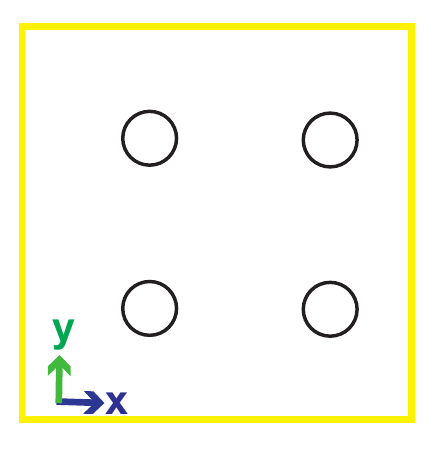}	}
	\subfigure[]{		\label{fig:pdmsyz}  \includegraphics[width=0.3\textwidth]{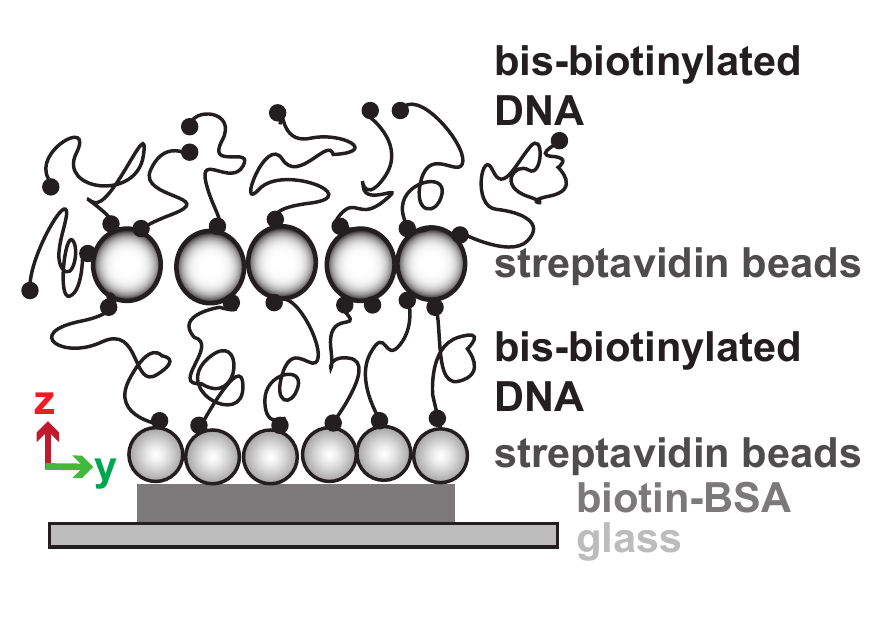}	}
	\subfigure[]{		\label{fig:asterSnap}  \includegraphics[width=0.3\textwidth]{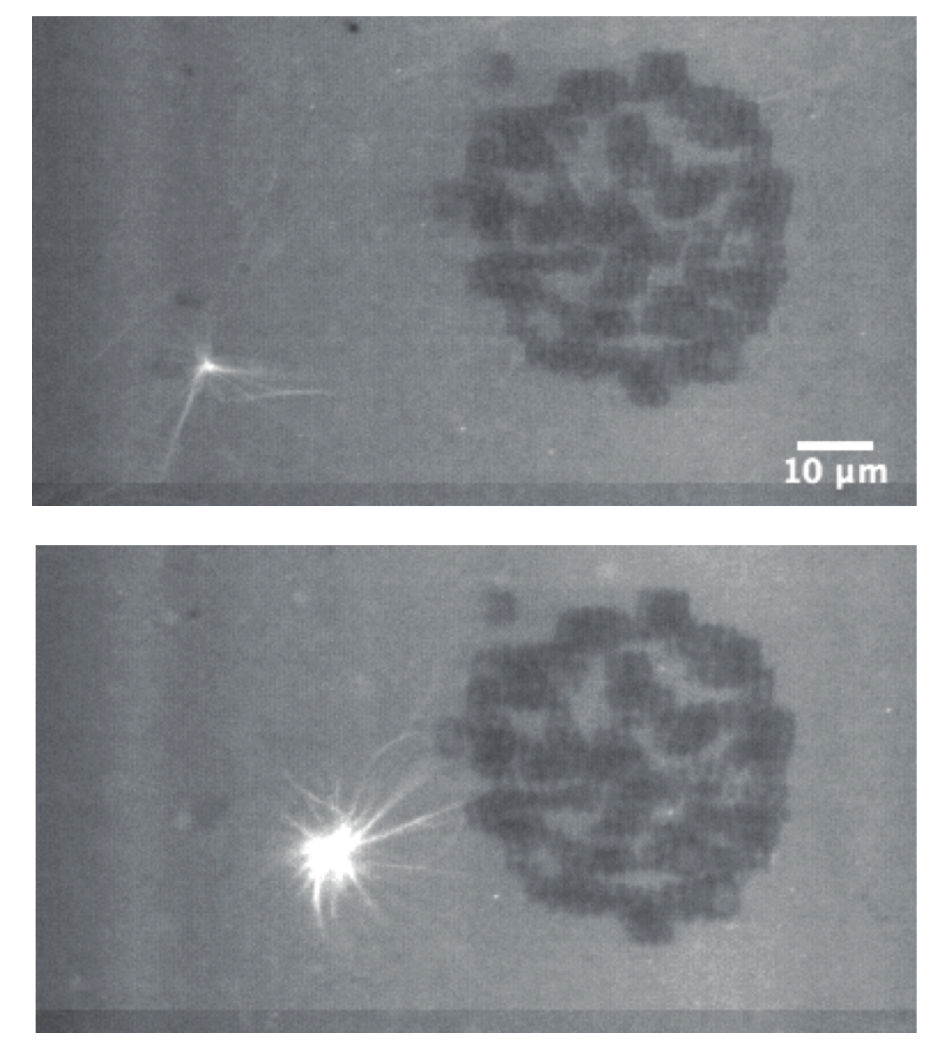} }
	\subfigure[]{		\label{fig:xytraj}  \includegraphics[width=0.45\textwidth]{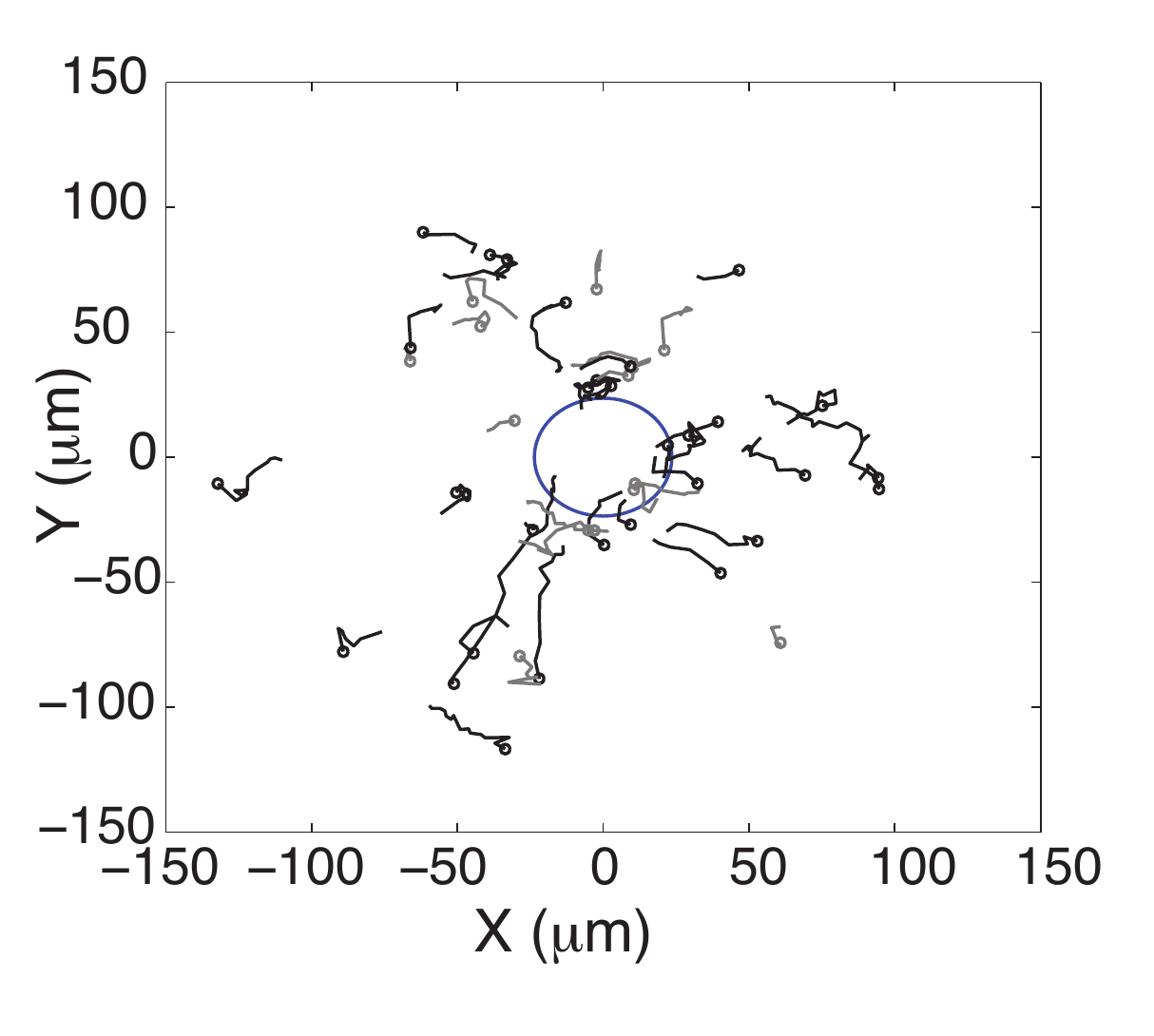} }
\caption{The DNA-micropatterned chamber for aster movement. \subref{fig:pdms3D} The PDMS chamber with vents for the addition of reagents is sealed with a cover glass. \subref{fig:pdmsxy} A 2D pattern of regularly spaced circular patches of biotinylated-BSA is made on the cover glass. \subref{fig:pdmsyz} The patterns bind to a layer of paramagnetic streptavidin-coated beads that sandwich a layer of bis-biotinylated DNA with a second layer of paramagnetic streptavidin coated beads. \subref{fig:asterSnap} Two representative images, separated by 2 minutes, depicting a fluorescently labelled aster moving towards the chromatin patch. \subref{fig:xytraj} The XY trajectories of multiple asters moving towards (black) and away from (gray) the nearest DNA pattern. Circles indicate the initial position of the asters. }
\label{fig:chamber}
\end{center}
\end{figure}

We observed centrosomal aster dynamics in a PDMS flow-chamber enclosing micron-sized DNA patterns (Figures \ref{fig:pdms3D},\subref{fig:pdmsxy}, \subref{fig:pdmsyz}). Fluorescent asters become visible within $\sim$1-2 minutes after the mixture containing centrosomes, Cy3-tubulin and {\it Xenopus} extract has been added. These asters move towards DNA (Figure \ref{fig:asterSnap}) with their lengths fluctuating. The directionality of the aster movement was evaluated using a directional motility coefficient (dmc) measure. The  analysis shows that the majority of asters move in the direction of DNA patterns (n = 45) as seen in Figure \ref{fig:xytraj}. These asters have $dmc > 0$ (black tracks) as compared to those moving randomly or away from DNA which have $dmc\leq 0$ (gray tracks). After $\sim90$ minutes the asters appeared to aggregate and could not be  tracked anymore.

\subsection{Statistics of centrosomal aster movement}
\begin{figure}[ht!]
\begin{center}
\leavevmode
	\subfigure[]{		\label{fig:freq_dtV}  \includegraphics[width=0.33\textwidth]{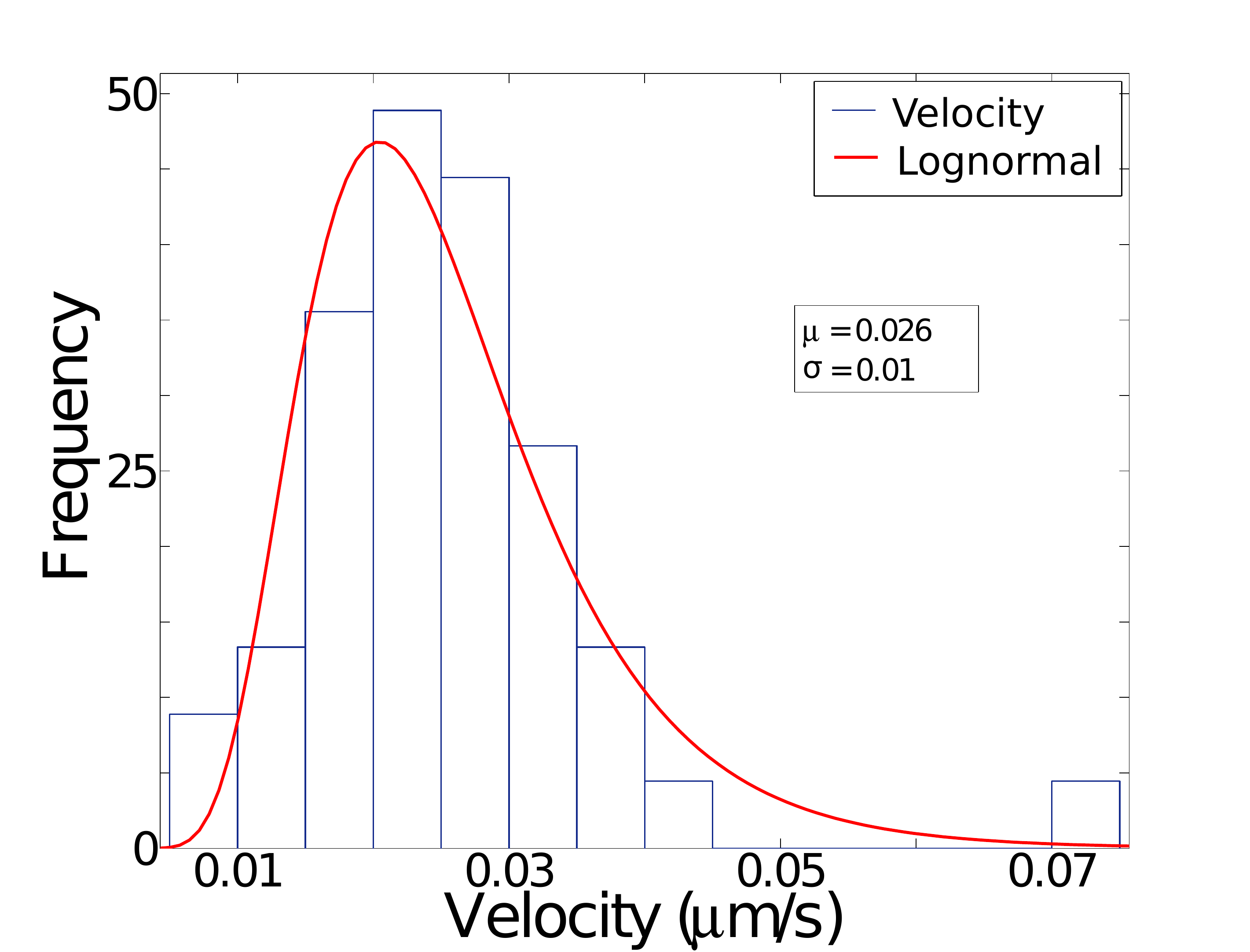} }
	\subfigure[]{		\label{fig:dmcsort_vel}  \includegraphics[width=0.35\textwidth]{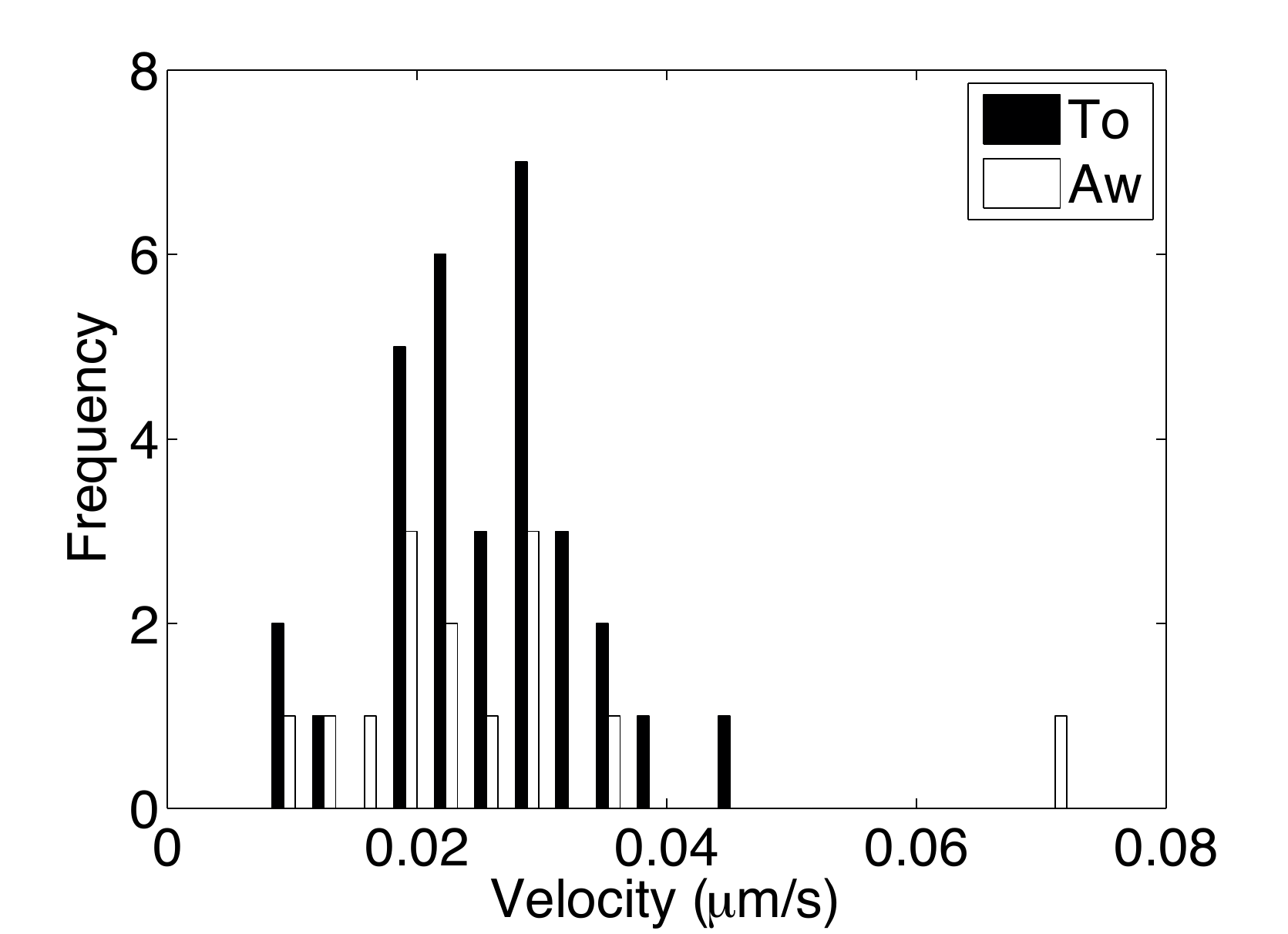} }
	\subfigure[]{		\label{fig:freq_netD}  \includegraphics[width=0.32\textwidth]{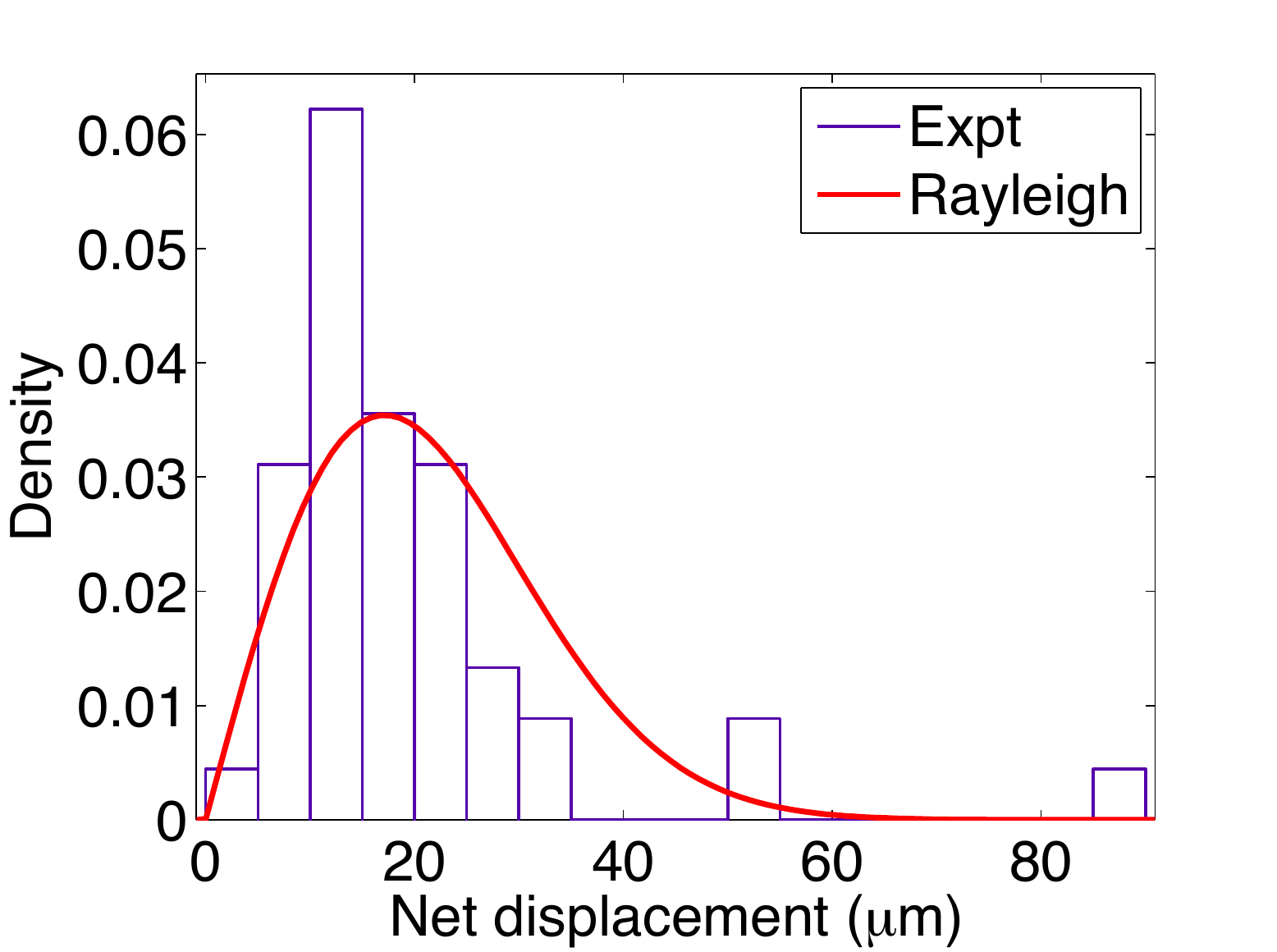}}
	\subfigure[]{		\label{fig:freq_pLen}  \includegraphics[width=0.3\textwidth]{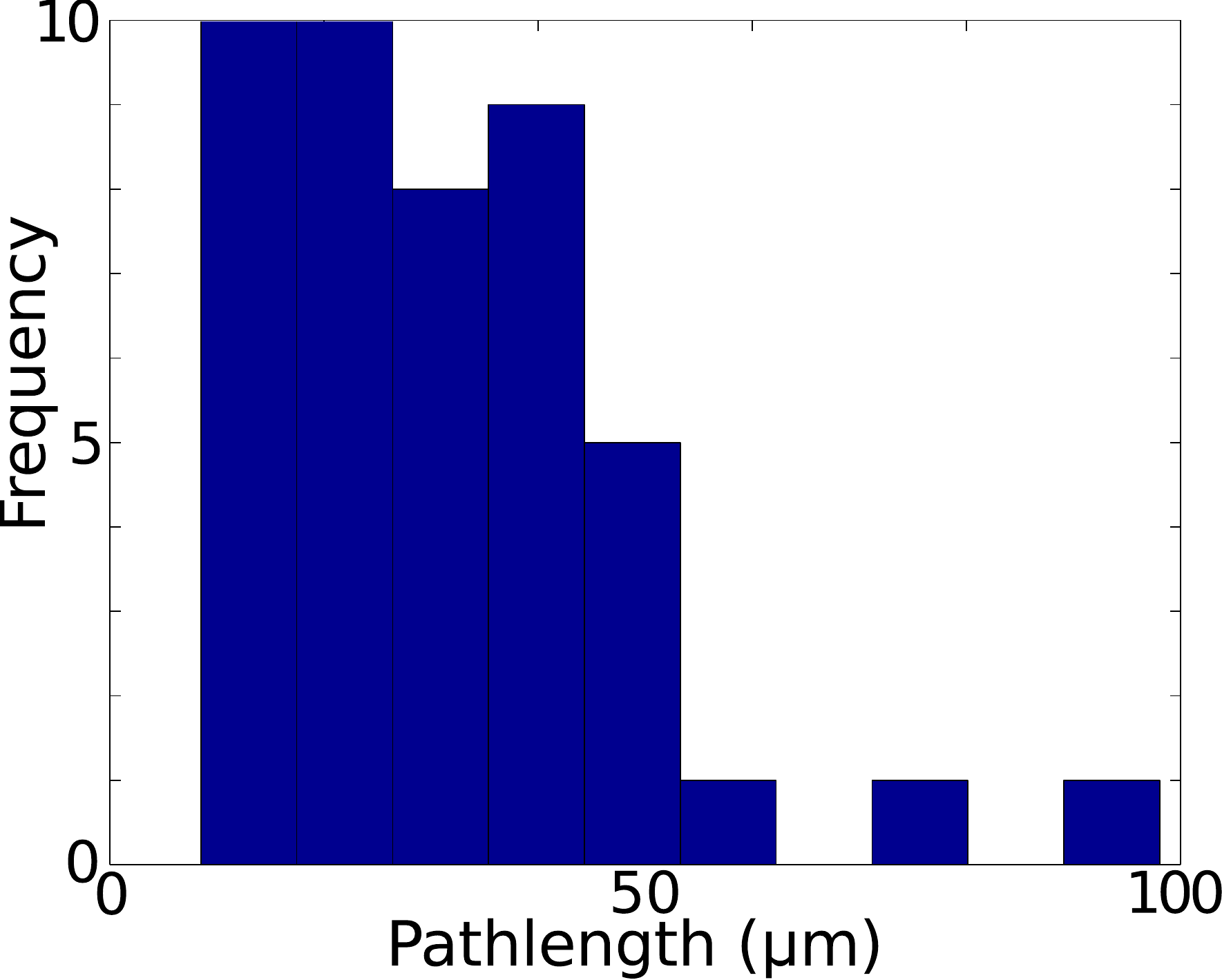}	}
	\subfigure[]{		\label{fig:freq_chi}  \includegraphics[width=0.3\textwidth]{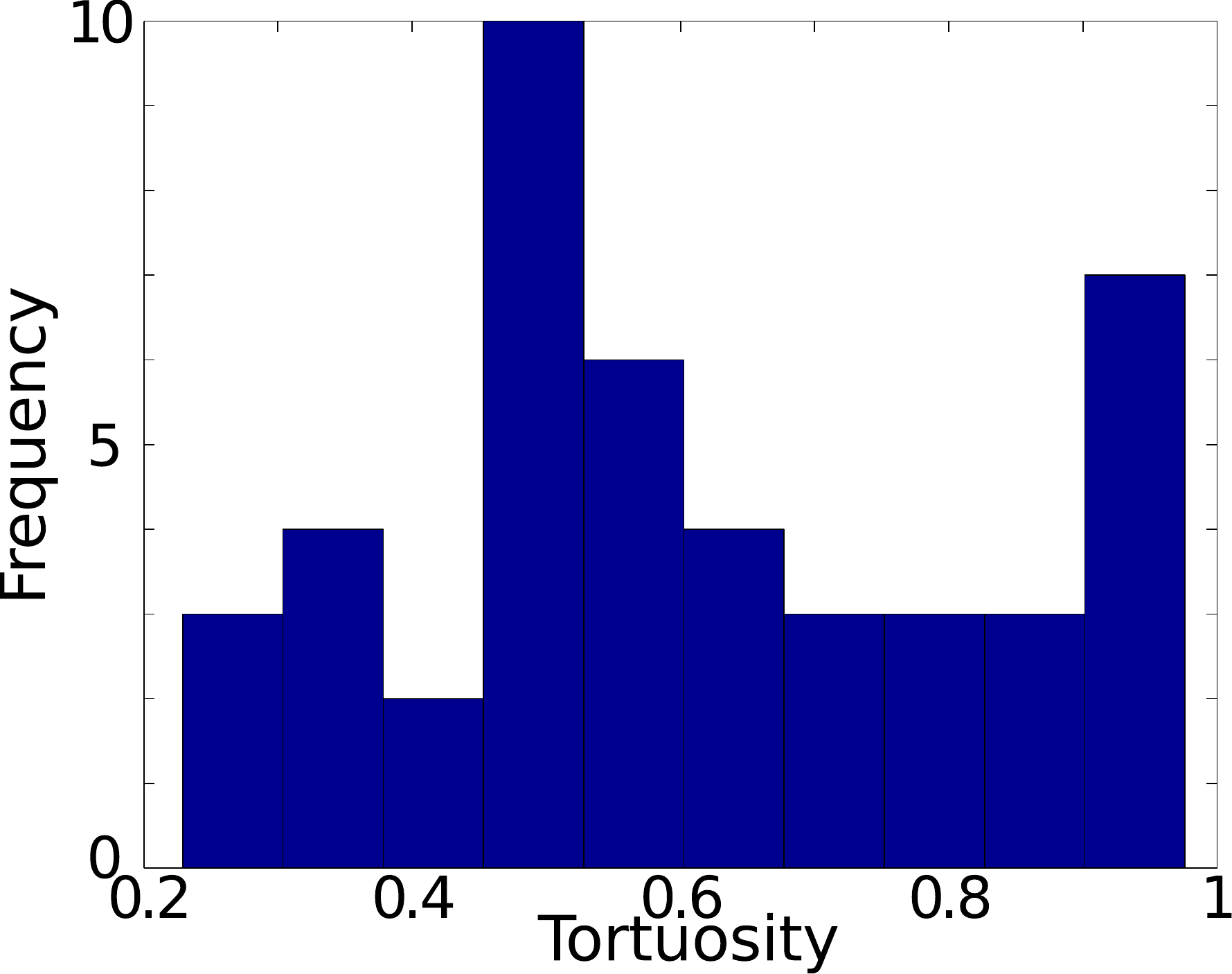}	}
	\subfigure[]{		\label{fig:freq_costheta}  \includegraphics[width=0.3\textwidth]{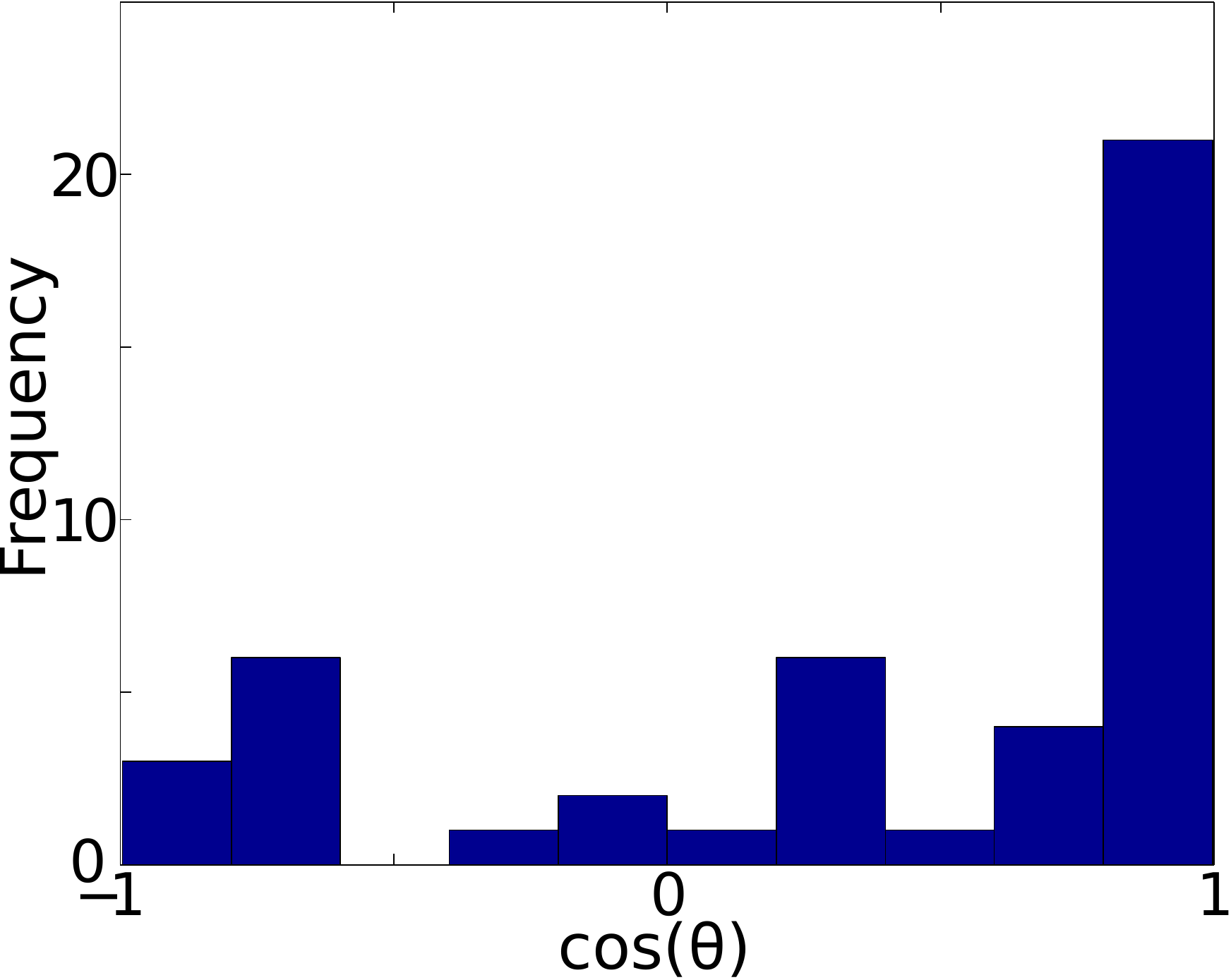} } 
	\subfigure[]{		\label{fig:freq_dmc}  \includegraphics[width=0.3\textwidth]{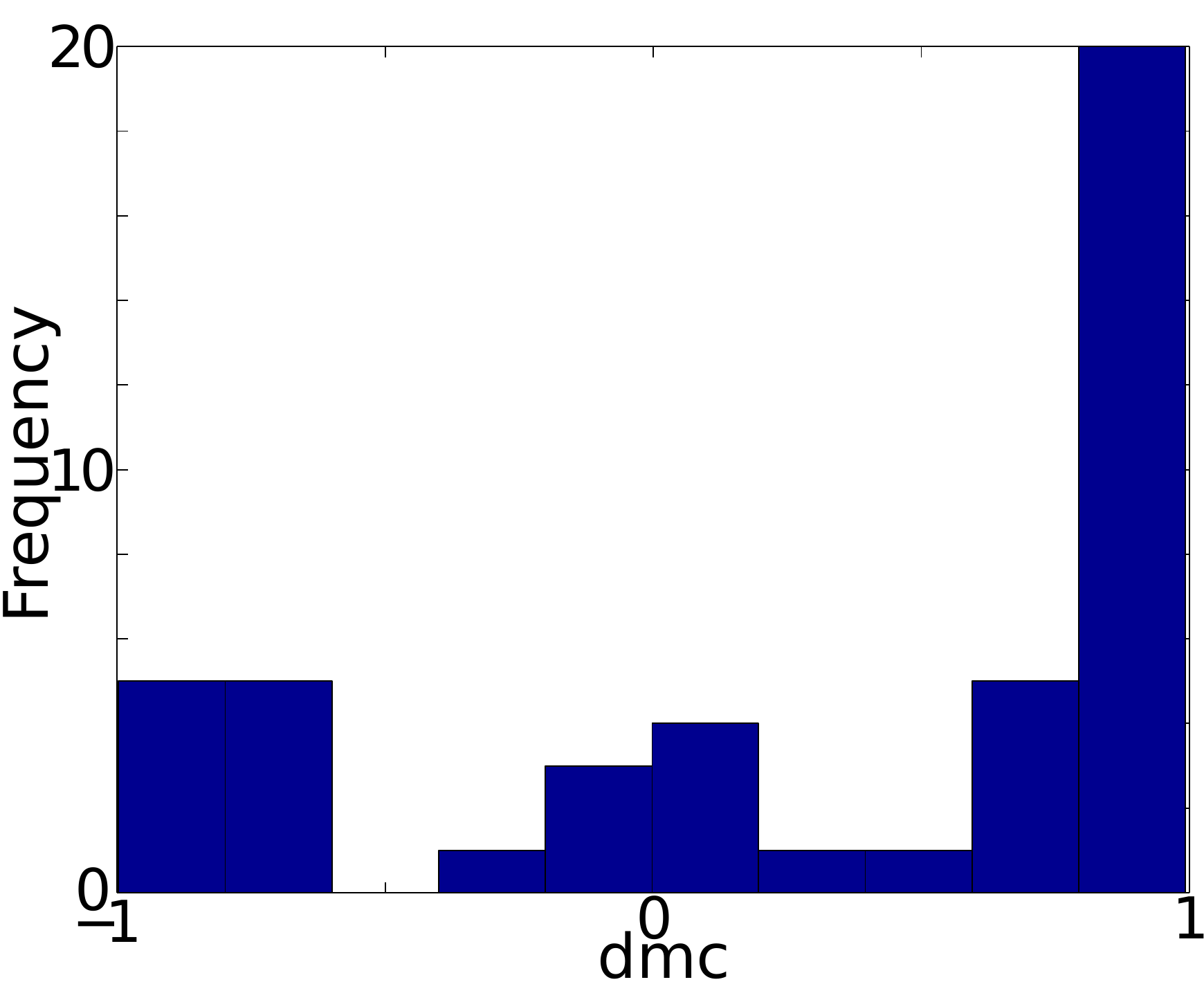} } 

\end{center}
\caption{Movement statistics of asters. \subref{fig:freq_dtV} A  log-normal (mean $\mu = 0.025 \mu m/s$, standard deviation $\sigma = 0.01$) function (red-line) is fit to the frequency distribution of instantaneous velocities (blue-bars). \subref{fig:dmcsort_vel} The frequencies of velocities sorted by their movement towards (+ve dmc, black) or away from (-ve dmc, white) DNA. \subref{fig:freq_netD} The density distribution of net displacement (blue-bars) fit to a Rayleigh distribution (red) with $\mu=21.46 \mu m$, $\sigma=11.22$ and b=17.12. The frequency distribution of \subref{fig:freq_pLen} path-lengths,  \subref{fig:freq_chi} tortuosity ($\mu=0.62$ and $\sigma= 0.22$),  \subref{fig:freq_costheta} $\cos(\theta)$ ($\mu=0.39, \sigma=0.68$) and \subref{fig:freq_dmc} dmc ($\mu = 0.32, \sigma= 0.7$).  }
\label{fig:exptfreq}
\end{figure}

Using the trajectories from SPT, we calculate the frequency distribution of the mean velocity (Figure \ref{fig:freq_dtV}) and fit a log-normal distribution. The frequency distribution of aster velocities sorted according to movement towards DNA ($dmc>0$) and away ($dmc\leq0$) shows that a smaller number of asters move away as compared to towards. However the velocities of the two populations are comparable (Figure \ref{fig:dmcsort_vel}). The frequency distribution of net displacement (i.e. displacement from start point) of the experimental trajectories was fit to a Rayleigh probability distribution ($\mu= 21.46 \mu m$, $\sigma = 11.22 \mu m$ and b=17.12) (Figure \ref{fig:freq_netD}), characteristic of a random walk. Considering the net displacement distributions and the zig-zag movement of the XY tracks of the asters (Figure \ref{fig:xytraj}) we analyzed the aster movement further as a modified random-walk model. The path lengths of the trajectories measured range between 10 to 50 $\mu m$ with a few outliers (Figure \ref{fig:freq_pLen}). To quantify the randomness, the tortuosity ($\chi$) was evaluated. It shows that most trajectories have $\chi \geq 0.5$ indicating relatively directed movement, while a few have $\chi \sim 0.3$ indicating tortuous or random movement (Figure \ref{fig:freq_chi}). Evaluating the trajectories for directed movement towards the nearest DNA patch showed that the $74\%$ of the evaluated trajectories moved towards the DNA, while $26 \%$ move away, as quantified by $\cos(\theta)$ (Figure \ref{fig:freq_costheta}) and dmc (Figure \ref{fig:freq_dmc}). Taken together the data indicates that the majority of the asters move vectorially towards the DNA patch, with a random component to the trajectory.
\subsection{Evaluating the measures of directionality}
In order to test our interpretation of the measures of directionality, we simulated trajectories of particles undergoing random walks with varying degrees of drift. The input weight of directionality ($W_d$) determines the degree of randomness (Figure \ref{sup:randwk}\subref{sfig:rwplt}). At the same time the mean velocity of the simulated trajectory is maintained constant (Figure \ref{sup:randwk}\subref{sfig:vel}). The range of input parameters from random ($W_d=0$) to completely directed motion ($W_d=1$) produced a mean tortuosity ($\chi$) that also changes from $\sim0.1$ to 1 (Figure \ref{sup:randwk}\subref{sfig:chi}). The distribution of $\chi$ for a given value of $W_d$ is tightly distributed around its mean (Figure \ref{distrChi}). In contrast while dmc (Figure \ref{sup:randwk}\subref{sfig:dmc}) and $\cos(\theta)$ (Figure \ref{sfig:costheta}) also increase with increasing $W_d$, they are more widely distributed. Even for $W_d \sim 0.4$, the dmc (Figure \ref{distrDmc}) and  $\cos(\theta)$ (Figure \ref{distrCos}) measures are highly spread.

\subsection{Dependence of  MT aster movement on distance from chromatin}
\begin{figure}[ht!]
\begin{center}
	\subfigure[]{		\label{fig:ex_vect}  \includegraphics[width=0.45\textwidth]{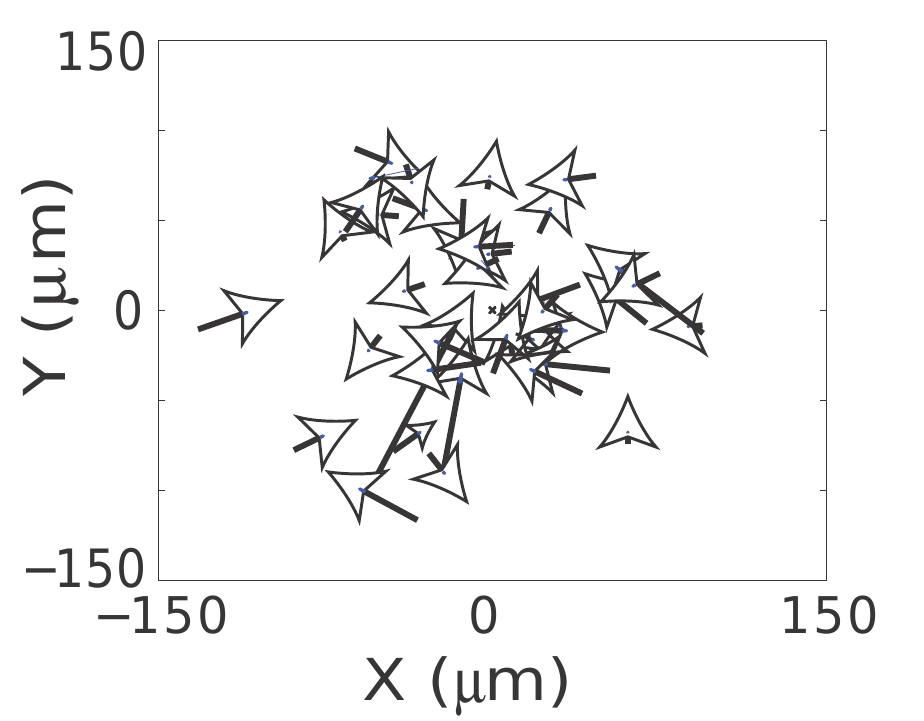}	}
	\subfigure[]{		\label{fig:ex_dmc}  \includegraphics[width=0.4\textwidth]{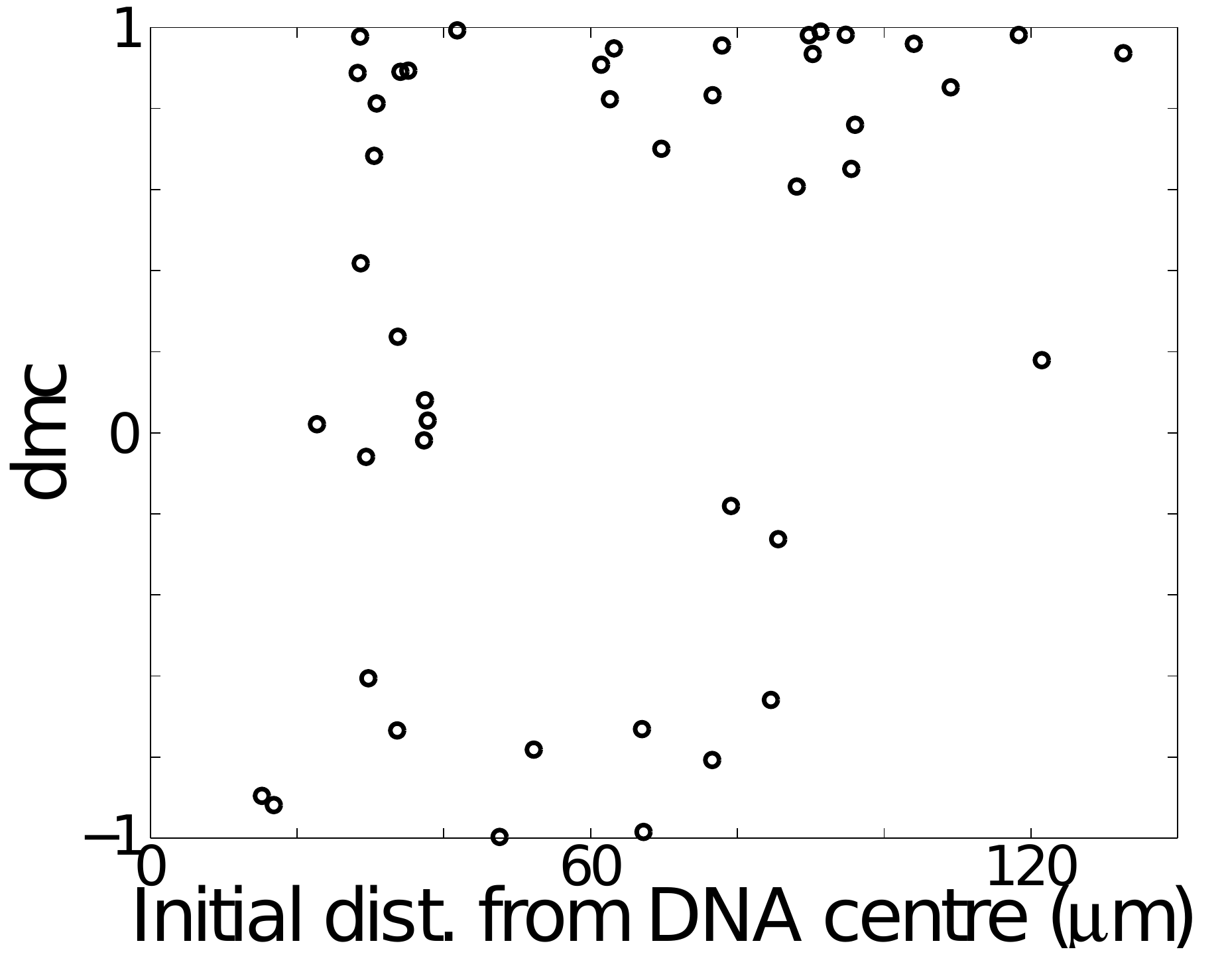}} 
	\subfigure[]{		\label{fig:ex_cos}  \includegraphics[width=0.45\textwidth]{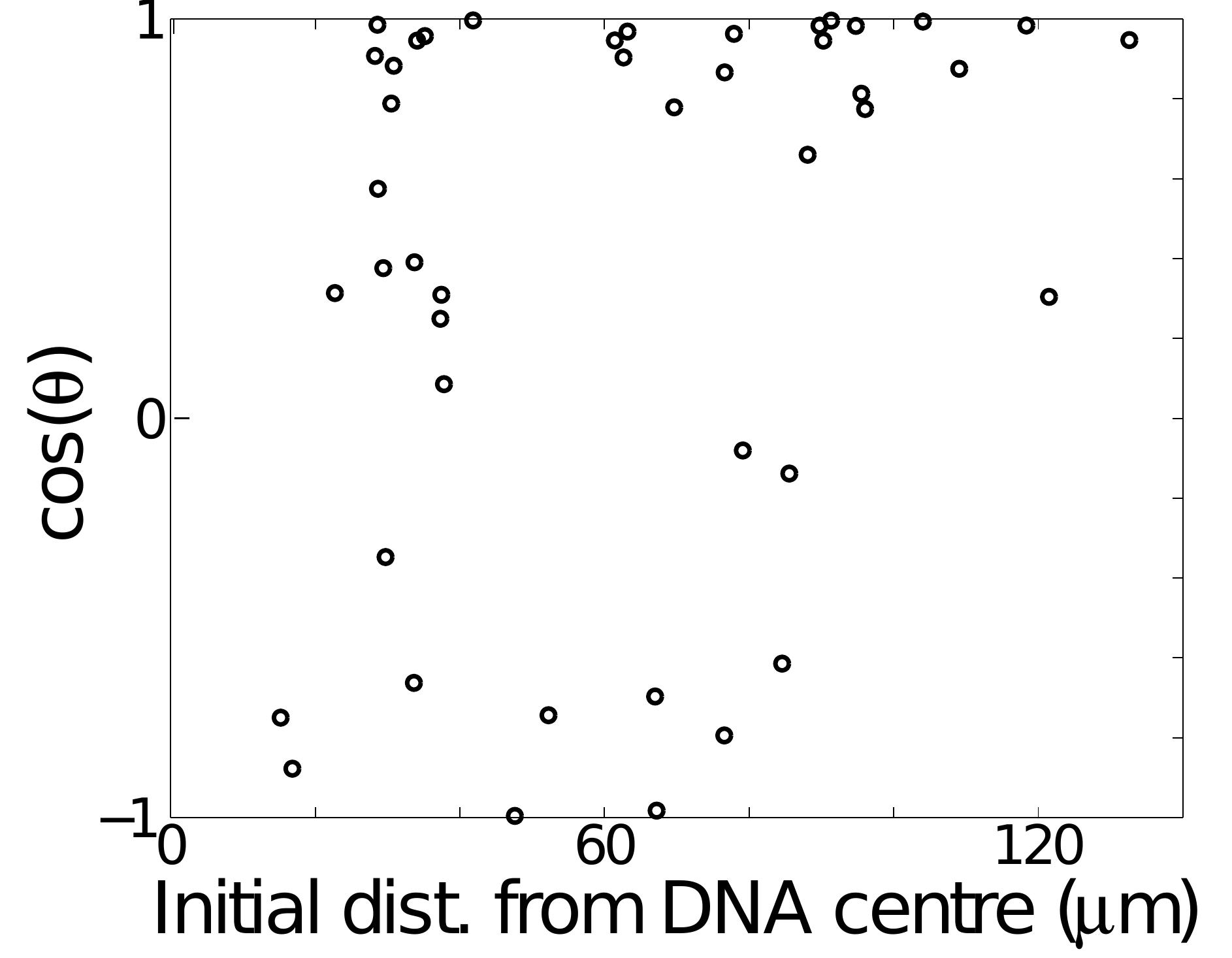}  }			
	\subfigure[]{		\label{fig:ex_chi}  \includegraphics[width=0.45\textwidth]{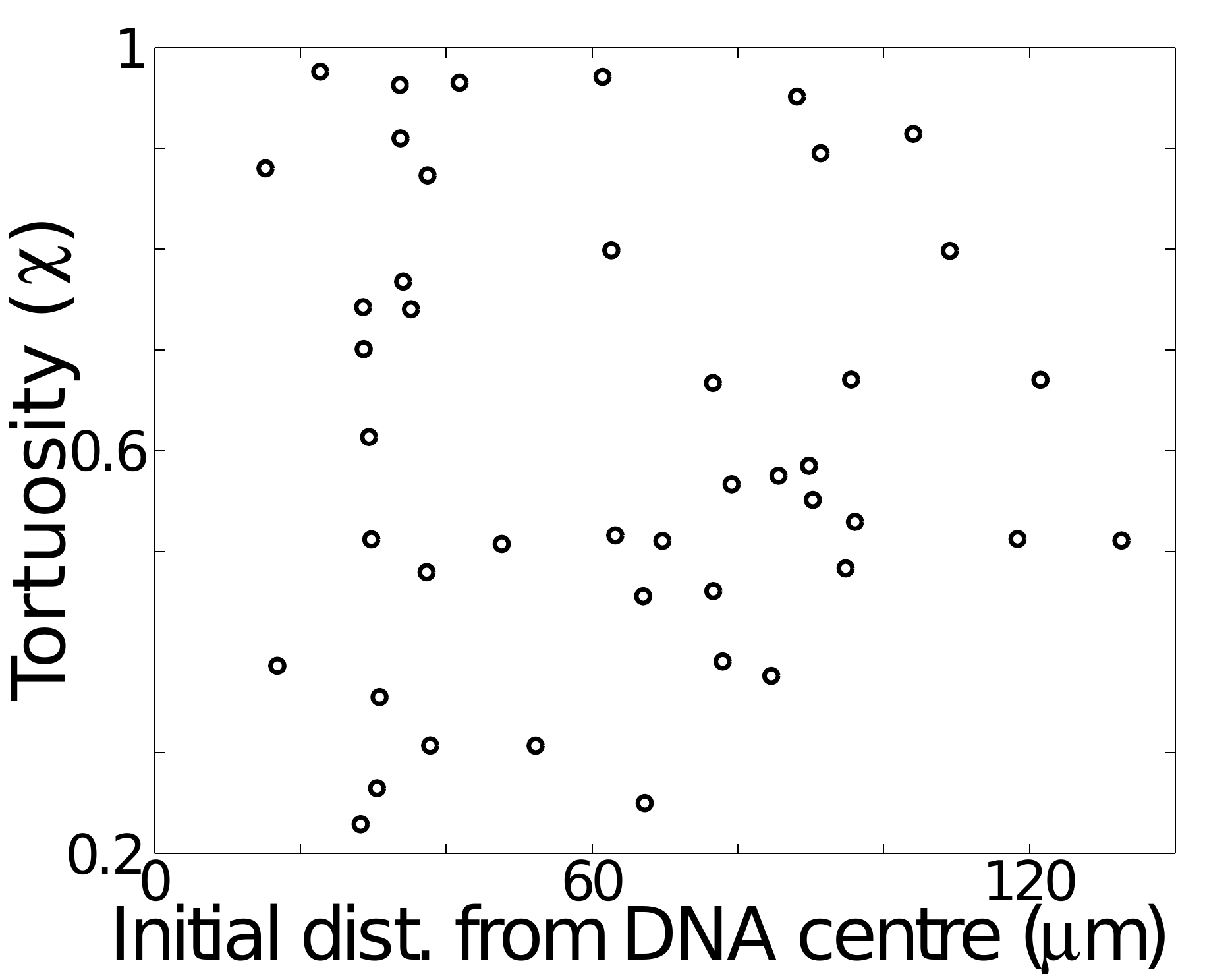}}   
	\subfigure[]{		\label{fig:ex_vel}  \includegraphics[width=0.45\textwidth]{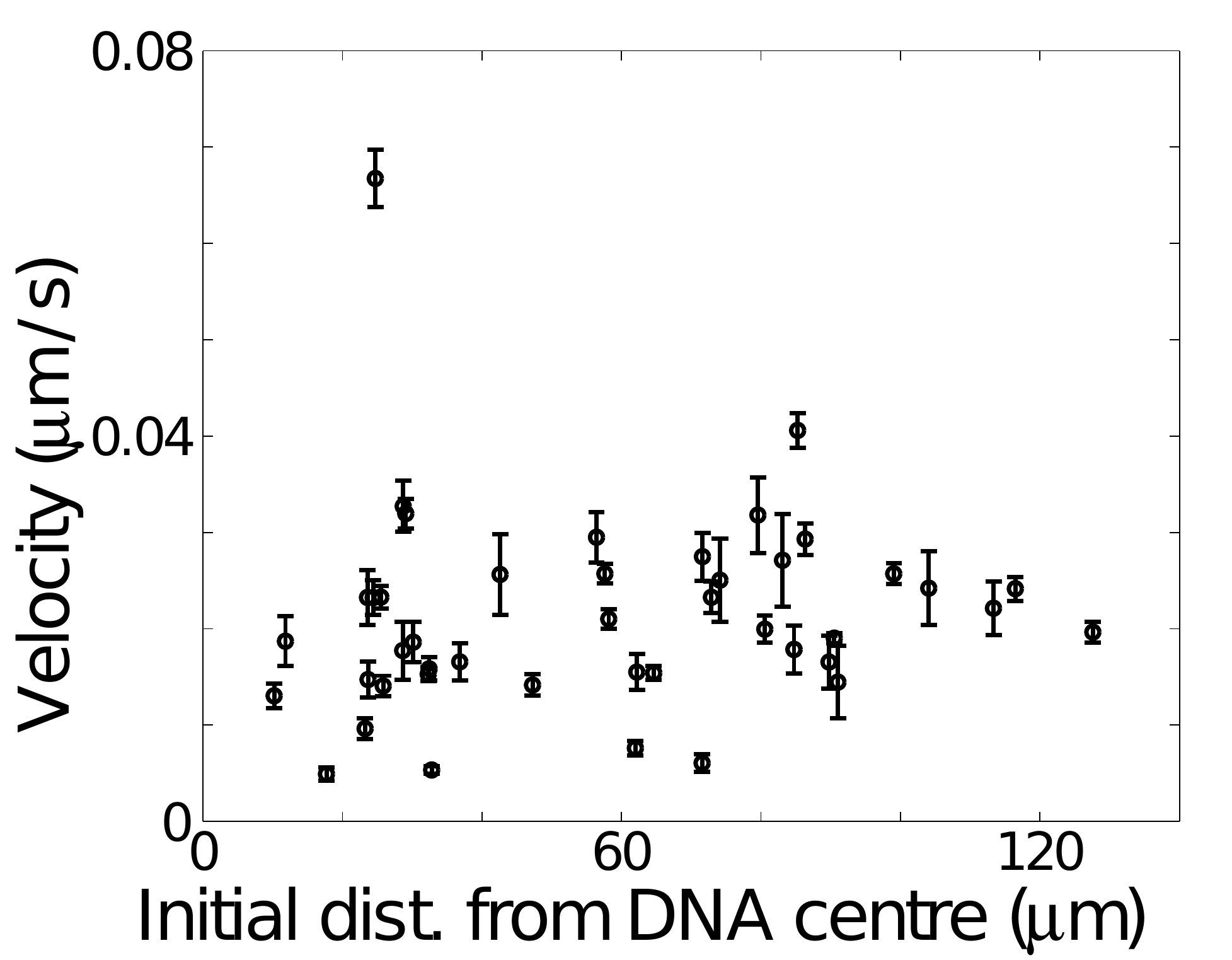}} 
\caption{ Distance-dependence of aster motility. \subref{fig:ex_vect} The net vectors connecting the start and end-point of each trajectory are plotted around the center of the nearest DNA patch. \subref{fig:ex_dmc} The directional motility coefficient (dmc), \subref{fig:ex_cos} $\cos(\theta)$, \subref{fig:ex_chi} the tortuosity and \subref{fig:ex_vel} mean velocities (error bars indicate standard error of mean) are plotted as a function of the start point of the trajectory.}
\label{fig:exptdirected}
\end{center}
\end{figure}

The simplest explanation for the observed movement of microtubule asters towards chromatin DNA is length dependent pulling of the aster by minus-ended motors in the cytoplasm anchored on the glass coverslip. Previous work has shown that mitotic aster MTs  are longer in the direction towards chromatin. The length asymmetry is due to a distance dependent effect mediated by a reaction-diffusion gradient of RanGTP and downstream components \cite{Carazo-Salas2003, Carazo2001, Wilde2001,Wollman2005, Caudron2005}. The effective gradient of microtubule `stabilization' (lengthening) is long range and step-like. Most net displacement vectors of the trajectories plotted around the center of the DNA patch point inwards towards the origin, i.e. the center of the DNA pattern, with no obvious distance-dependence (Figure \ref{fig:ex_vect}). Further quantification was performed by measuring the directionality measures dmc (Figure \ref{fig:ex_dmc}), $\cos(\theta)$ (Figure \ref{fig:ex_cos}) and $\chi$ (Figure \ref{fig:ex_chi}) as a function of initial distance of the aster from DNA. The values of dmc and $\cos(\theta)$ show no obvious distance-dependence although a larger proportion of dmc and $\cos(\theta)$ values are $\sim 1$. The values of $\chi$ are widely distributed for most values of initial distance. The mean instantaneous velocity varies between 0.01 to 0.04 $\mu m/s$ with no apparent dependence on distance from DNA (Figure \ref{fig:ex_vel}).

\subsection{Simulation of radial MT aster motility: resolving a tug-of-war}
\begin{figure}[ht!]
\begin{center}
	\subfigure[]{		\label{fig:simsnap}  \includegraphics[width=0.6\textwidth]{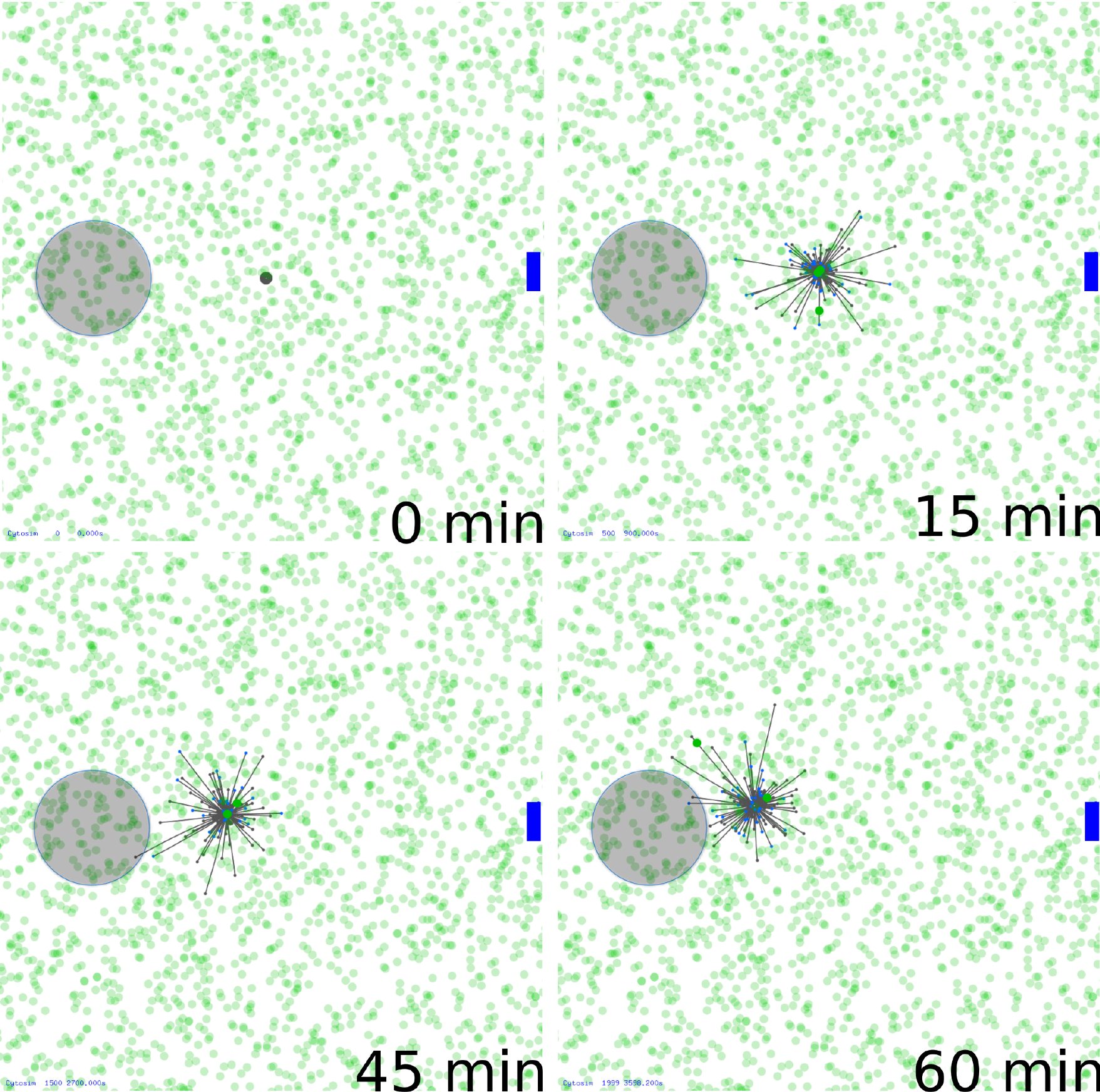}       }			
	\subfigure[]{		\label{fig:simgrad}  \includegraphics[width=0.5\textwidth]{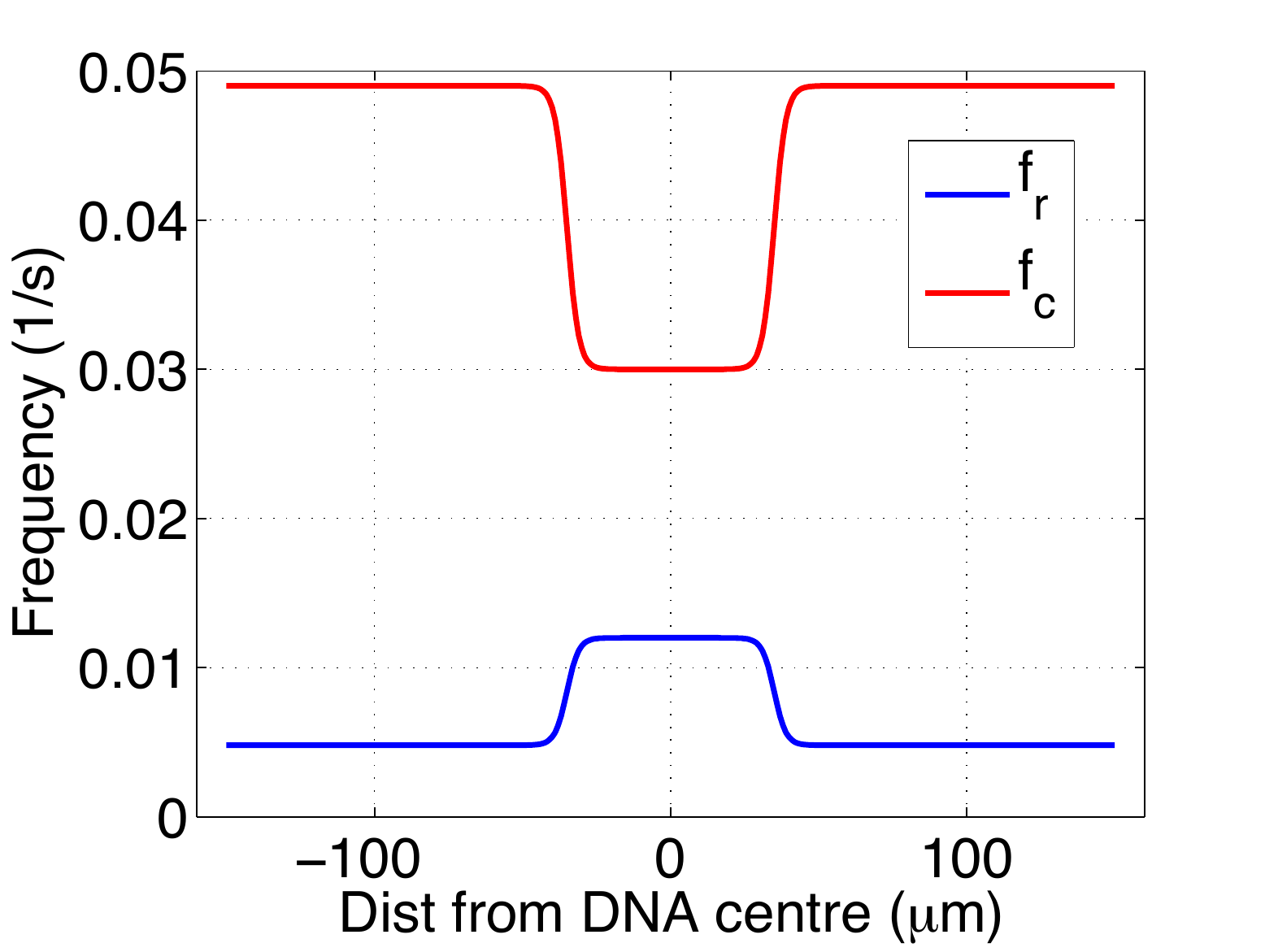} } 
\caption{Schematic representation of the simulations. \subref{fig:simsnap} Langevin dynamic simulations of microtubule asters (gray) initialized $45 \mu m$ away from the center of a simulated DNA patch (gray-circle) are run for 60 min ($\delta t=0.05 s$). The simulation snapshot includes immobilized motors (green) and growing (blue-ends) and shrinking (gray-ends) microtubules. The scale bar in the right corner corresponds to 10 $\mu m$. \subref{fig:simgrad} The simulated gradients of frequency of rescue ($f_r$) and catastrophe ($f_c$) (Equations \ref{eq:gradres} and \ref{eq:gradcat}) plotted as a function of distance from the center of DNA.}
\label{fig:simschem}
\end{center}
\end{figure}

We have developed a minimal model simulation of centrosomal movement by surface-attached motors (Figure \ref{fig:simsnap}) to test if the experimental data can be reproduced, as well as allow us to consider general properties of the system. In the absence of any directional cues, these asters are expected to move randomly. The tug-of-war that arises out of geometric considerations is resolved in our model by MT length asymmetry. Experimentally derived gradients of $f_{r}$ (Equation \ref{eq:gradres}) and $f_{c}$ (Equation \ref{eq:gradcat}) as a function of the distance from the DNA (Figure \ref{fig:simgrad}) affect growth of MTs at the plus-tips. Assuming a spatially uniform motor-density, when MTs come within the field of `stabilization', they become longer. The length of MTs attached to motors and consequently asymmetric forces result in directional movement. MTs are occasionally observed to bend when multiple motors bind the same fibre in different orientations (Figure \ref{fig:simsnap}), Video \ref{sup:simvid}). The movement of an aster results from a force-asymmetry due to either a length-asymmetry or stochastic fluctuations in bound motors, thus resolving the tug-of-war. 

\subsection{Comparing simulation with experiment}
\begin{figure}[ht!]
\begin{center}
       		\subfigure[]{		\label{fig:simexp_velDist}  \includegraphics[width=0.4\textwidth]{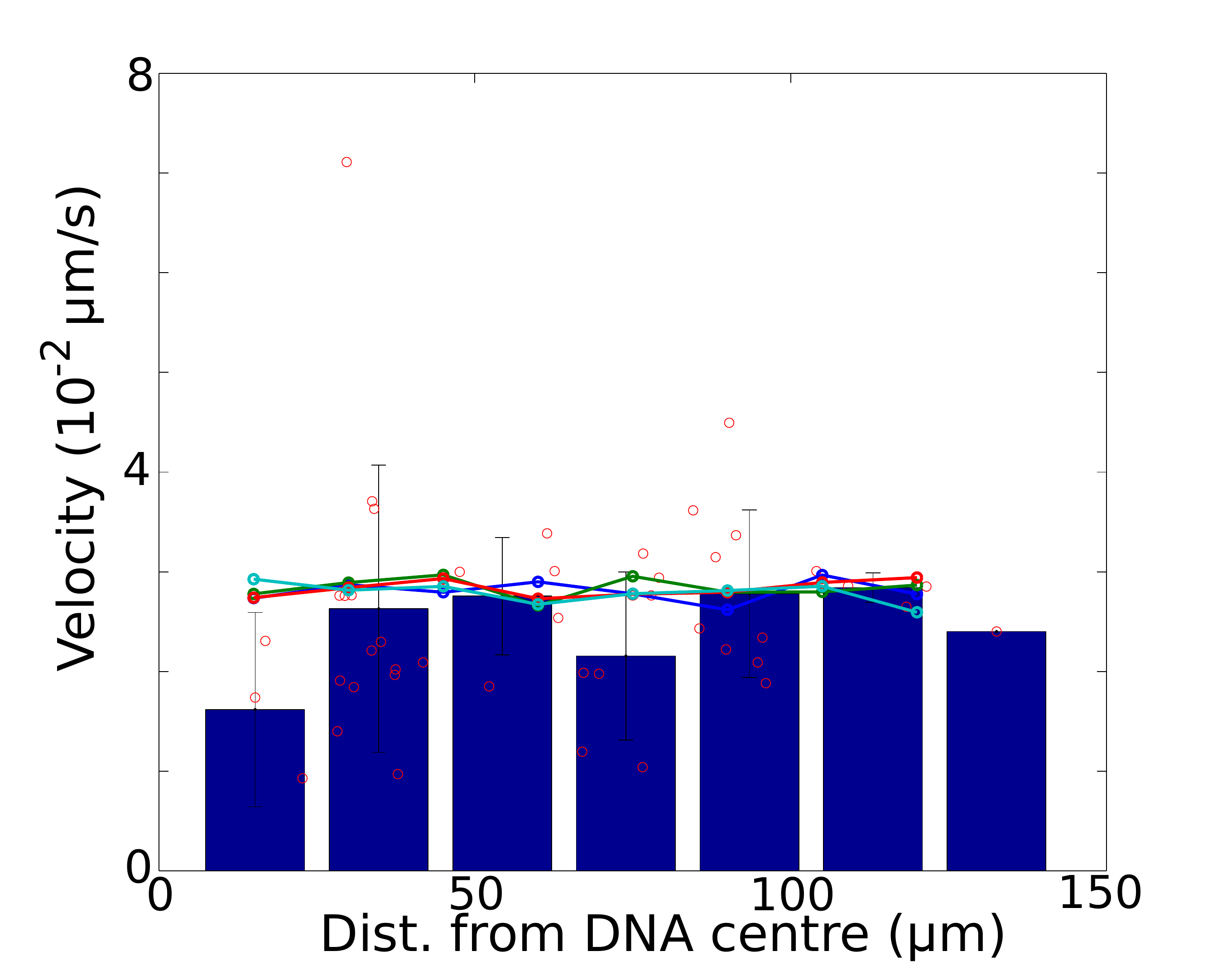}} 
		\subfigure[]{		\label{fig:simexp_costhDist}  \includegraphics[width=0.4\textwidth]{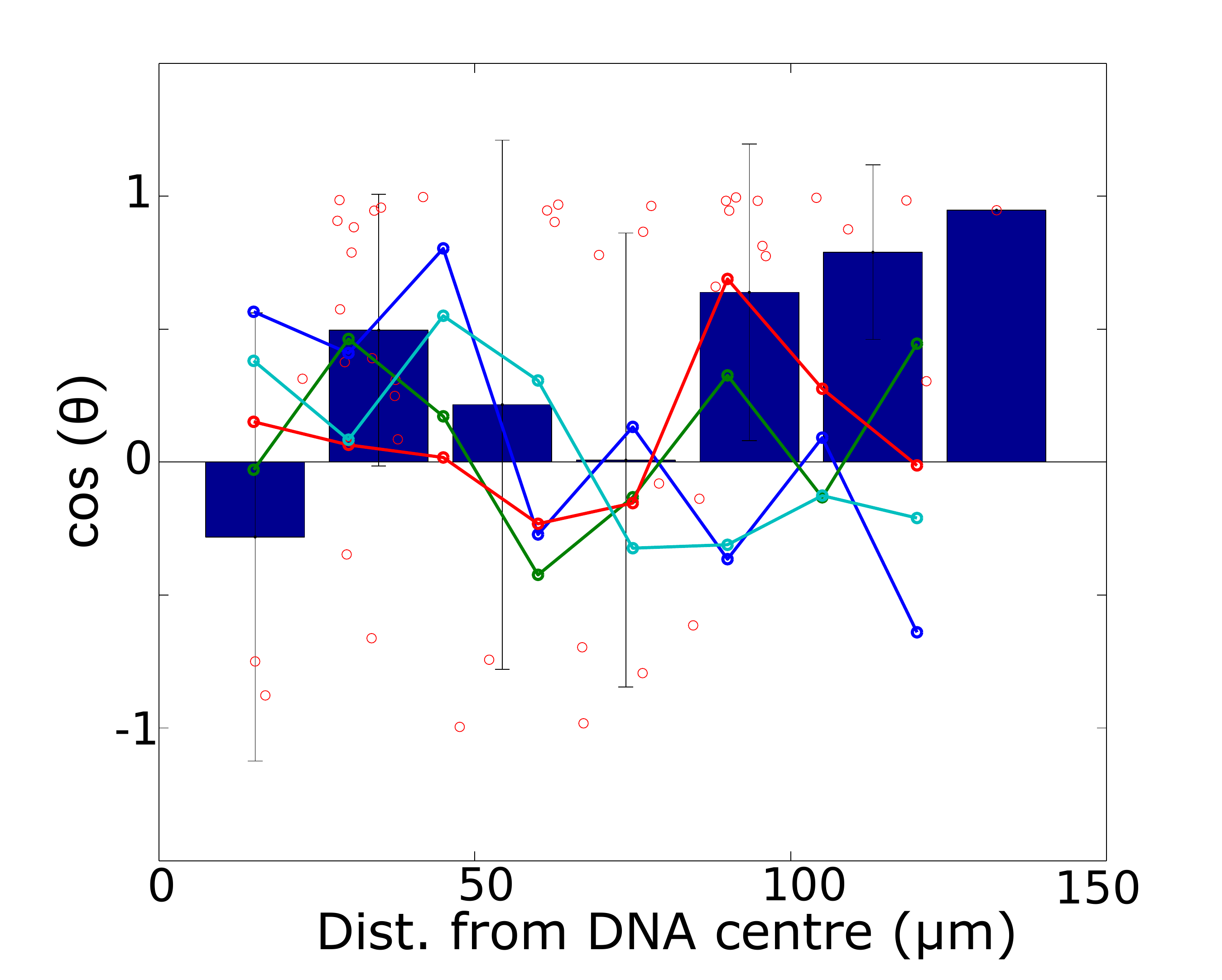}} 
		\subfigure[]{		\label{fig:simexp_dmcDist}  \includegraphics[width=0.4\textwidth]{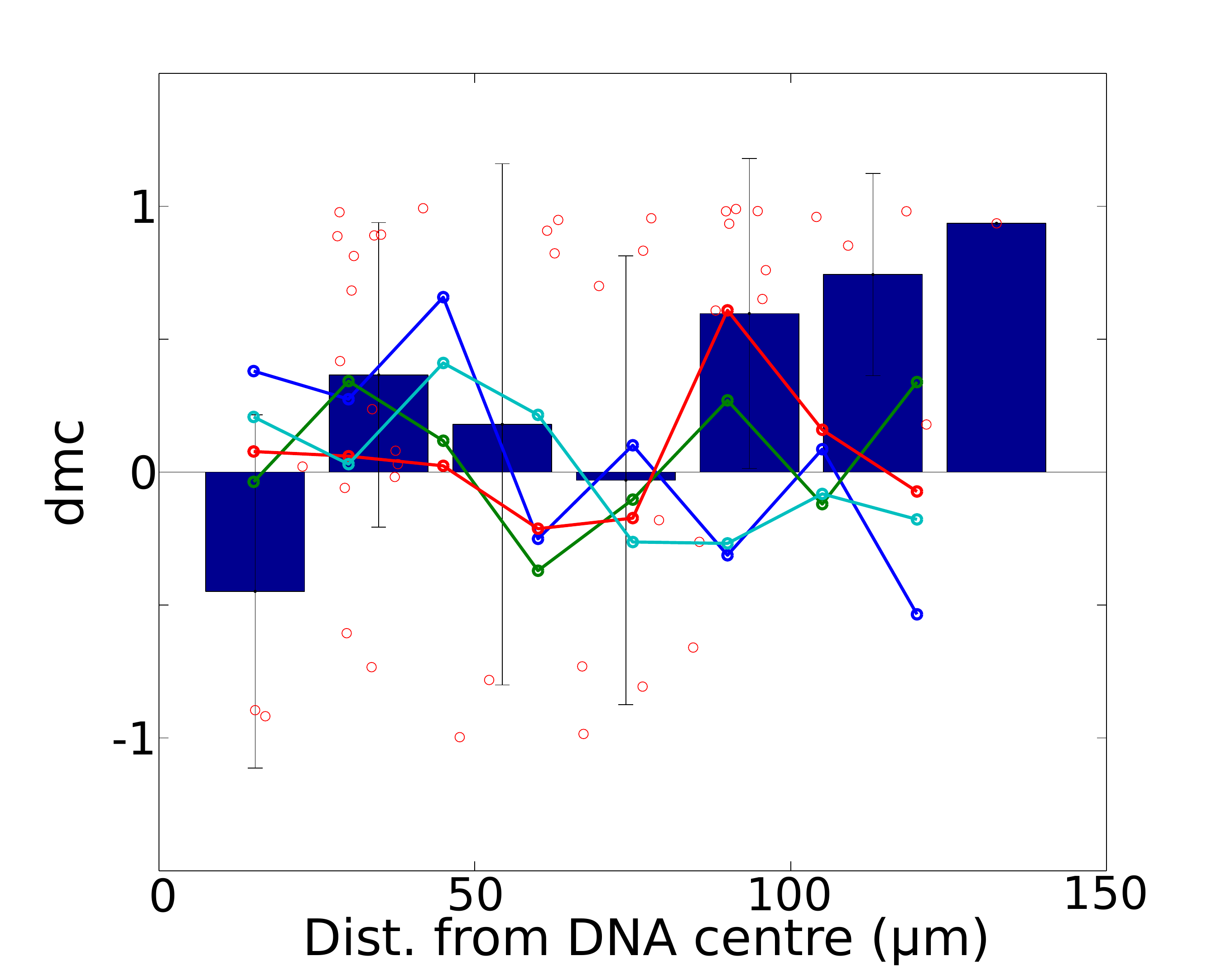}} 
\caption{Comparing experiment and simulation. Experimentally measured values (red circles) of \subref{fig:simexp_velDist} velocity, \subref{fig:simexp_costhDist} $\cos(\theta)$ and \subref{fig:simexp_dmcDist} dmc plotted against initial distance from DNA are distance-binned (blue-bars). The simulated velocity, $\cos(\theta)$ and $dmc$ means (10 runs) for motor densities $10^{-3}$ (blue), $10^{-2}$ (green), $10^{-2}$ (red) and 1 (cyan) $motors/\mu m^2$ are plotted as lines as a function of initial aster position. Error bars denote standard deviations.}
\label{fig:simexp}
\end{center}
\end{figure}

Our model is constrained by experimentally measured parameters of motor mechanics and the spatial dynamic instability gradient. Hence, to compare the model against our experimental measurements, we vary only the (a) initial distance of asters from DNA and (b) the minus-ended motor density. The simulated mean velocity is comparable to the distance binned data from experiments and shows no change as a function of distance from DNA (Figure \ref{fig:simexp_velDist}) over three orders of magnitude of motor density ($10^{-3}$ to 1 $motors/\mu m^{2}$).The distance binned directionality measures from experiment- $\cos(\theta)$ (Figure \ref{fig:simexp_costhDist}) and  dmc (Figure \ref{fig:simexp_dmcDist})- both show an initial increase with distance up to $50 \mu m$, a drop at $\sim 75 \mu m$ and a further increase at greater distances. In the simulations, the trend of an initial increase in directionality is seen at the lowest motor density ($10^{-3}$ $motors/\mu m^2$), while only the calculations at high motor densities (1 $motor/\mu m^2$) show an increase in directionality at $75 \mu m$, dropping off again at greater distances. At distances $>100 \mu m$ we could only measure 5 experimental tracks, which could potentially skew the distribution. Experimental track lengths are variable as evidenced by the path-length distribution (Figure \ref{fig:freq_pLen}) as a result movement of out of the plane of focus and aster coalescence. In our analysis we consider distance from DNA based on initial positions alone, so a long track of asters moving directionally can skew the distribution. The combination of these effects results in certain regions of the simulation data not matching experiments. Simulations indicate that the directionality measures will peak at different distances for different motor densities. Although mean measures of directionality do show agreement between simulation and experiment, better sampling of experimental data and motor density perturbations are needed to validate the distance dependence of movement seen in simulations.

%
\subsection{Transition from vectorial transport to diffusion}
The initial positions of the simulated asters determines whether the aster will experience any length asymmetry. Asters located at d=15 $\mu m$ by virtue of being inside the `stabilization' zone and at d=120 $\mu m$ far from any directional cue are both symmetric. Asters at d=30 to 45 $\mu m$ lie at the boundary of the stabilization zone and are asymmetric. The distribution of motors and their attachment rates ($r_{attach}$) are both spatially homogeneous. Thus, we hypothesize that changes in aster mobility will result from: 
\begin{itemize}
\item[(i)] MT lengths that determine the mean number of bound motors, \textit{i.e.} asters inside the `stabilization' zone will show greater displacement and higher $D_{eff}$
\item[(ii)] aster MT asymmetry will result in increasing values of the anomaly parameter ($\alpha$)
\end{itemize}  

\begin{figure}[ht!]
\begin{center}
     \subfigure[]{		\label{fig:msd1}  \includegraphics[width=0.4\textwidth]{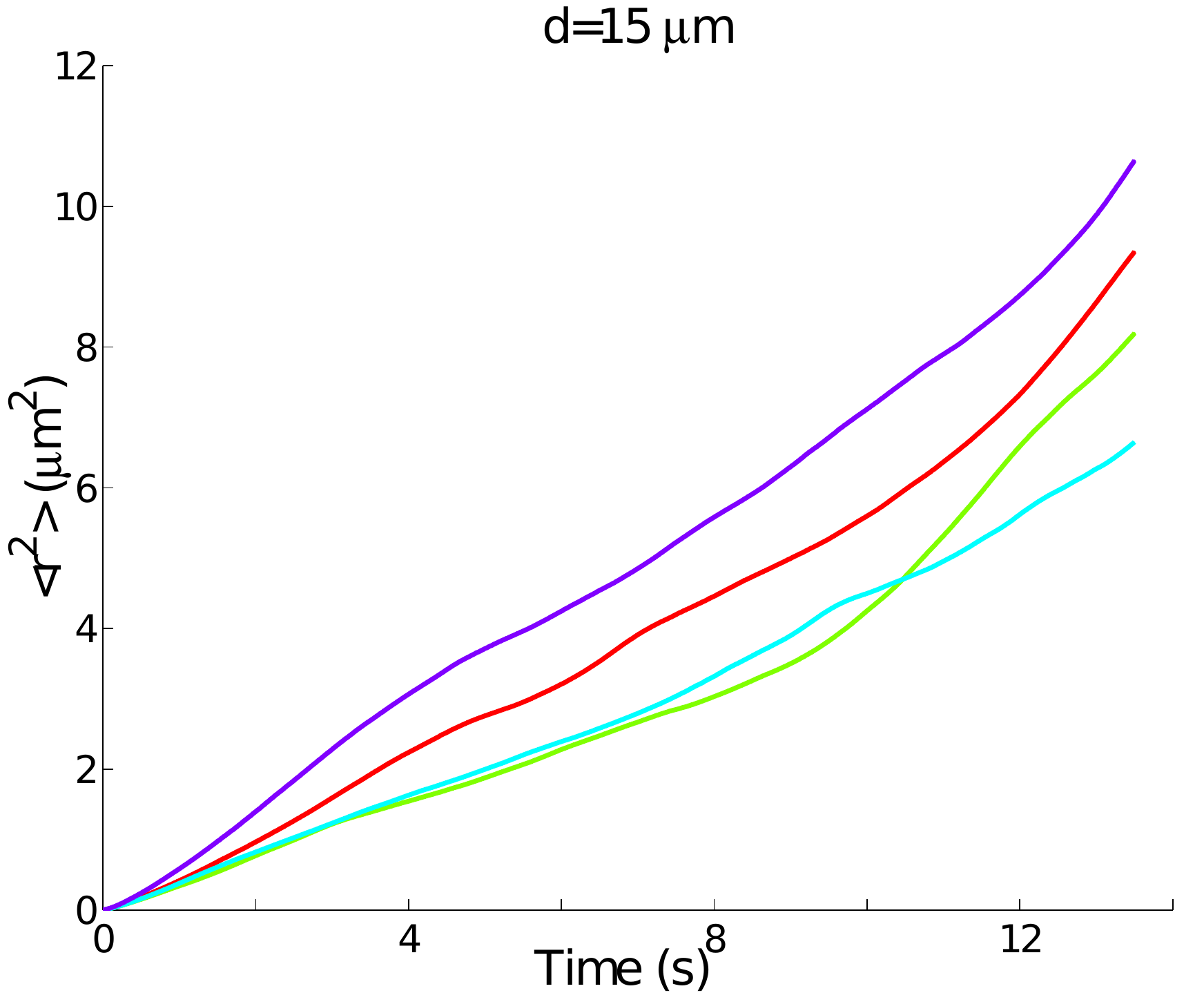}  }
     \subfigure[]{		\label{fig:msd2}  \includegraphics[width=0.4\textwidth]{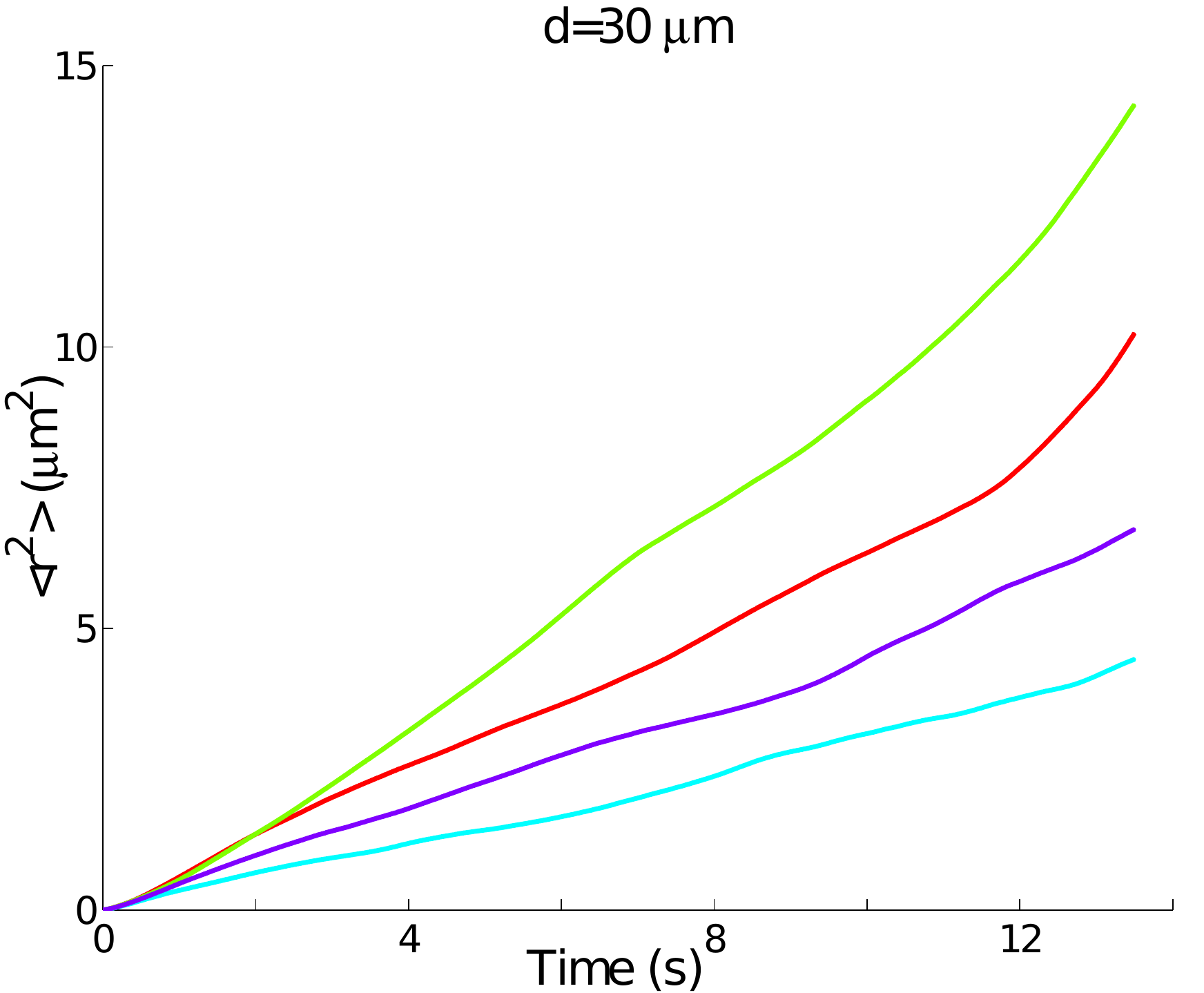}  }
     \subfigure[]{		\label{fig:msd3}  \includegraphics[width=0.4\textwidth]{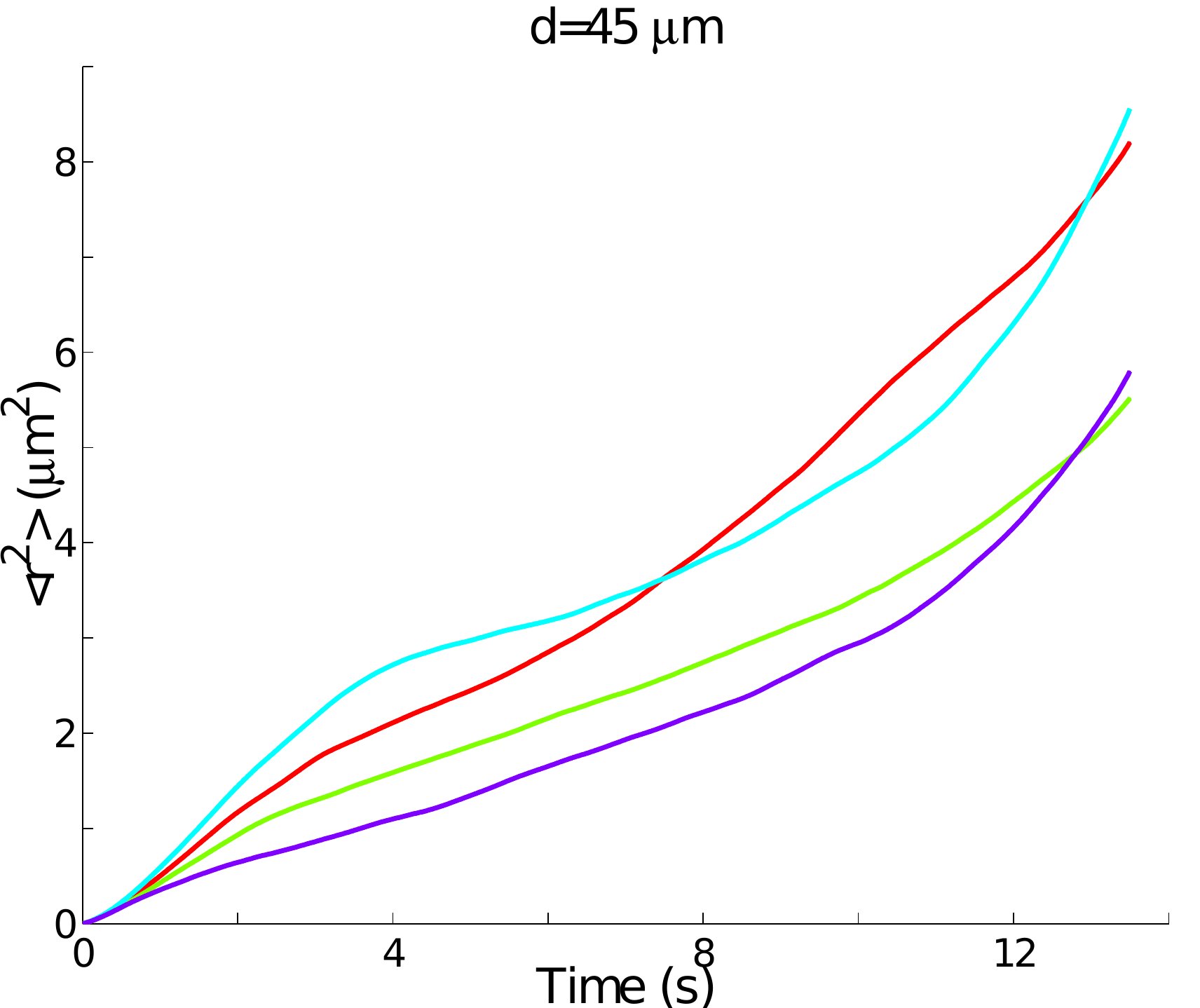}  }
     \subfigure[]{		\label{fig:msd4}  \includegraphics[width=0.4\textwidth]{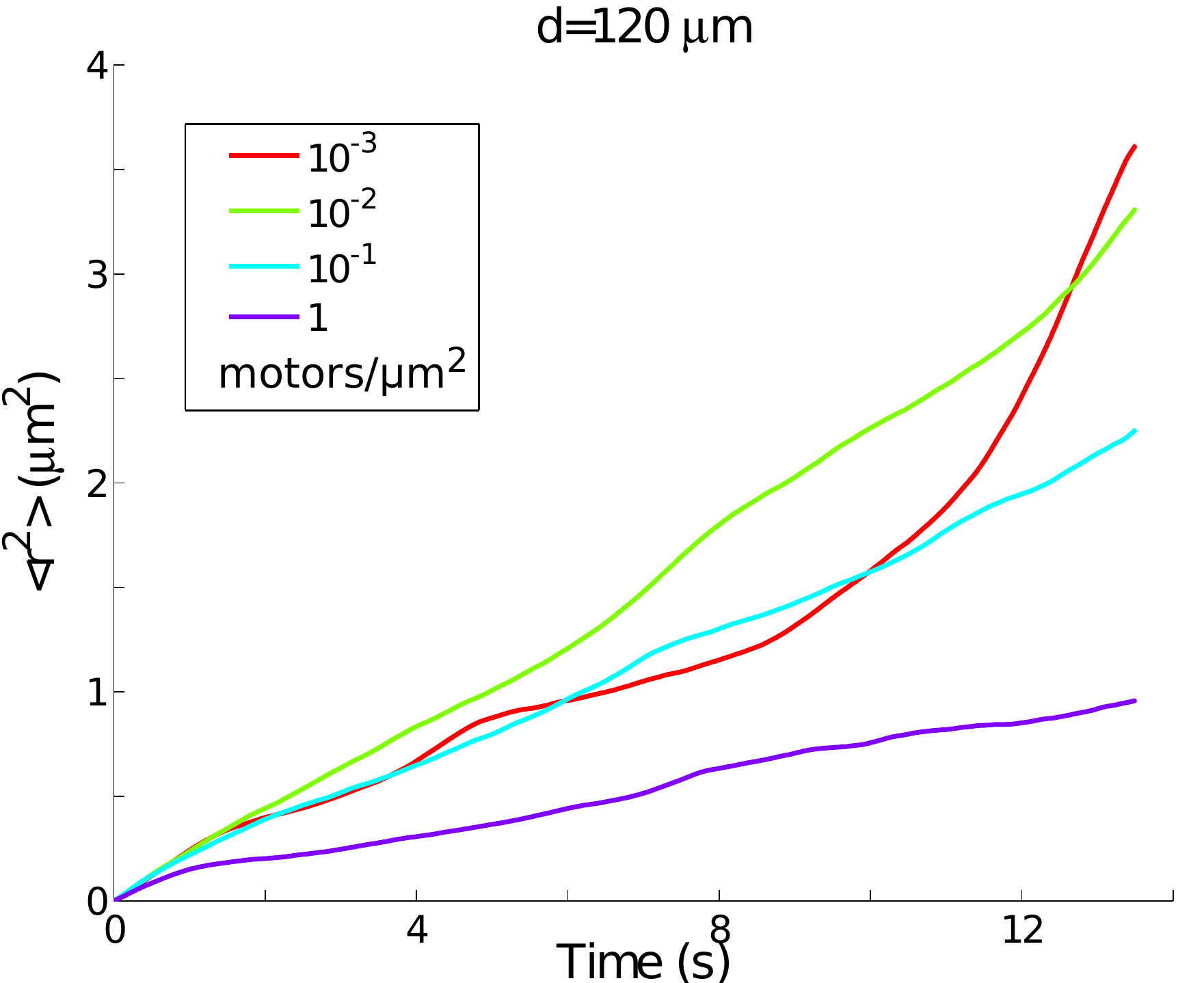}  }
     \subfigure[]{		\label{fig:alpha}  \includegraphics[width=0.4\textwidth]{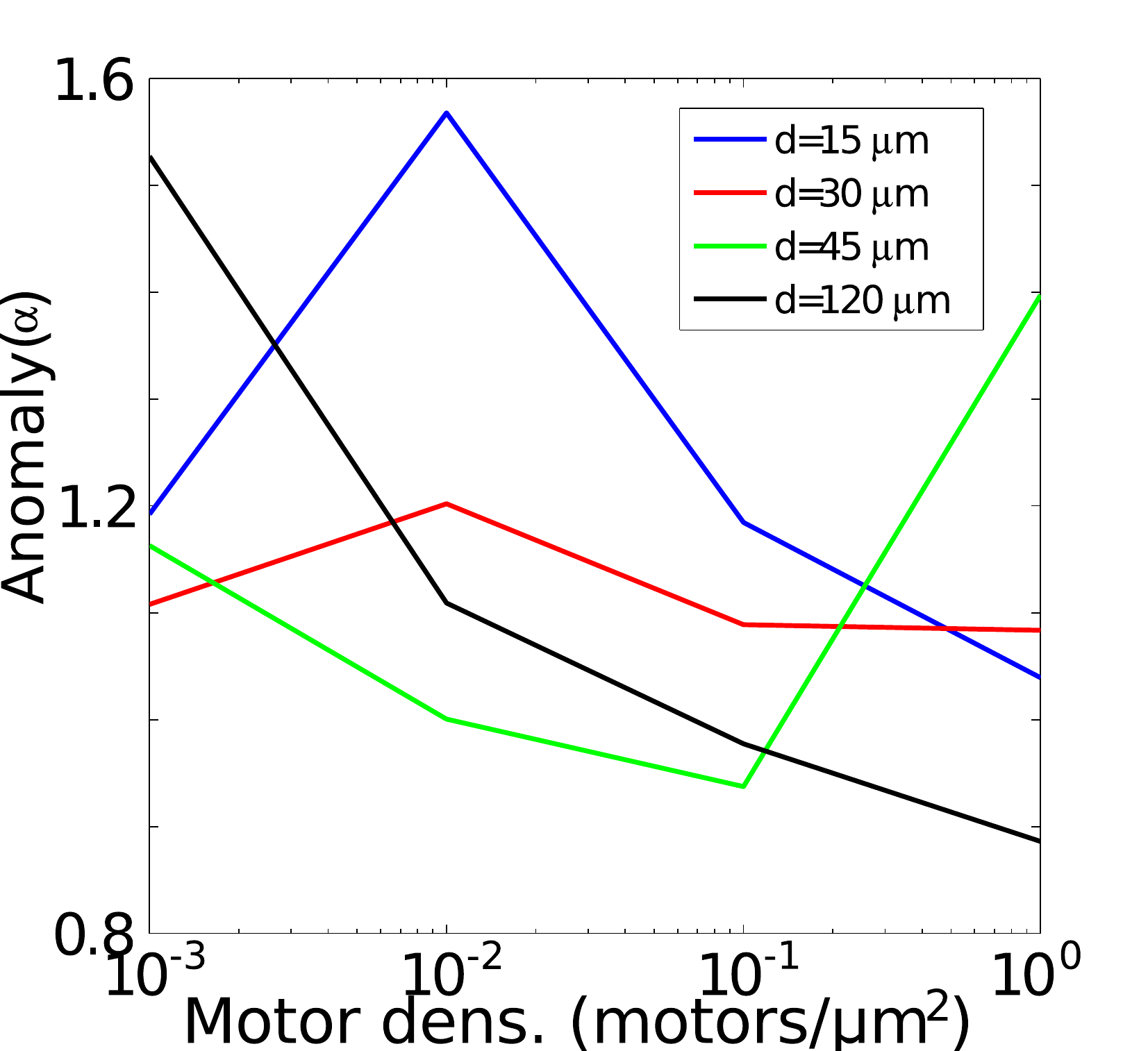}  }
     \subfigure[]{		\label{fig:Deff}  \includegraphics[width=0.4\textwidth]{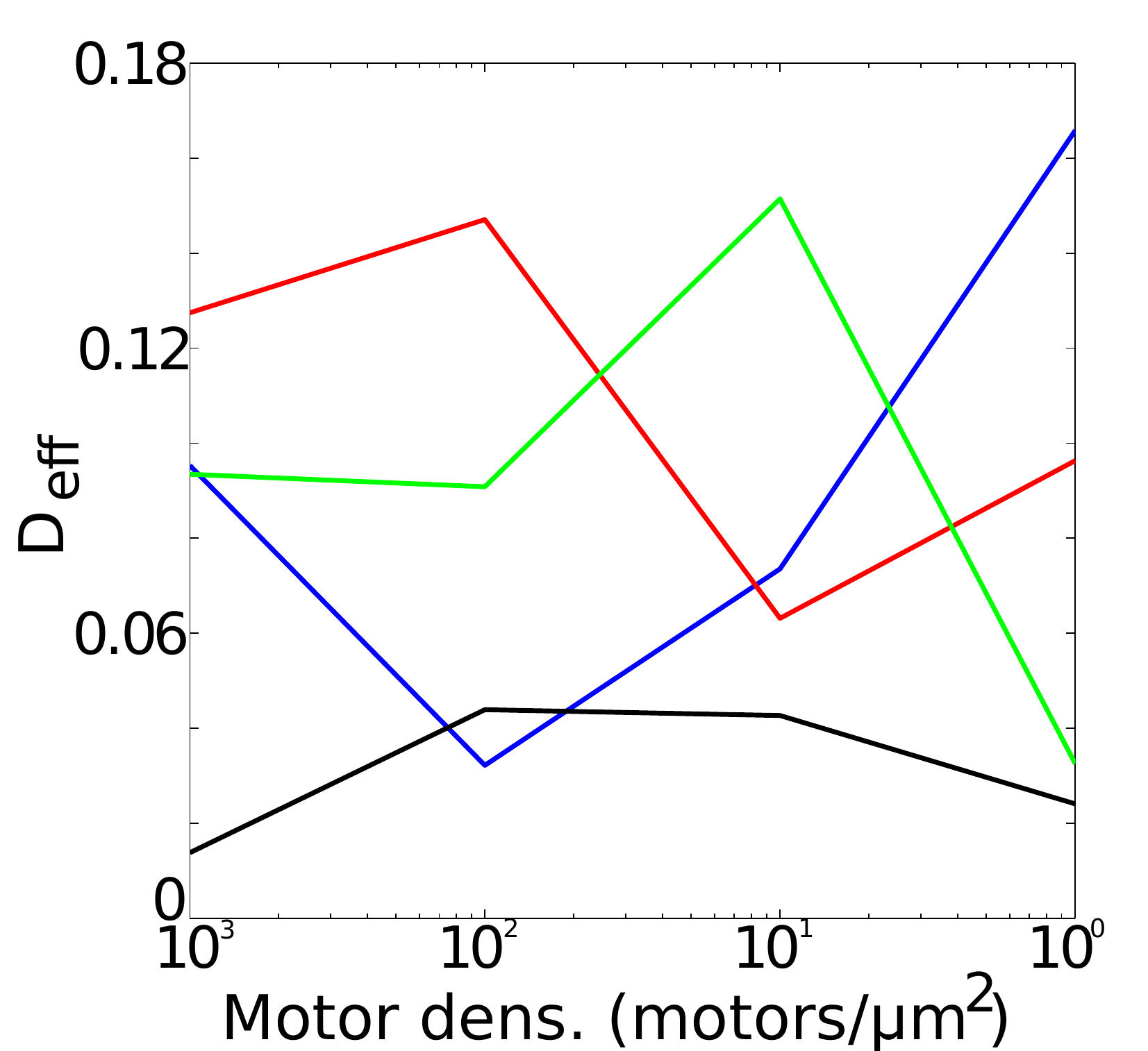}}
   %
\caption{Analysis of diffusion. MSD profiles of asters initialized at \subref{fig:msd1} d=15, \ref{fig:msd2} 30, \subref{fig:msd3} 45 and \subref{fig:msd4} 120 $\mu m$ are calculated for increasing motor densities. The motor densities in \subref{fig:msd1}-\subref{fig:msd4} correspond to $10^{-3}$ (red), $10^{-2}$ (green), $10^{-1}$ (cyan) and $1$ (purple) motors/$\mu m^2$, as shown in the inset in Figure \ref{fig:msd4}. 
\subref{fig:alpha} The anomaly parameter $\alpha$ and \subref{fig:Deff} $D_{eff}$ are plotted as a function of motor density for different initial aster distances from DNA: d=15 (blue), 30 (red), 45 (green) and 120 $\mu m$ (black). All calculations were averaged over 10 asters.}
\label{fig:densmsd}
\end{center}
\end{figure}

In order to test these predictions, we fit the MSD to the anomalous diffusion model (Equation \ref{eq:msd}). We find the magnitude of the MSD is greatest for highly asymmetric asters (Figure \ref{fig:msd1},\subref{fig:msd2}), less for those at the edge of the gradient (Figure , \ref{fig:msd3}) and half for those far away from the gradient (Figure \ref{fig:msd4}). The difference in MSD validates hypothesis (i), i.e. the length dependence of motor binding and resultant displacement. The $\alpha$ value indicates the extent of deviation from diffusive movement. At $d = 45 \mu m$, asters are expected to be maximally asymmetric and $\alpha$ initially decreases with increasing motor density (up to $10^{-1} motors/\mu m^2$) and then increases. The initial decrease is due to the increase in diffusive movement due to stochastic binding (Figure \ref{fig:Deff}). Beyond a critical density  of motors ($>10^{-1}$ motors/$\mu m^2$), the anomaly parameter increases based on hypothesis (ii), i.e. a high asymmetry with sufficient force generators results in directed motion. The dependence of $\alpha$ with increasing motor density (Figure \ref{fig:alpha}) for asters completely inside the zone of asymmetry ($d = 15 \mu m$) surprisingly shows a peak at $10^{-2} motors/\mu m^2$ dropping off in either direction. This suggests that a critical density of motors leads to increased super-diffusive movement. Below and above this value of motor density asters either have too few motors or too many bound motors, respectively. The anomaly parameter of asters $\alpha$ remains $\sim 1.1$  at a distance $d = 30 \mu m$ independent of motor density. To our surprise, an increase in motor density for symmetric asters ($d=120 \mu m$) leads to a decrease in the anomaly parameter. The mean velocities for a given initial distance from DNA, show little difference with changing motor density (Figure \ref{motdens-vel}). Thus our model simulations make experimentally testable predictions of transitions from diffusive to vectorial transport for both symmetric and asymmetric asters, dependent on motor densities.

\section{Conclusions}

We have found that centrosomal MT asters move preferentially towards DNA in experiments with meiosis II arrested {Xenopus} egg extracts. Our  quantitative study maintains a realistic uniform size and density of DNA and substantially improves on previous work \cite{Carazo-Salas2003}. The 2D time-lapse tracks and net displacement distributions suggest a degree of randomness in the motion. Both qualitatively and using multiple measures of directionality, we find that asters show a directional bias in their motion. However, aster velocities (Figure \ref{fig:ex_vel}) and directionality measures (Figure \ref{fig:ex_dmc}, \subref{fig:ex_cos}) lack an obvious distance dependence. The analysis of experimental data is also more detailed than similar work on {\it in vivo} movements of smaller MTOC asters in mouse oocytes \cite{schuh2007} and results in improved statistics over previous work in {\it Xenopus} oocytes \cite{Karsenti1984b} and extracts \cite{Carazo-Salas2003}. 

In our model the centering movement of asters is the result of: (i) a gradient MT dynamic instability, and (ii) MT-length dependent forces generated by molecular motors immobilized on the substrate. In this scenario, a tug-of-war emerges from the geometry of the system and asters will move only if the symmetry is broken. Gradients of RanGTP dependent MT dynamic instability in mitotic {\it Xenopus} extracts and cells \cite{Caudron2005, Wilde2001} are know to generate a length asymmetry in asters \cite{Athale2008, Caudron2005}. The mechanics of force generation by minus-ended motors is based on experimentally measured single-molecule properties of dynein which include: (a) load-dependent step-size reduction, (b) reversals by super-stall forces and (c) force dependent detachment rates. Using this setup we have found that aster gliding movement is 100-fold slower than the intrinsic velocity of single dynein motors (Table \ref{tab:motpar}, Figure \ref{fig:simexp_velDist}) as a consequence of the 2D tug-of-war. The two-fold length asymmetry in asters \cite{Athale2008, Carazo-Salas2003} is sufficient to result in directional motion in experiment and simulation (Figure \ref{fig:simexp_costhDist}, \subref{fig:simexp_dmcDist}). In additional calculations, we find that a force independent model of motor mechanics can also move asters and qualitatively reproduce the velocity and dmc profiles from experiment (Figure \ref{forceindepModel}). However insights from single molecule experiments suggest such a simple model is physically unrealistic and do not analyze it further.

The qualitative agreement of our simulated velocities with experiments demonstrates that the expected tug-of-war can be resolved with biologically realistic parameters resulting in small (upto two-fold) length asymmetries. This is in contrast to the extreme degrees of asymmetry (10-100 fold) examined in work by Leduc et al. \cite{Leduc2010} where they examined a tug-of-war in a 1D anti-parallel MT array gliding on a sheet of kinesin. The simulated distance dependent velocity profiles are independent of motor densities in the range of $0.001$ to 1 $motors/\mu m^2$. Models and experiments with kinesin based microtubule doublet gliding \cite{Leduc2010} had used a motor density value orders of magnitude higher. Indeed, work on mitotic aster rotation in sea urchin embryos has shown that a low density of force generators ($< 1$ $motors/\mu m^2$) are sufficient to reorient asters in a length-depdenent manner \cite{Minc2011} suggesting our choice of motor densities is in the physiological range. 

While the velocity profiles of experiment and simulation appear to match quite well, the trend in the distance binned average directionality measures (dmc, $\cos(\theta)$) from experiment differs from the simulated values. We believe this difference is potentially due to the low sampling density of the experimental data, combined with the nature of the measures of dmc and $\cos(\theta)$. The frequency distribution of dmc (Figure \ref{distrDmc}) and $\cos(\theta)$ (Figure \ref{distrCos}) in directed random-walk simulations is broad, even when the weight of directionality is at 40\% ($W_d \sim 0.4$). A better sampling in future experiments could help to test our model predictions. The simulated mean directionality measures also show peaks at different distances from the DNA as a function of the motor density, providing yet another test of the model.

Our model also predicts that changes in motor density result in a transition of aster movement from effective diffusion to vectorial transport. When the aster asymmetry is highest ($d = 45 \mu m$), the anomaly parameter ($\alpha$) increases with increasing motor density as expected indicating an increasingly vectorial movement (Figure \ref{fig:msd3}, \subref{fig:alpha}). However in the case of asters very close to DNA ($d=15 \mu m$) $\alpha$ is highest at $10^{-2} motors/\mu m^2$, dropping off on either side (for higher and lower densities). Such a dependence on motor density suggests that when the MT lengths in the aster are radially symmetric,  (i) the `duty ratio' \cite{Howard2001book} characteristic of the motor and (ii) the average MT lengths together determine the density at which $\alpha$ is the highest. This is the critical density of motors which results in super-diffusive motion. At densities higher than the critical density, the asters are in a mostly bound state, while at lower densities the force generators are very few- both scenarios lead to effectively diffusive motion. When the aster is far from any gradient ($d=120 \mu m$), the MT lengths are also radially symmetric \cite{Athale2008} but shorter than in asters inside the stabilization zone. Given a constant `duty ratio', the value of $\alpha$ is the highest at $10^{-3} motors/\mu m^2$ and decreases with increasing density (Figure \ref{fig:alpha}). This suggests that at low densities of motors, random imbalances in forces cause vectorial movement. If a motor attaches, it tugs at the aster moving it vectorially. As motor densities increase, the number of binding events increase in all directions, resulting in more effectively diffusive movement. This suggests that in the absence of a directional cue, simply modulating the motor density can regulate the qualitative nature of aster movement. Thus, modeling dynein mechanics based on single-molecule experimental insights produces new insights into the collective behaviour of aster-motility, as compared to implicit models of motors in {\it C. elegans} aster motility \cite{Kimura2007, Kimura2010}. The dynamics of movement of asymmetric asters at distances of 30 and 45 $\mu m$ from the DNA center suggest that they are qualitatively different. Their movement results from a combination of length asymmetry and motor density driven effects. As a result small increases in motor density lead to a slight drop in directionality, while higher densities lead to an increase. These predictions for both symmetric and asymmetric asters can be tested using an {\it in vitro} reconstitution approach \cite{Laan2012}, and would serve as an additional means of understanding the collective behaviour of dynein.

We find that the standard deviation of $\alpha$ is orders of magnitude smaller than the differences between conditions. To examine the heterogeneity in the MSD plots, we examined the complete distribution for representative cases of asters at different distances from DNA (Figure \ref{distrMSD}). Distance from DNA results in differences in magnitude of MSD, a prediction that can be experimentally tested.

The observed convergence of all asters onto the DNA patterns is a result of the anti-TPX2 antibody treatment to inhibit Tpx2 activity. Tpx2 mediates nucleation of MTs from the DNA \cite{bird2008} essential for stable spindle assembly \cite{Heald1997}. In its absence, the interaction between centrosomal MTs and DNA is through a chemical reaction-diffusion gradient of stabilization \cite{Caudron2005, Wilde2001,Kalab2006} and kinesins associated with chromosomes, i.e. chromokinesins (reviewed in \cite{Tanenbaum2010}). Despite the relative simplicity of our model and experimental setup, the qualitative agreement between theory and experiment suggests is could serve as a module in a more complex model of spindle assembly. Such a model would include additional interactions with Tpx2-nucleated MTs around DNA and chromokinesin activity.

In conclusion, we have shown that the centrosomal aster movement in {\it Xenopus} extracts with chromatin-DNA is directional. A model of MT asters gliding on dynein motors has been developed that includes detailed motor mechanics and stochastic binding and unbinding. This model can reproduce the qualitative trends in our experimental data. Additionally, the model also predicts that in the absence of a directional cue, aster movement will undergo a transition from vectorial to effective diffusive movement, based on motor density alone. This tradeoff has relevance to the control of centrosomal and MTOC aster positioning. The integration of MT dynamics and motor-mechanics in the model suggests a role for a reaction-diffusion gradient to resolve a tug-of-war resulting from a single motor species. 

\section{Acknowledgements}
CA acknowledges IISER Pune core-funding. We thank Hemangi Chaudhari for re-evaluating the single particle tracking data.

\section*{References}
\bibliographystyle{unsrt}
\bibliography{aster-motility}

\newpage

\section{Tables}


\begin{table}[ht!]
\begin{tabular}{|p{6cm}|p{2cm}|p{2cm}|p{2cm}|}
  \hline
  {\bf Parameter} & {\bf Value}     & {\bf Units} \\
  \hline
  \hline
 Time step & $0.05$ & s \\
  \hline
 Filament section & 0.5  & $\mu m$ \\
  \hline
 Simulation box (2D square) & 400x400 & $\mu m$ x $\mu m$ \\
  \hline
 Total simulated time & 60  & min  \\
  \hline
 Thermal energy ($k_B$T) & $4.2$  & $pN \cdot nm$ \\
  \hline
 Microtubule bending modulus & 20 & $pN \cdot \mu m^2$ \\
  \hline
 Fluid viscosity & 0.05 & $pN \cdot s /\mu m^{2}$  \\
  \hline
 Number of MTs per aster & 100 & - \\
 \hline\hline
\end{tabular}
\begin{flushleft}
\end{flushleft}
\caption{\bf{The simulation parameters.} }
\label{tab:simpar}
\end{table}


\begin{table}[ht!]
\begin{tabular}{|p{6cm}|p{2cm}|p{2.5cm}|p{2cm}|}
  \hline
  {\bf Motor parameter} & {\bf Value}     & {\bf Units} & {\bf Reference} \\
  \hline
  \hline
Motor link rigidity ($k$) & 0.1 & $pN/nm$ & \cite{Oiwa2005} \\
  \hline
Motor stall force ($f_{0}$) & 1.75 & $pN$ & \cite{Mallik2005, Mallik2004} \\
  \hline
    Density of motors & 0 to 1 & $motors/\mu m^{2}$ & estimate \\
  \hline 
    Motor speed ($v_{mot}$) & 2 & $\mu m/s$ & \cite{Howard2001book}, \cite{Nedelec2002} \\
  \hline
  Motor attachment rate ($r_{attach}$) & 12 & $1/s$ & \cite{Howard2001book}, \cite{Nedelec2002} \\
  \hline  
   Motor basal detachment rate ($r^{'}_{detach}$) & 1 & $1/s$ &  \cite{Nedelec2002} \\
  \hline  
   Motor detach rate from tips ($k^{tip}_{detach}$) & 1 & $1/s$ & \cite{Howard2001book}, \cite{Nedelec2002}\\
     \hline  
   Motor attachment distance ($d_{attach}$) & 0.02 & $\mu$m & \cite{Nedelec2002} \\
     \hline  
  \hline\hline
\end{tabular}
\begin{flushleft}
\end{flushleft}
\caption{\bf{Molecular motor parameters.} }
\label{tab:motpar}
\end{table}

\begin{table}[ht!]
\begin{tabular}{|p{4cm}|p{5cm}|p{1.5cm}|p{2cm}|}
  \hline
  {\bf MT Dynamics parameter} &  {\bf Cytoplasm; Chromatin}     & {\bf Units} & {\bf Reference} \\
  \hline
  \hline
  Growth velocity ($V_g$)  & 0.196;  0.196 & $\mu m/s$ &   \cite{Carazo2001, Wilde2001}\\
 \hline  
  Shrinkage velocity ($V_s$)  & 0.325;  0.325 & $\mu m/s$ & " \\
 \hline  
Rescue frequency ($f_r$)  & 0.0048 ($f_r^{min}$); 0.012 ($f_r^{max}$) & $1/s$ & " \\
 \hline  
Catastrophe frequency ($f_c$)  & 0.049 ($f_c^{max}$);  0.03 ($f_c^{min}$) & $1/s$ & " \\
  \hline\hline
\end{tabular}
\begin{flushleft}
\end{flushleft}
\caption{\bf{MT dynamic instability parameters.} }
\label{tab:mtdynpar}
\end{table}

\newpage
\clearpage
\section{Figures}

\clearpage
\newpage
\section*{Supporting Figures}
\setcounter{figure}{0}
\makeatletter 
\renewcommand{\thefigure}{S\@arabic\c@figure}

\begin{figure}[ht]
\begin{center}
     \subfigure[]{ \label{sfig:rwplt}          \includegraphics[width=0.8\textwidth]{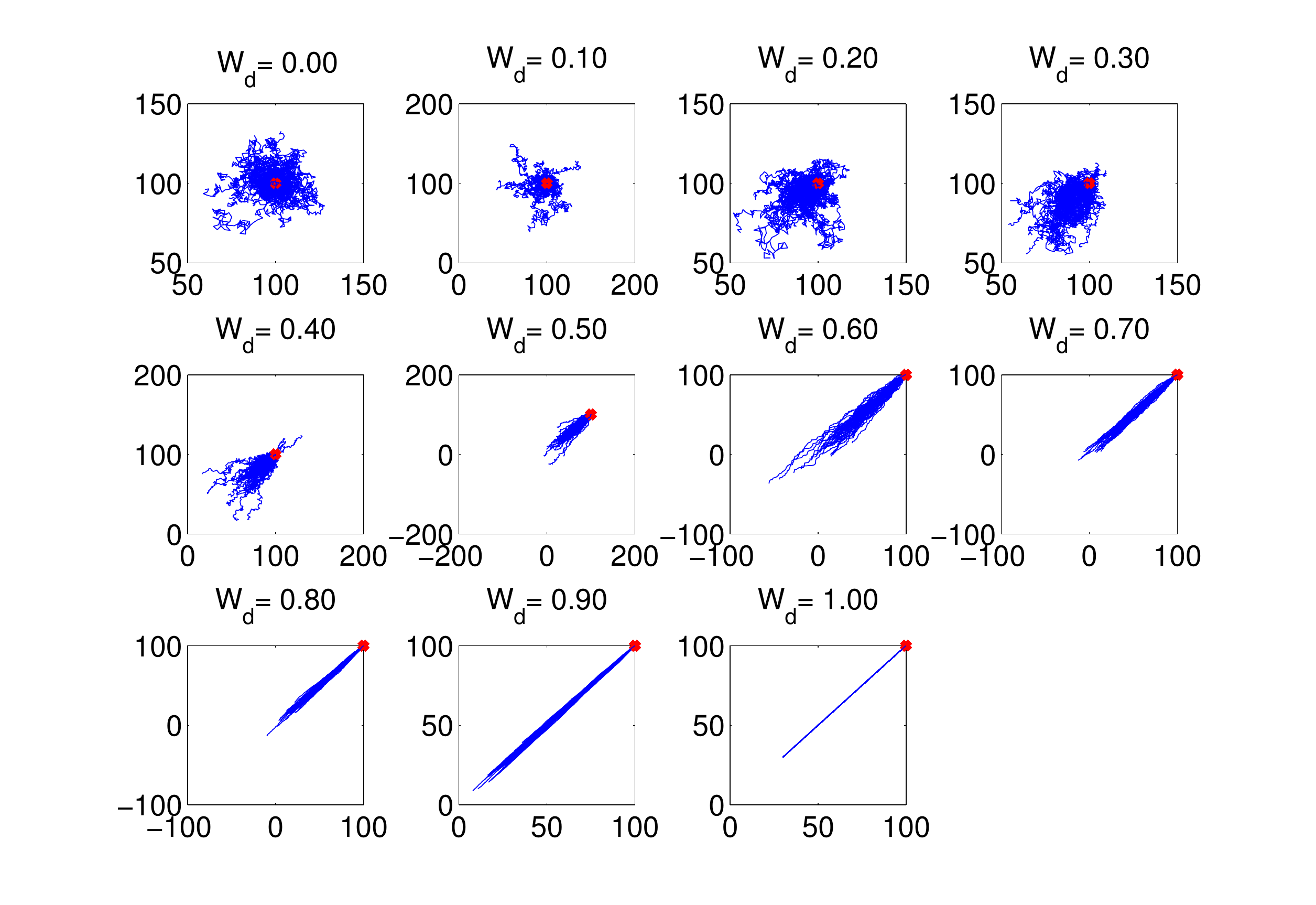}}\\
     \subfigure[]{  \label{sfig:vel}              \includegraphics[width=0.4\textwidth]{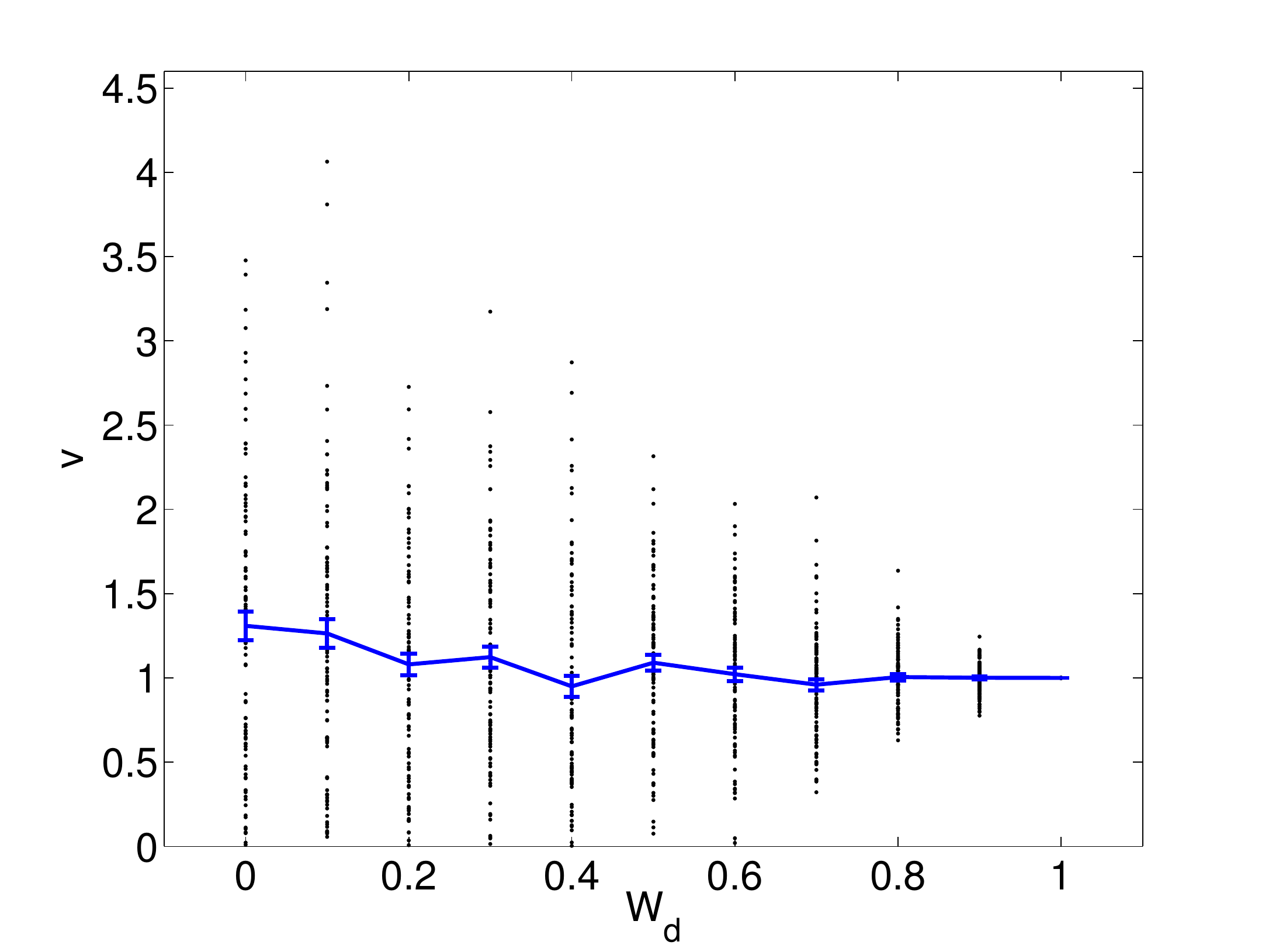}}
     \subfigure[]{  \label{sfig:chi}              \includegraphics[width=0.42\textwidth]{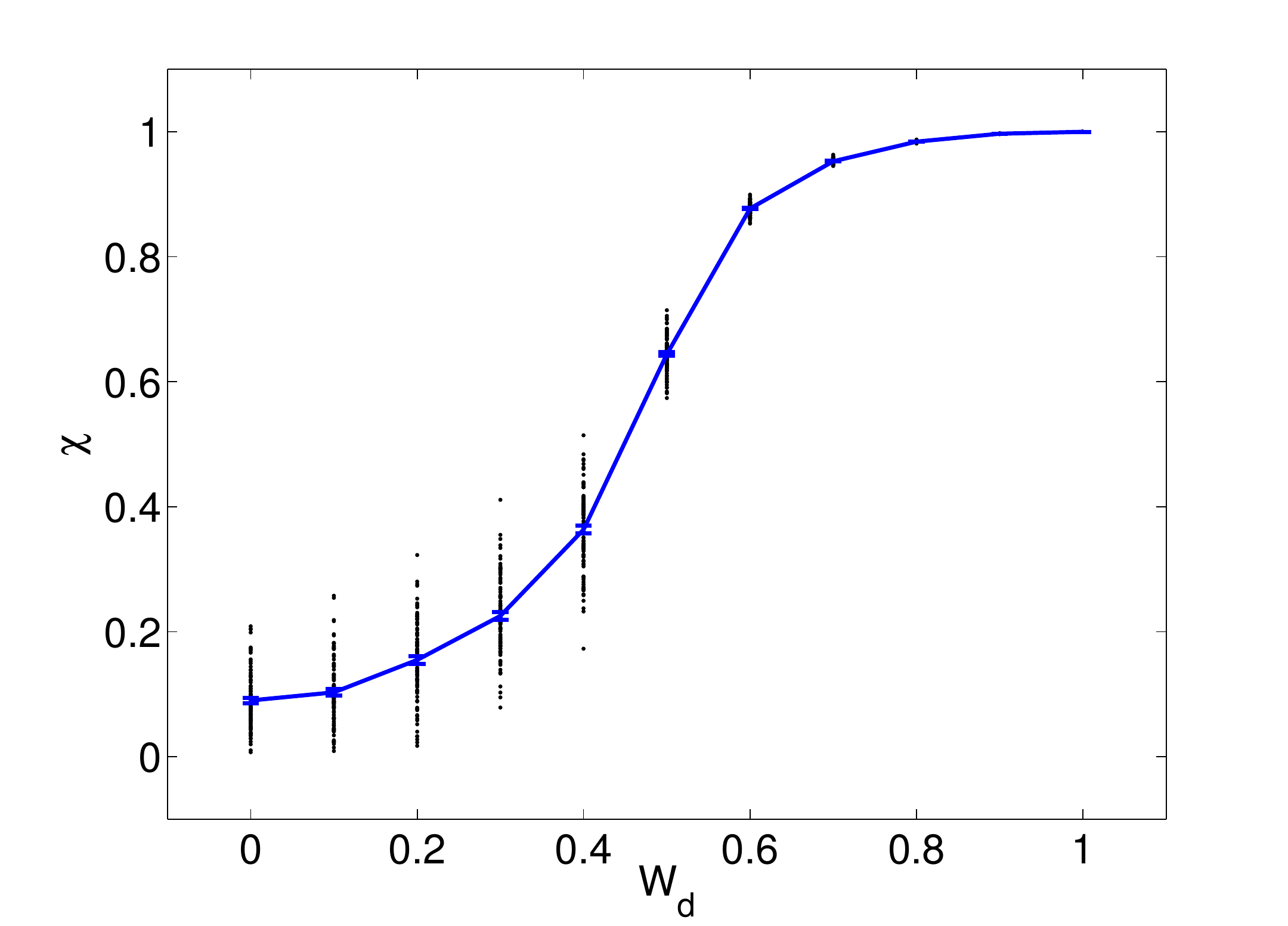}}\\
     \subfigure[]{  \label{sfig:dmc}            \includegraphics[width=0.4\textwidth]{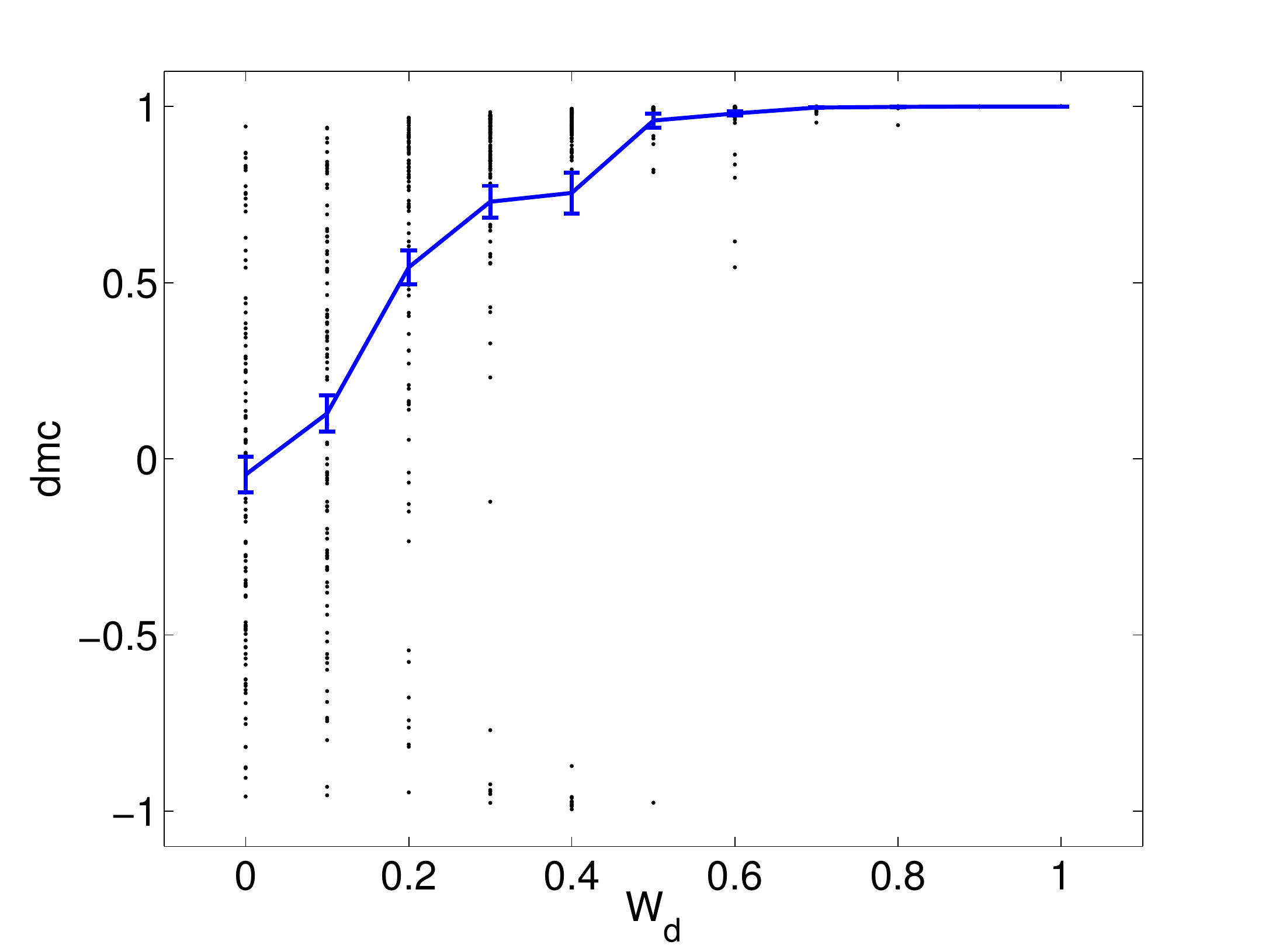}}
     \subfigure[]{  \label{sfig:costheta}  \includegraphics[width=0.42\textwidth]{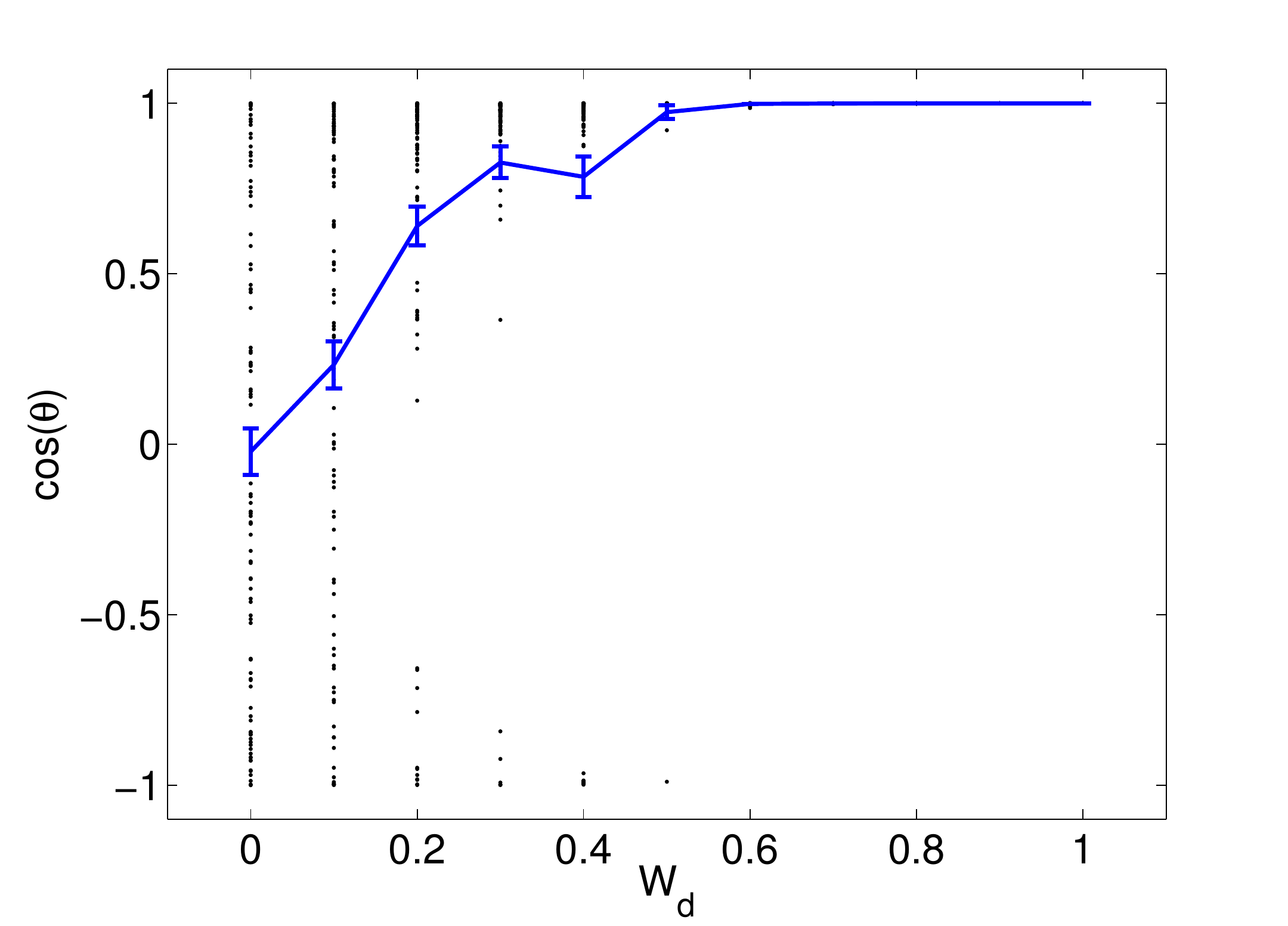}}
\end{center}
\caption{ \subref{sfig:rwplt} The XY-trajectories (n= 100 particles) for increasing weights of directional movement ($W_d$). The particles are all initialized at the same position (red) and the target is the origin. \subref{sfig:vel} The magnitude of the velocity, \subref{sfig:chi} tortuosity ($\chi$), \subref{sfig:dmc}directional motility coefficient (dmc) and \subref{sfig:costheta} $\cos(\theta)$ are plotted against $W_d$. The raw values (black) are superimposed with the mean (blue). The error bars indicate the standard error of mean (s.e.m.).}
\label{sup:randwk}
\end{figure}

\begin{figure}[ht]
\begin{center}
\includegraphics[width=0.9\textwidth]{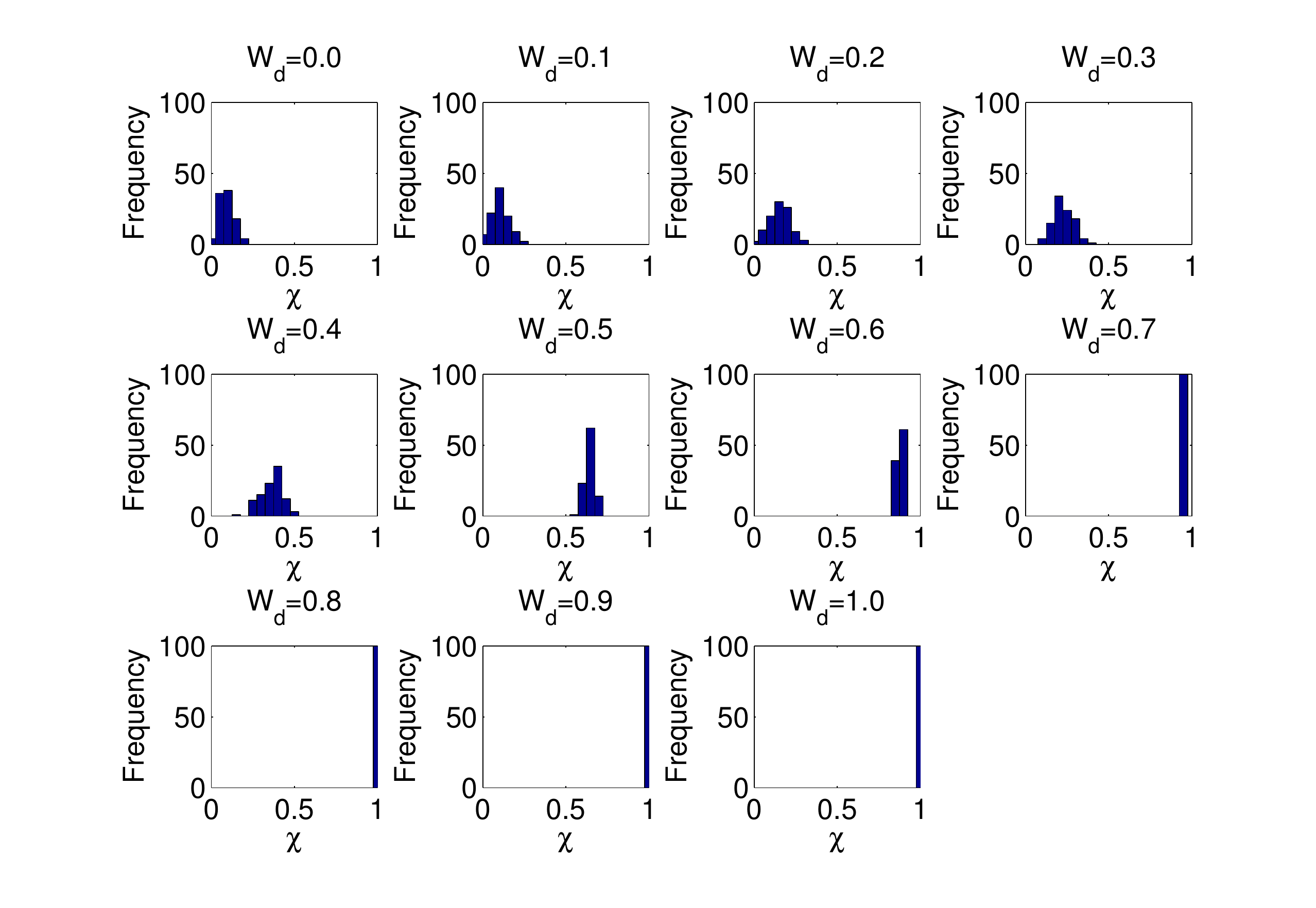}  
\caption{The frequency distributions of tortuosity for increasing values of $W_d$ are plotted for 100 particles based on data from the directed random-walk simulations.}
\label{distrChi}
\end{center}
\end{figure}

\begin{figure}[ht]
\begin{center}
	\includegraphics[width=0.9\textwidth]{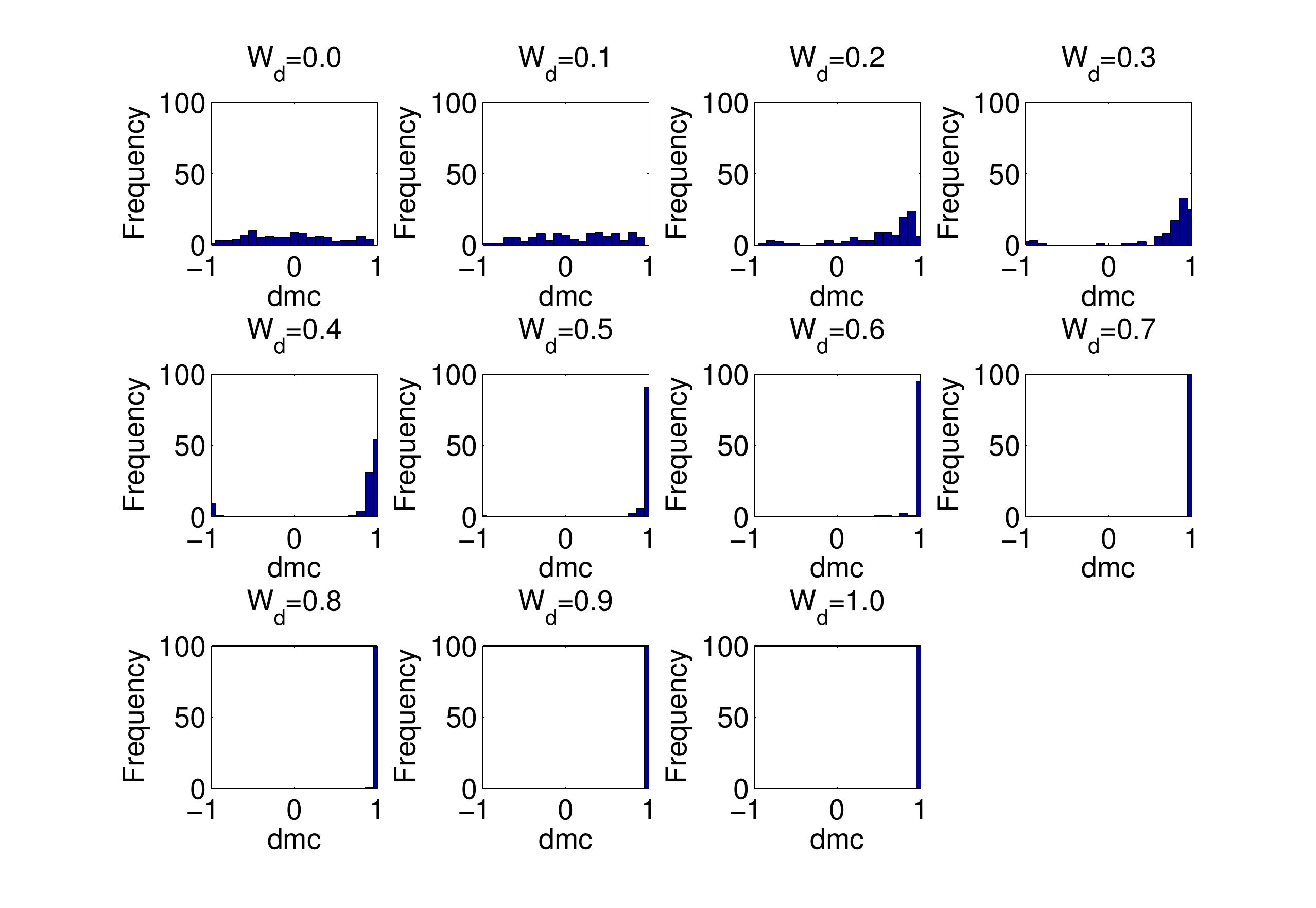} 
\caption{The frequency distributions of dmc for increasing values of $W_d$ are plotted for 100 particles based on the directed random-walk simulations.}
\label{distrDmc}
\end{center}
\end{figure}

\begin{figure}[ht]
\begin{center}
	\includegraphics[width=0.9\textwidth]{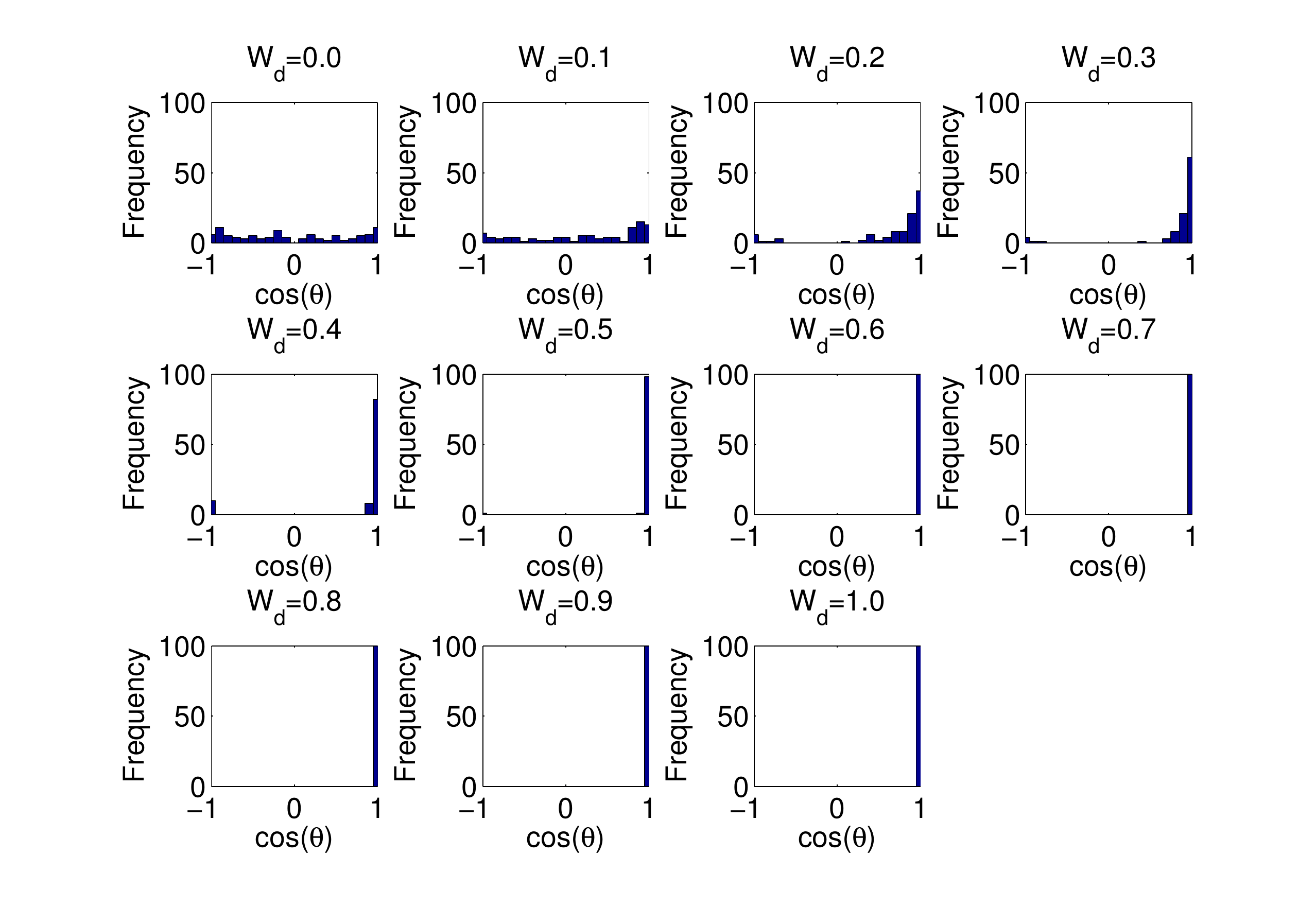} 
\caption{The frequency distributions of $\cos({\theta})$ for increasing values of $W_d$ are plotted for 100 particles based on the directed random-walk simulations.}
\label{distrCos}
\end{center}
\end{figure}

\begin{figure}[ht]
\begin{center}
	\subfigure[]{  \label{sfig:findep-snap}  \includegraphics[width=0.5\textwidth]{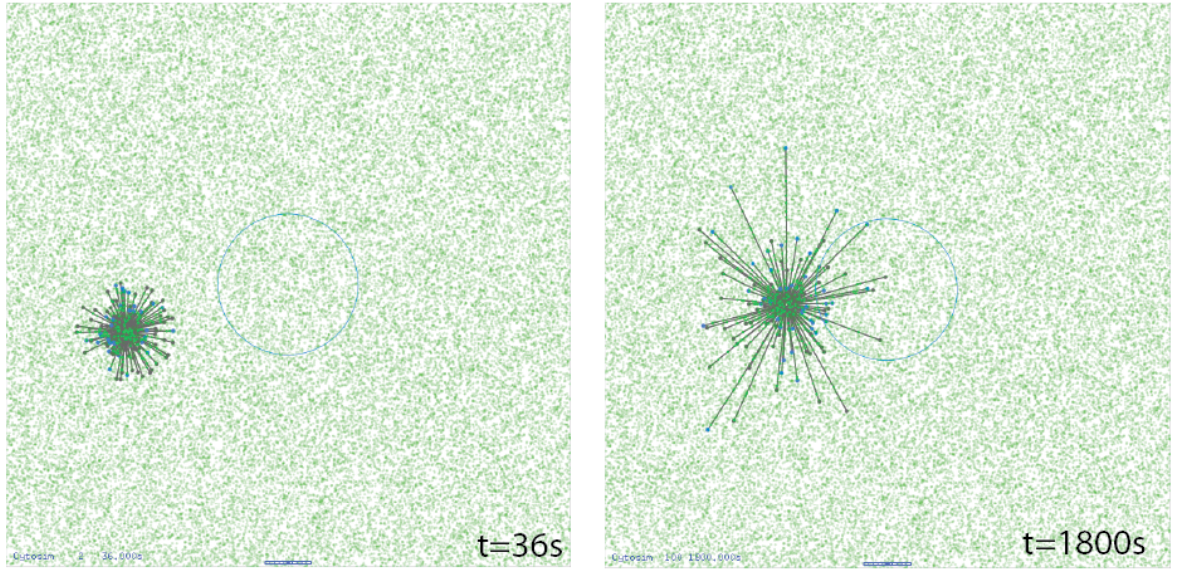}}
	\subfigure[]{  \label{sfig:findep-vel}     \includegraphics[width=0.5\textwidth]{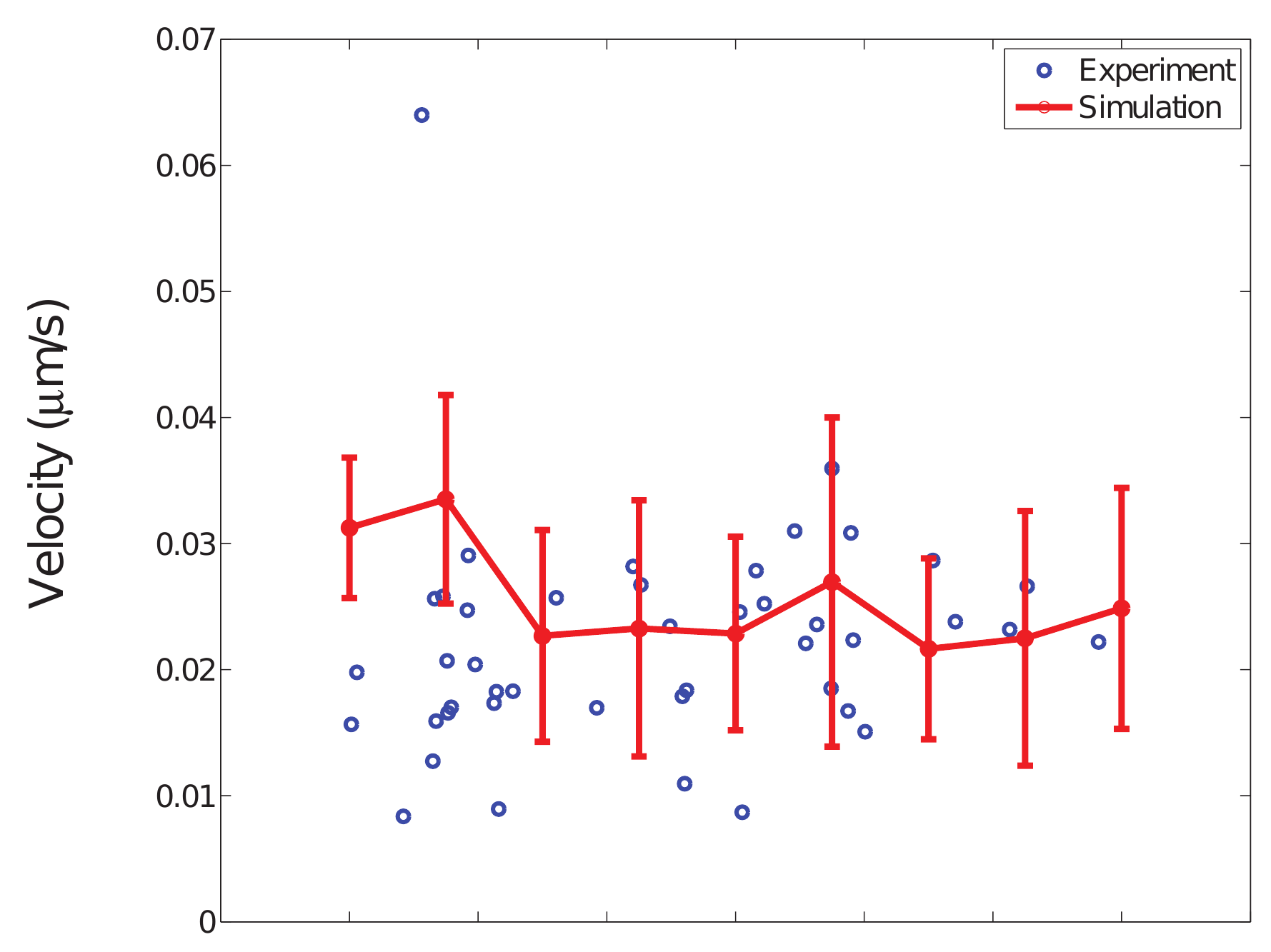} }
	\subfigure[]{  \label{sfig:findep-dmc}   \includegraphics[width=0.5\textwidth]{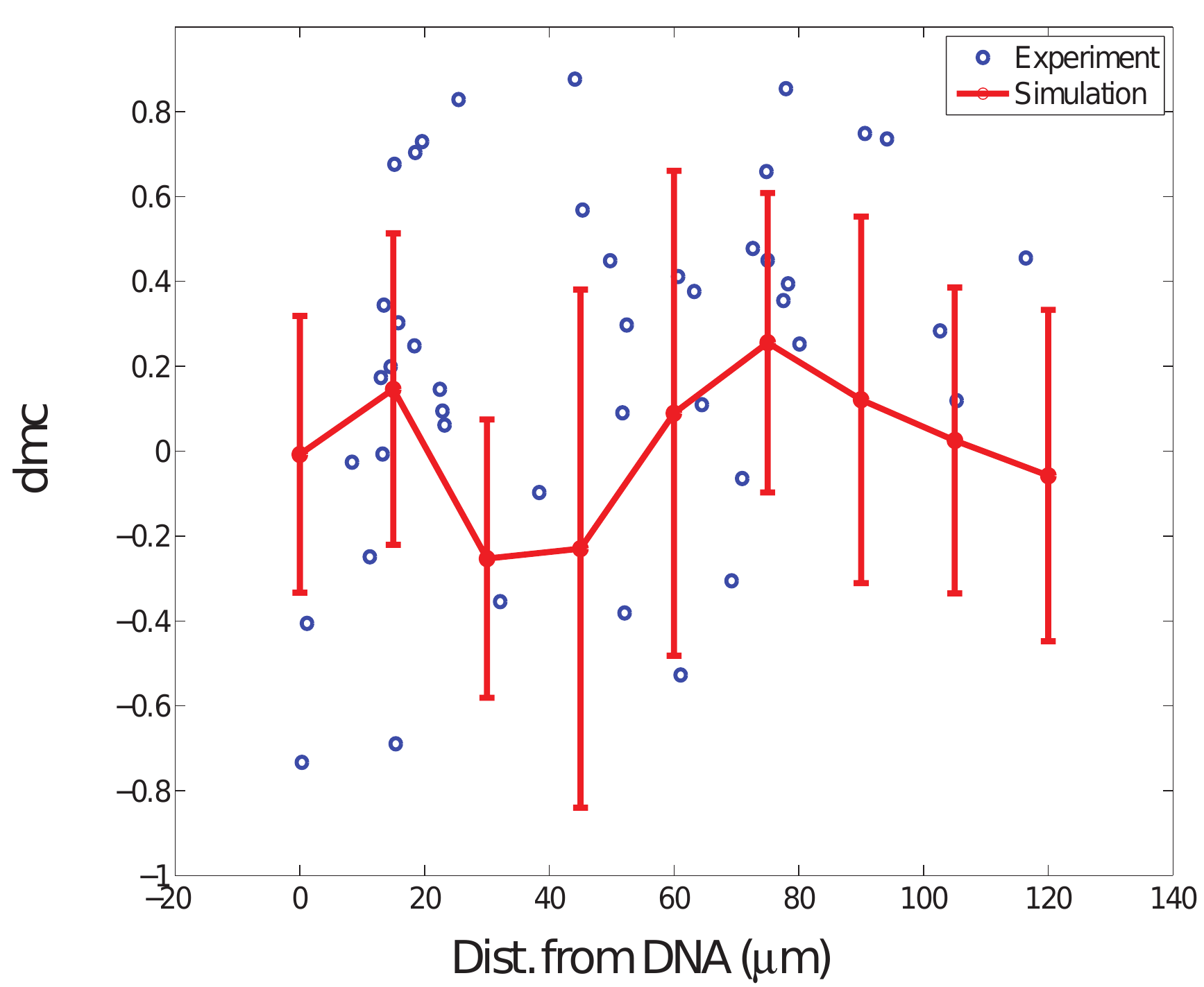} }
\caption{\subref{sfig:findep-snap} A simplified model of motors with a constant attachment ($r_{attach}=12 s^{-1}$) and detachment rate ($r_{detach}=1.5 s^{-1}$), fixed step sizes and no backward stepping was simulated and shows aster movement towards chromatin. \subref{sfig:findep-vel} The mean velocity and \subref{sfig:findep-dmc} dmc profiles for the simulated asters initialized over a range of distances from the chromatin center with a motor density of $4\cdot 10^{-3}$ $motors/\mu m^2$ (red) compared to experimental (black circles) data show qualitative agreement. The error bars are standard deviations. }
\label{forceindepModel}
\end{center}
\end{figure}

\begin{figure}[ht]
\begin{center}
	\subfigure[]{  \label{sfig:msd-d015}  \includegraphics[width=0.4\textwidth]{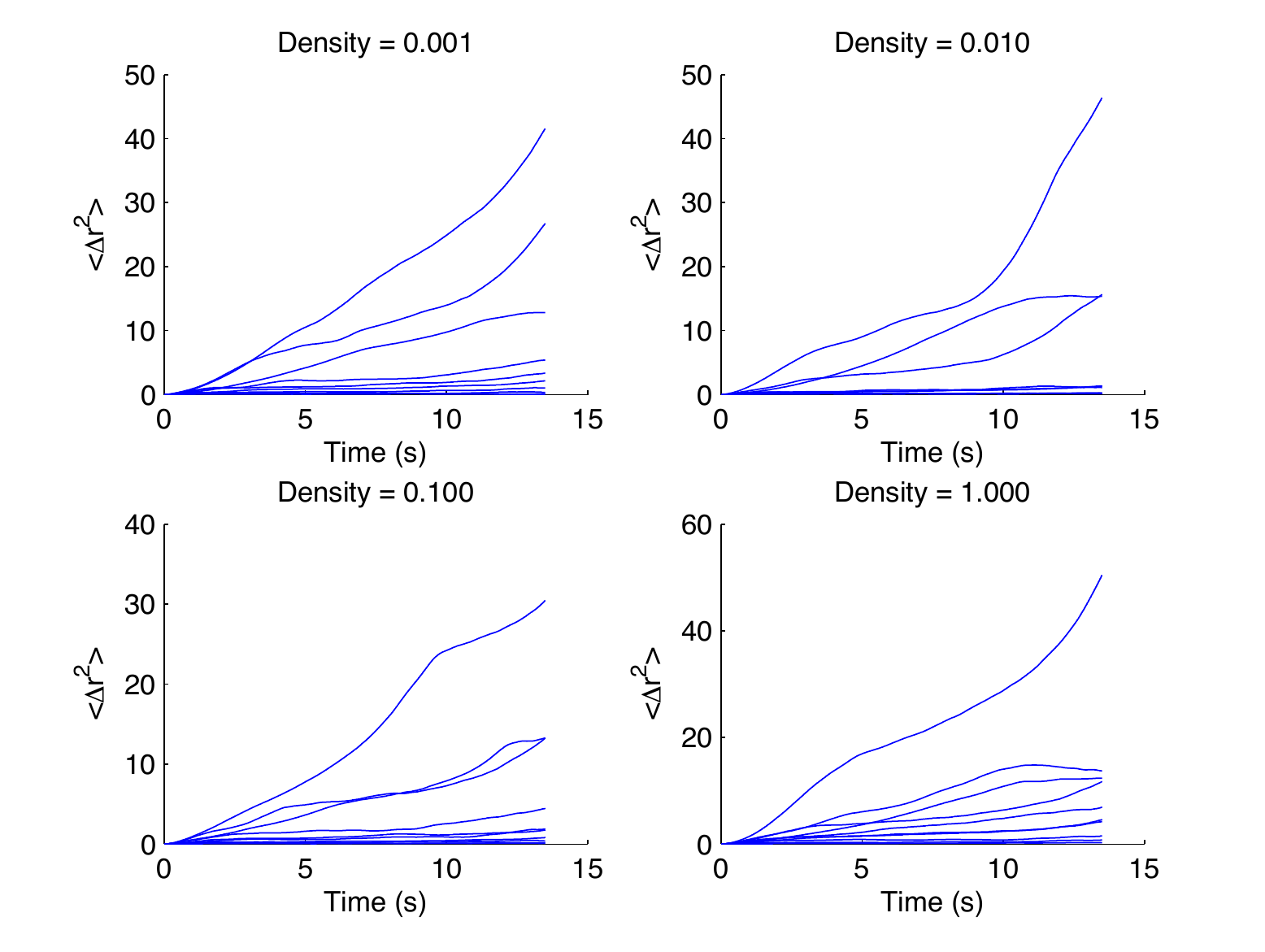}}
	\subfigure[]{  \label{sfig:msd-d030}  \includegraphics[width=0.4\textwidth]{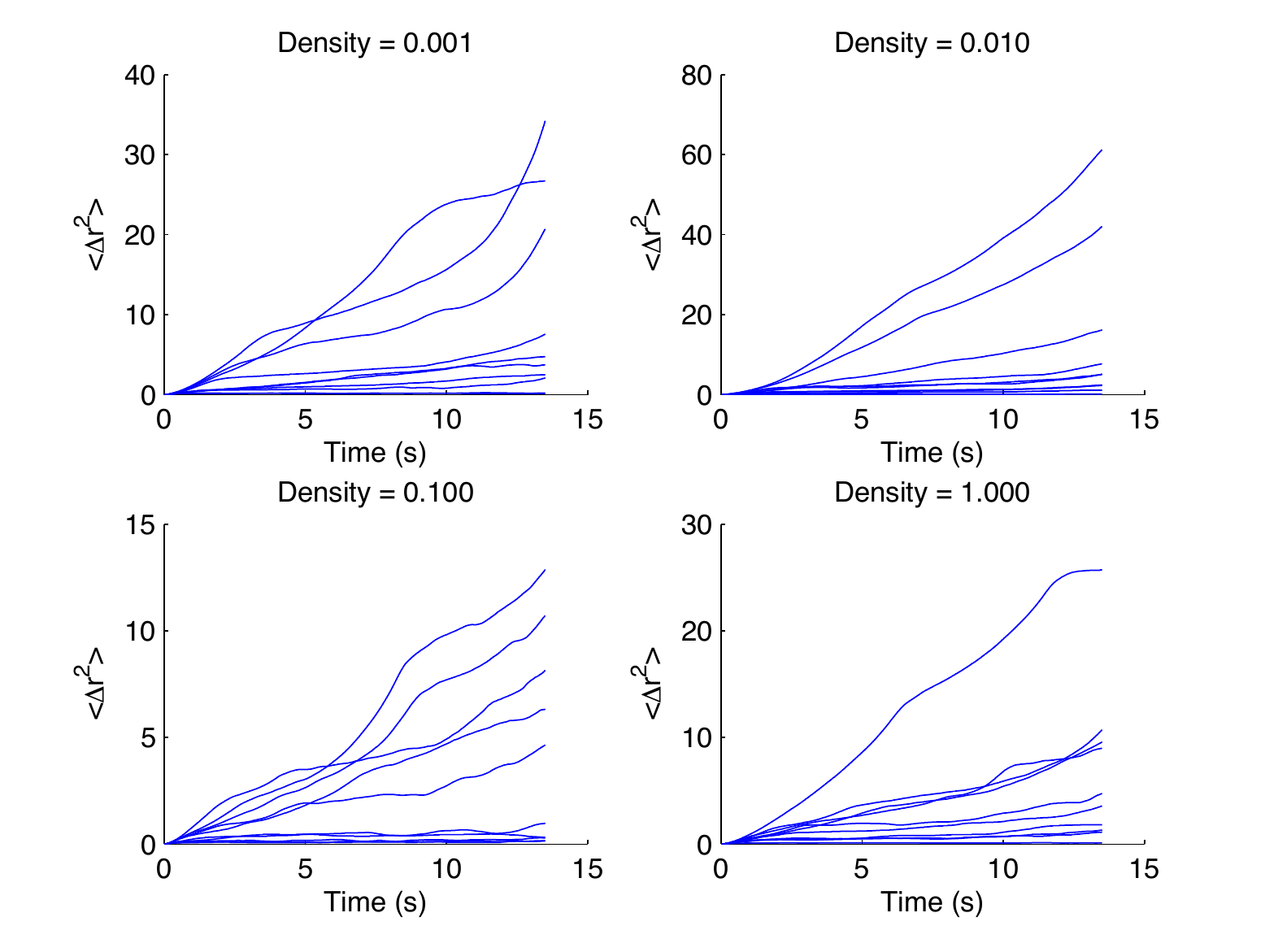}}
	\subfigure[]{  \label{sfig:msd-d045}  \includegraphics[width=0.4\textwidth]{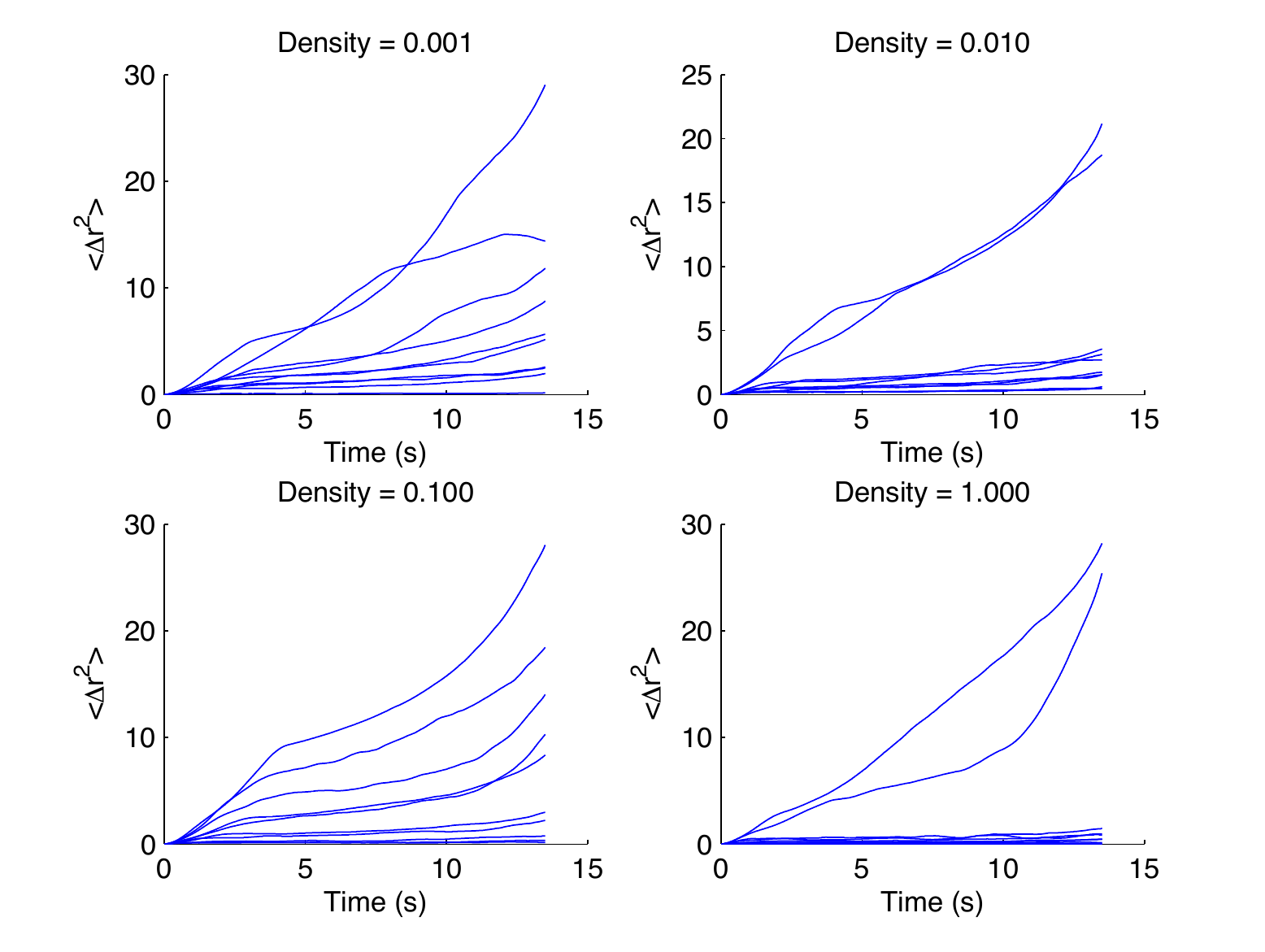}}
	\subfigure[]{  \label{sfig:msd-d120}  \includegraphics[width=0.4\textwidth]{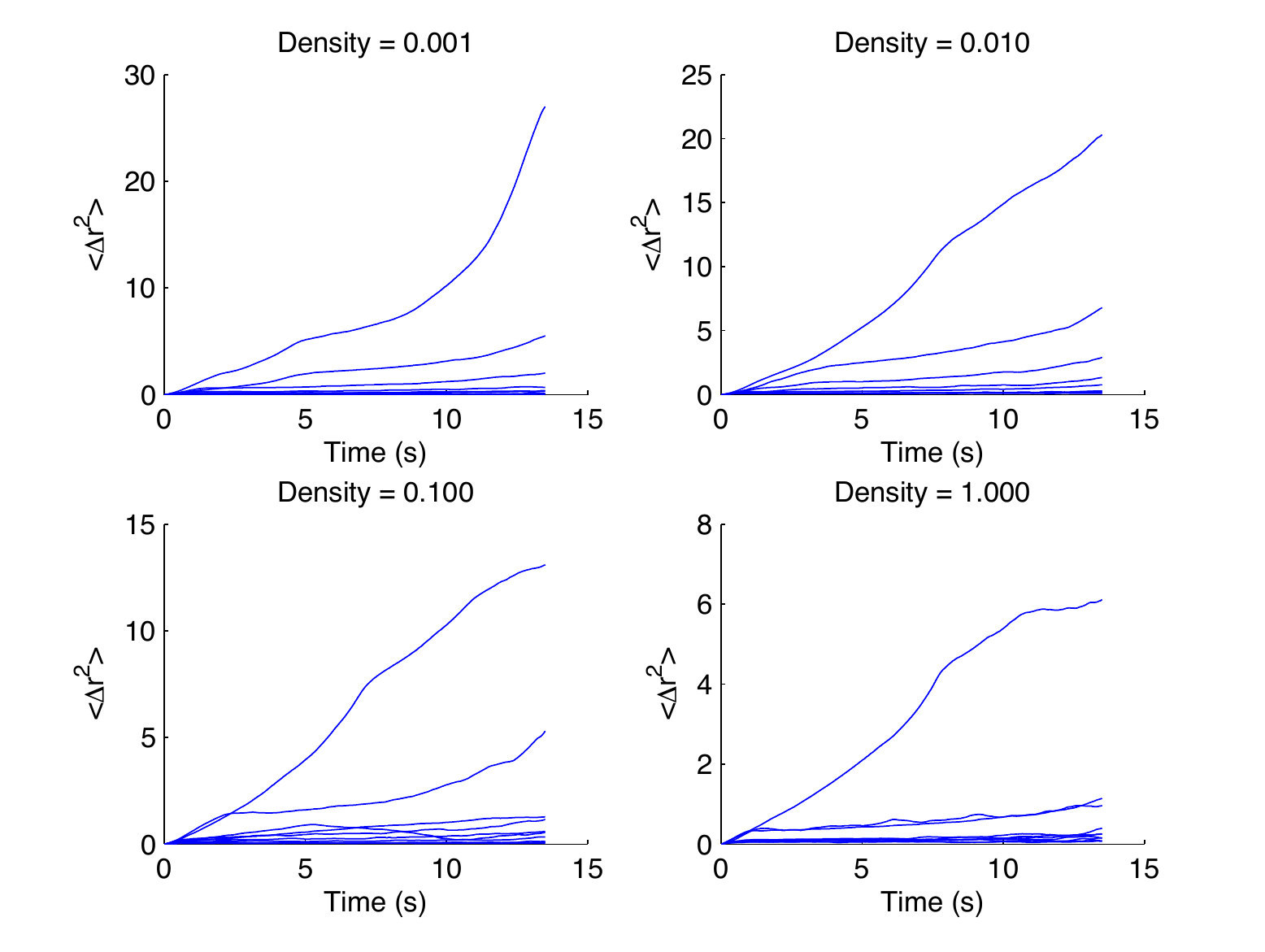}}
\caption{The MSD plots of simulated aster trajectories are plotted for all asters (10) simulated for 3600 s for asters initialized at \subref{sfig:msd-d015} 15, \subref{sfig:msd-d030} 30, \subref{sfig:msd-d045} 45 and \subref{sfig:msd-d120} 120 $\mu m$ for increasing motor densities as indicated. These datasets were used to generate the mean MSD profiles in Figure \ref{fig:densmsd}.}
\label{distrMSD}
\end{center}
\end{figure}

\begin{figure}[ht]
\begin{center}
  \includegraphics[width=0.5\textwidth]{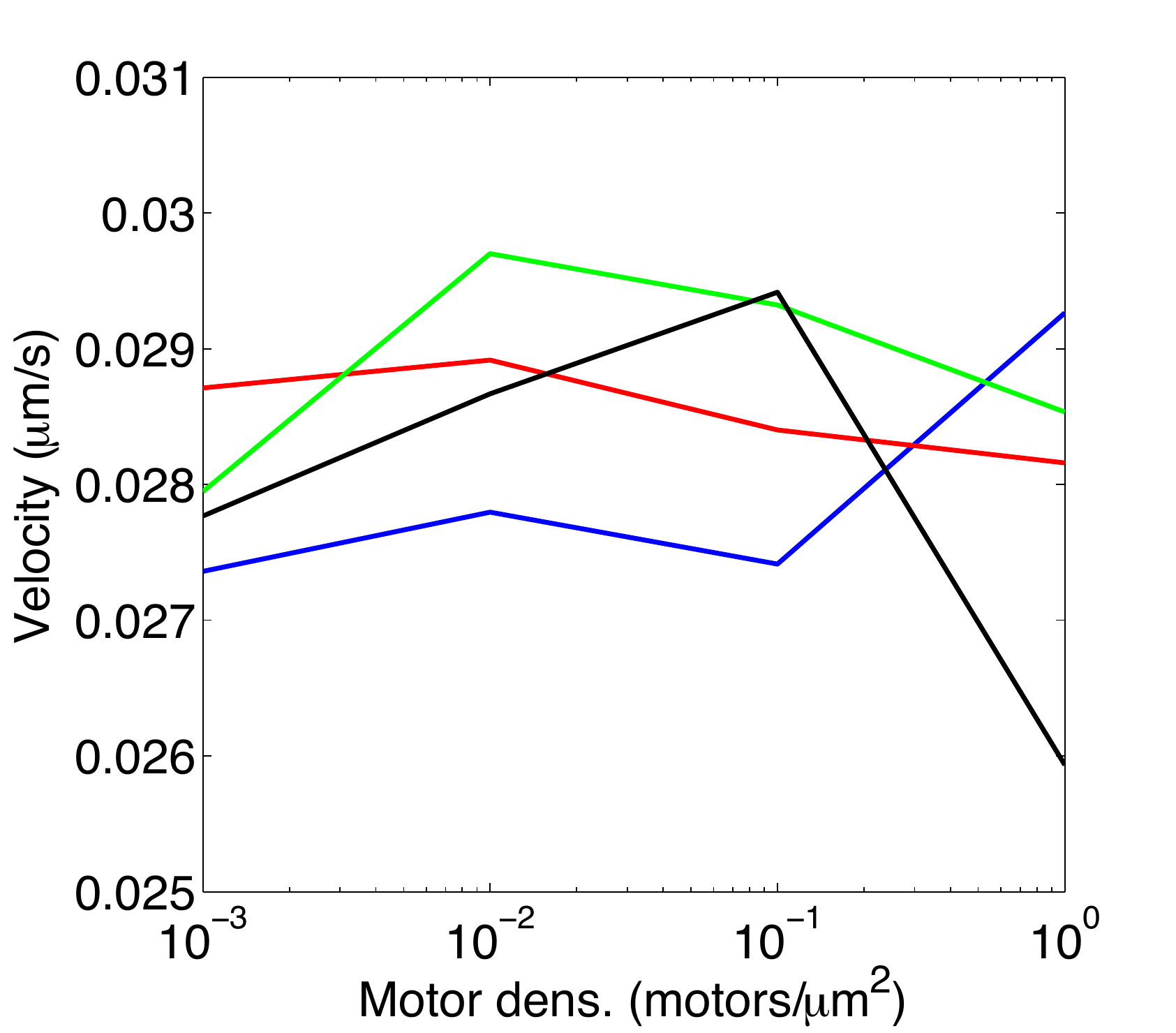} 
\caption{The mean velocities of the simulated asters initialized at 15 (blue), 30 (red), 45 (green) and 120 $\mu m$ (black) are plotted as a function of increasing motor densities.}
\label{motdens-vel}
\end{center}
\end{figure}

\begin{figure}[ht]
\begin{center}
	\includegraphics[width=0.4\textwidth]{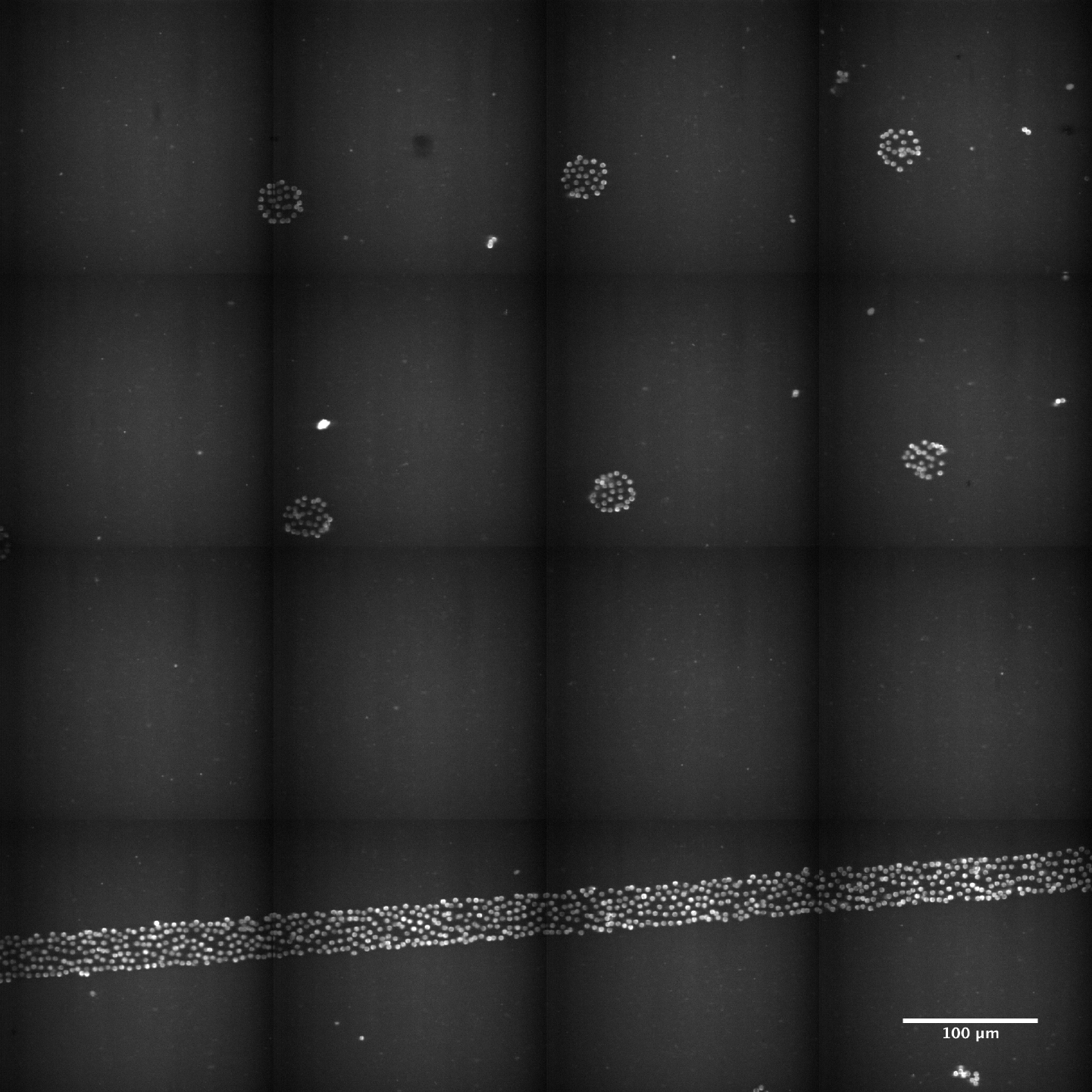} 
\caption{A single tiled (4x4) image of the Hoechst labelled DNA micro-patterns was acquired before imaging the microtubules. The scale-bar corresponds to $100 \mu m$}
\label{sfig:dnaPatt}
\end{center}
\end{figure}

\clearpage
\newpage
\section*{Supporting Videos}
\setcounter{figure}{0}
\makeatletter 
\renewcommand{\thefigure}{SV\@arabic\c@figure} 
\renewcommand{\figurename}{Video }

\begin{figure}[ht]
\begin{center}
      \includegraphics[width=0.4\textwidth]{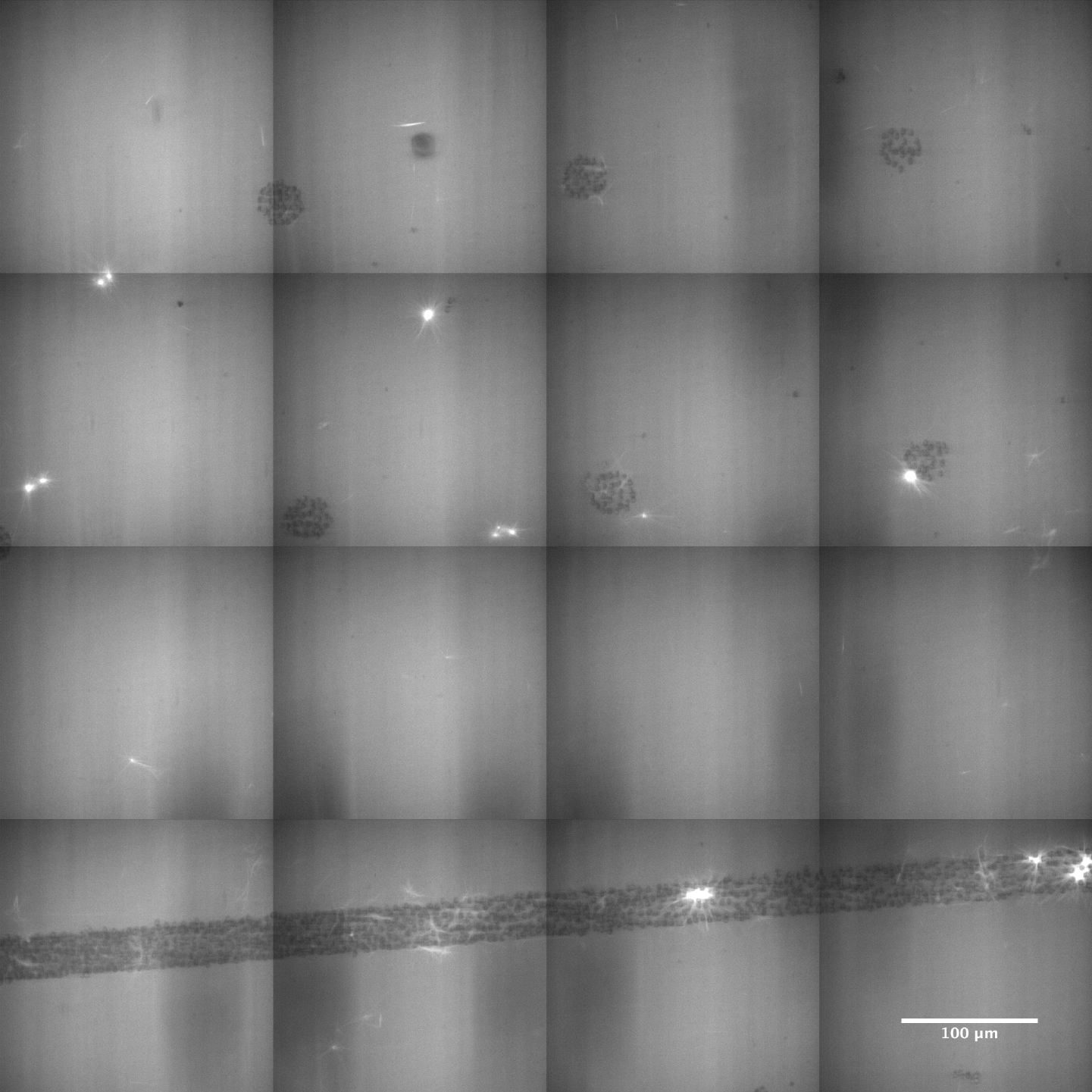} 
   \end{center}
\caption{A time-series of 4x4 tiled images (dt = 117 s, 35 frames) was acquired. The tubulin was labelled with Cy3-Tubulin in centrosome containing {\it Xenopus} extract. The dark regions correspond to DNA patterns (Figure \ref{sfig:dnaPatt}). Scale-bar = $100 \mu m$.}
\label{sup:asterdna}
\end{figure}

\begin{figure}[ht]
\begin{center}
      \subfigure[]{  \label{sim:1}        \includegraphics[width=0.4\textwidth]{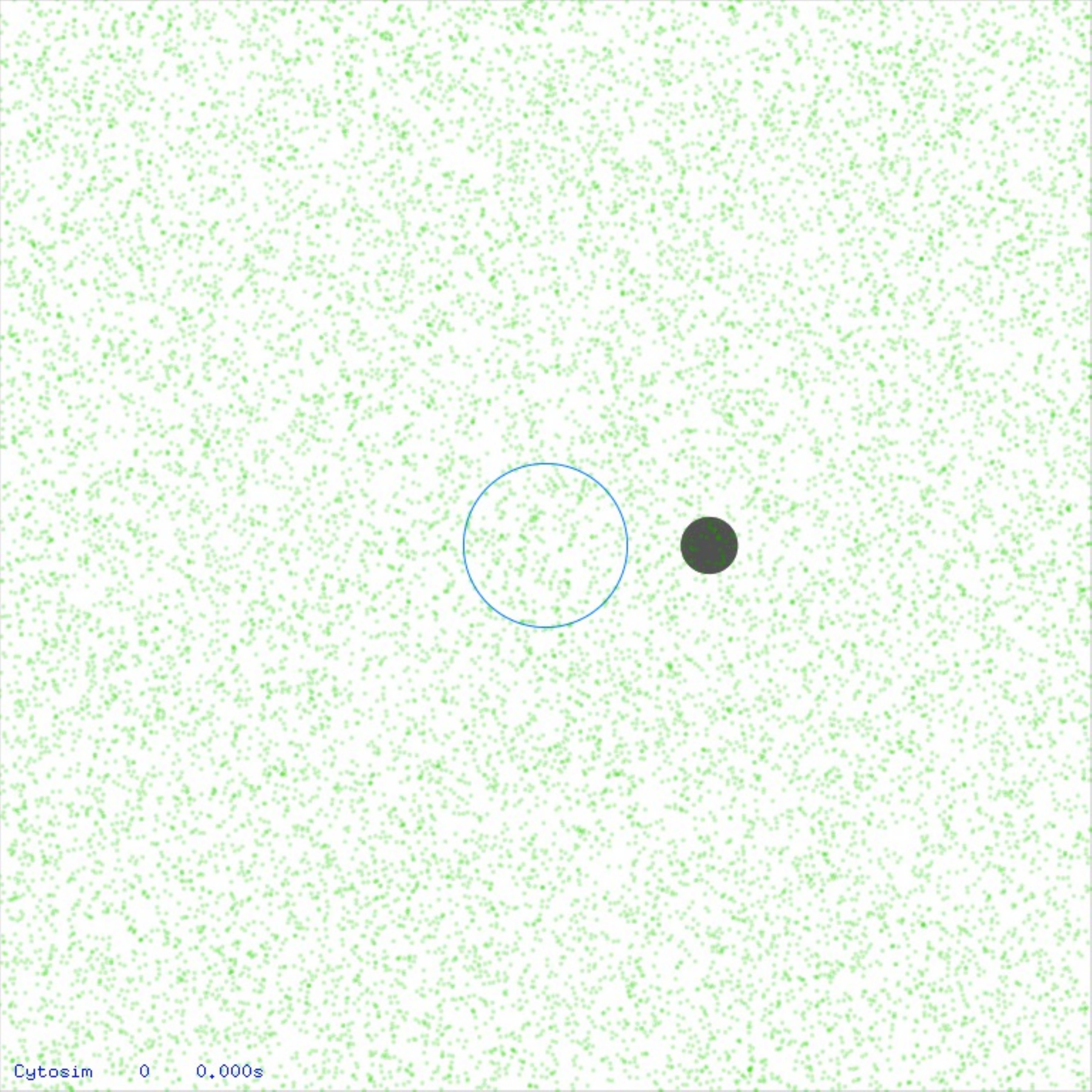} }
       \subfigure[]{  \label{sim:2}        \includegraphics[width=0.4\textwidth]{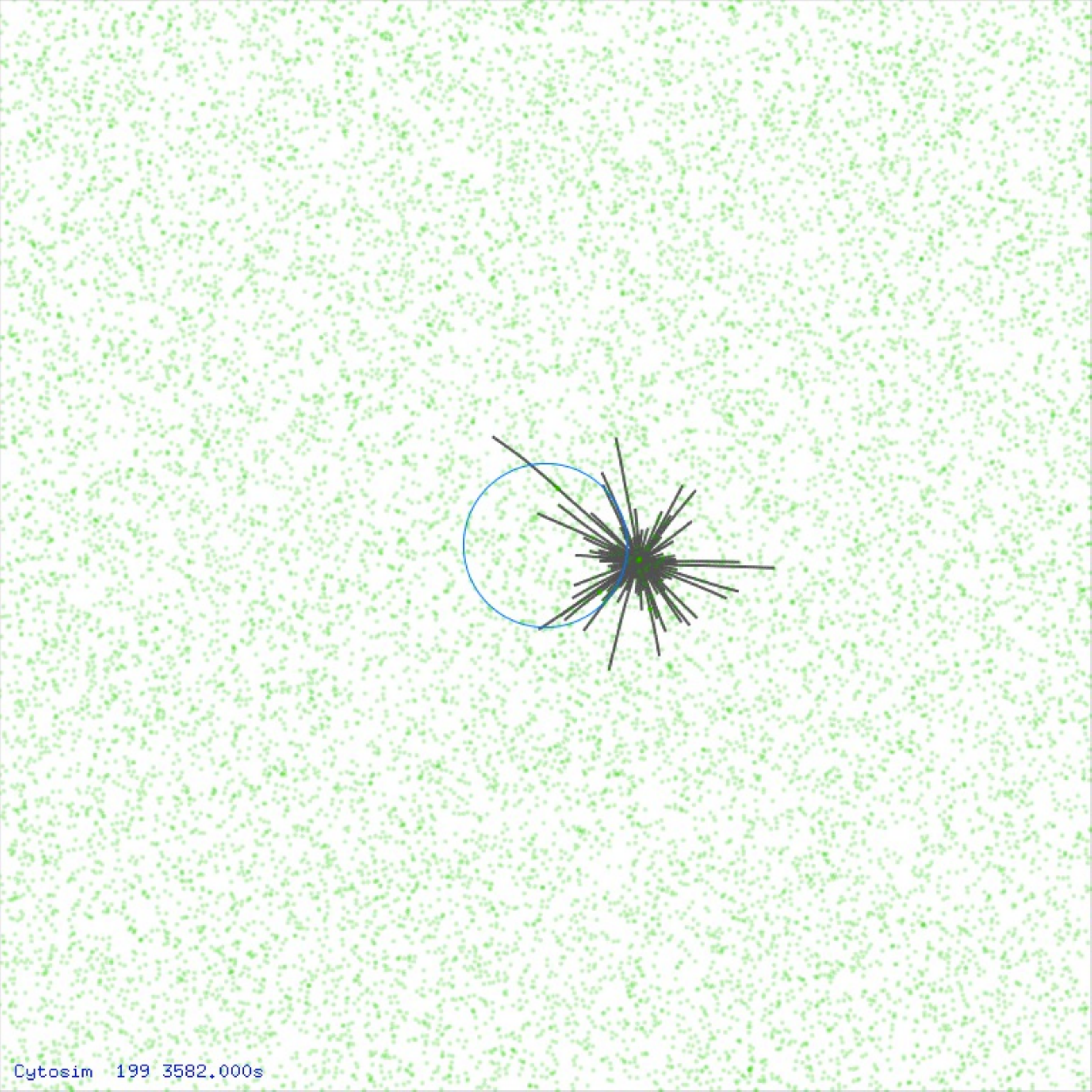} }
   \end{center}
\caption{Snapshots of the simulation at time \subref{sim:1} 0 s and \subref{sim:2} 59.7 min. The blue circle is the DNA patch (diameter $30 \mu m$) , the green dots are the minus-ended motors and the gray lines are microtubules of the aster.}
\label{sup:simvid}
\end{figure}

\end{document}